\documentclass{jfm}
\usepackage{graphicx}
\usepackage{epstopdf, epsfig}
\usepackage[caption=false]{subfig} 
\usepackage{braket}
\usepackage{amsmath}
\usepackage{soul}

\shorttitle{Linear stability of 3D baroclinic vortices}

\shortauthor{Mahdinia et al.}

\title{Stability of 3D Gaussian vortices in an unbounded, rotating, vertically-stratified, Boussinesq flow: Linear analysis}

\author{Mani Mahdinia\aff{1}, Pedram Hassanzadeh\aff{2,3}, Philip S. Marcus\aff{1}\corresp{\email{pmarcus@me.berkeley.edu}}, 
 \and Chung-Hsiang Jiang\aff{1}}

\affiliation{\aff{1}Department of Mechanical Engineering, University of California, Berkeley, CA 94720, USA  
\aff{2}Department of Mechanical Engineering, Rice University, Houston, TX 77005, USA
\aff{3}Center for the Environment and Department of Earth and Planetary Sciences, Harvard University, Cambridge, MA 02138, USA}

\begin{document}

\maketitle

\begin{abstract}
The linear stability of three-dimensional (3D) vortices in rotating, stratified flows has been studied by analyzing the non-hydrostatic inviscid Boussinesq equations. We have focused on a widely-used model of geophysical and astrophysical vortices, which assumes an axisymmetric Gaussian structure for pressure anomalies in the horizontal and vertical directions. For a range of Rossby number ($-0.5 < Ro < 0.5$) and Burger number ($0.02 < Bu < 2.3$)
relevant to observed long-lived vortices, the growth rate and spatial structure of the most unstable eigenmodes have been numerically calculated and presented as a function of $Ro-Bu$. We have found neutrally-stable vortices only over a small region of the $Ro-Bu$ parameter space: cyclones with $Ro \sim 0.02-0.05$ and $Bu \sim 0.85-0.95$. However, we have also found that anticyclones in general have slower growth rates compared to cyclones. In particular, the growth rate of the most unstable eigenmode for anticyclones in a large region of the parameter space (e.g., $Ro<0$ and $0.5 \lesssim Bu \lesssim 1.3$) is slower than $50$ turn-around times of the vortex (which often corresponds to several years for ocean eddies). For cyclones, the region with such slow growth rates is confined to $0<Ro<0.1$ and $0.5 \lesssim Bu \lesssim 1.3$. While most calculations have been done for $f/\bar{N}=0.1$ (where $f$ and $\bar{N}$ are the Coriolis and background Brunt-V\"ais\"al\"a frequencies), we have numerically verified and explained analytically,  using non-dimensionalized equations, the insensitivity of the results to reducing $f/\bar{N}$ to the more ocean-relevant value of $0.01$. The results of our stability analysis of Gaussian
vortices both support and contradict findings of earlier studies with QG or multi-layer models or with other families of vortices. 
The results of this paper  provide a steppingstone to study the more complicated problems of the stability of geophysical (e.g., those in the atmospheres of giant planets) and astrophysical vortices (in accretion disks).                 
\end{abstract}


\section{Introduction} \label{sec:introduction}
Coherent vortices are prominent features of geophysical and astrophysical turbulent flows. Examples include the oceanic vortices such as Gulf Stream rings \citep{olson1991rings} and Mediterranean eddies (Meddies) \citep{McWilliams1985,armi1988history} and similar vortices in other regions including the  Ulleung Basin, Red Sea, and bay of Biscay \citep{meschanov1998young,carton2001hydrodynamical,chang2004circulation}, as well as vortices in the atmosphere of gas giants such as Jupiter and Saturn \citep{marcus1993jupiter,vasavada2005jovian,o2015polar}, extreme-weather-causing blocking anticyclones in the Earth's atmosphere \citep{tyrlis2008aspects,hassanzadeh2014responses,hassanzadeh2015blocking}, and vortices in the protoplanetary disks where stars and planets form \citep{barge1995did,barranco2005three,marcus2013three}. Understanding the dynamics of these vortices, such as their formation, longevity, and stability, are of great interest as these vortices can strongly affect their surroundings, for example by efficiently mixing and transporting heat, momentum, and material \citep{gascard2002long,marcus2004prediction,Dong2014,marcuszombie}. Despite their widely different environments and time and length scales, a common aspect of these vortices is that their dynamics are predominantly controlled by the rotation, stratification, and (in some cases) shear of their environment.                   
 
The  linear and nonlinear (i.e., finite-amplitude) stability of vortices in rotating, stratified flows has been extensively studied in the past $30$ years. However,
the majority of those studies have used idealized models for the vortices or for the governing equations. For example,
\citet{ikeda1981instability}, \citet{helfrich1988finite}, and \citet{benilov2005stability} studied quasi-geostrophic
(QG) vortices in discrete two-layer flows;
\citet{Gent1986} studied columnar (i.e., with no variation in the vertical direction) QG vortices; \citet{Flierl1988} examined columnar and 3D QG vortices; \citet{Nguyen2012} studied 3D QG vortices;
\citet{carton89} investigated one and two-layer QG vortices; \citet{Dewar1995}, \citet{killworth1997primitive}, \citet{Dewar1999}, \citet{Baey2002}, \citet{benilov2004}, \citet{benilov2005thin}, \citet{benilov2008effect}, \citet{lahaye2015centrifugal}, and \citet{benilov1998} examined
two-layer ageostrophic vortices (the latter also studied geostrophic vortices); \citet{Katsman2003} examined multi-layer ageostrophic vortices; \citet{Smyth1998}, \citet{Billant2006}, and \citet{yim2015mechanism} studied columnar ageostrophic vortices; \citet{Stegner2000} examined
shallow-water ageostrophic vortices; 
\citet{lazar2013inertial2,lazar2013inertial} studied shallow-water inertially-unstable vortices; \citet{Sutyrin2015} examined two and three-layer ageostrophic vortices; \citet{Suzuki2012} investigated the evolution of 3D ageostrophic vortices (but this was not technically a stability study
because the initial vortices were created through geostrophic adjustment and thus  out-of-equilibrium); and \citet{Tsang2015} also studied the evolution, rather than the stability, of 3D ageostrophic vortices made from  piecewise-constant elements of potential vorticity that were
not exact equilibrium solutions of their equations of motion. One study focused on 3D equilibrium vortices using the full 3D Boussinesq equation is that of \citet{Yim2016} who examined the linear stability of a specific family of vortices with Gaussian angular velocity.

Two of the main motivations for some of the studies listed above have been (a) the observed stability of the long-lived, approximately axisymmetric vortices  in the oceans, and (b) the observed cyclone-anticyclone asymmetry in the oceans and planetary atmospheres. It has been observed through tracking individual vortices and by satellite observations  that coherent oceanic vortices with radii of tens to hundreds of kilometers can last for months and even years ($\sim1/2\,\text{-}\,3$) while remaining nearly axisymmetric \citep{lai1977distribution,Armi1989,olson1991rings,Chelton2011}. However, most theoretical studies of axisymmetric vortices in rotating stratified flows have found them to be linearly unstable (usually with fast growth rates that are incompatible with the observed longevity of these vortices), unless unrealistic parameters or vertical structures are assumed \citep[see the discussions in][]{Stegner2000,benilov2004,benilov2005stability,Sutyrin2015}. Observations of planetary atmospheres \citep{mac1986merging,cho1996morphogenesis}, and oceans at the mesoscales \citep{McWilliams1985,Chelton2007,Chelton2011,Mkhinini2014} show that long-lived vortices are predominantly anticyclones. Whether this asymmetry is due to differences between the stability (linear or nonlinear) properties of cyclones and anticyclones  requires a better understanding of how stability changes with the Rossby number. It should be noted that factors other than stability can be responsible for,  or at least contribute to, the observed cyclone-anticyclone asymmetry; for example the creation mechanisms might favor anticyclones \citep{perret2011large}, anticyclones might decay slower than cyclones (\citeauthor{hoskins1985use} \citeyear{hoskins1985use}, section 7; \citeauthor{Graves2006} \citeyear{Graves2006}), or coherent cyclones might be harder to observe in planetary atmospheres than anticyclones \citep{marcus2004prediction}.                                        

While valuable information on the stability of vortices in rotating stratified flows, vortices in planetary atmospheres, and oceanic eddies has been gained through the aforementioned studies, further investigation of the linear and nonlinear stability that extends beyond the simplifications and limitations of these studies is still needed. In the current study, we address the stability of isolated, 3D, axisymmetric vortices in rotating, stably-stratified, inviscid flows by analyzing the full non-hydrostatic Boussinesq equations with an $f$-plane approximation in a 3D domain with periodic boundary conditions (modified to simulate an unbounded flow). We focus on a widely-used model of geophysical and astrophysical vortices, which have pressure anomalies that are Gaussian in the radial and vertical directions and are in exact equilibrium.
\citep[e.g.,][]{McWilliams1985,van2009laboratory,Chelton2011,Hassanzadeh2012}. Our work extends the analyses of the previous studies in several ways, including:
\begin{enumerate}
\item By using the Boussinesq equations, we can study vortex dynamics with any Rossby number and internal stratification. Here we focus on cyclones and anticyclones in the geostrophic balance regime ($-0.5 < Ro < 0.5$), which is the range of $Ro$ relevant to most long-lived geophysical and astrophysical vortices \citep[e.g.,][]{olson1991rings,Aubert2012} (all parameters and dimensionless numbers are defined in \S\ref{sec:problem_formulation}). The vertical stratification inside the 3D equilibrium vortices that are studied here can be much stronger or much weaker compared to the stratification of the background (i.e., far from the vortex) flow, which is also the case for many oceanic and atmospheric vortices \citep[e.g.,][]{Aubert2012}. Considering vortices with finite Rossby numbers and with internal stratifications that significantly differ from the stratification of the background flow extends the stability analysis well beyond the QG approximation.                        
\item Geophysical and astrophysical vortices that are far from both horizontal  and vertical boundaries (e.g., free surfaces or solid surfaces) and that are in quasi-equilibrium have been observed to be three-dimensional (rather than 2D Taylor columns); examples include Jupiter's Great Red Spot \citep{marcus1993jupiter}, Meddies \citep{Aubert2012,bashmachnikov2015properties}, and zombie vortices in the protoplanetary disks \citep{barranco2005three,marcus2013three,marcuszombie}. The vertical length scales of these vortices are finite and usually much smaller than their horizontal length scales, which can be understood as a direct consequence of the gradient-wind balance \citep[see][]{Hassanzadeh2012}. The present study extends the rigorous stability analysis of Boussinesq vortices beyond barotropic Taylor columns.             
\item Exploiting the universal scaling law of \citet{Hassanzadeh2012} and \citet{Aubert2012}, the 3D baroclinic vortices studied here are exact equilibrium solutions of the full 3D non-hydrostatic Boussinesq equations (see \S\ref{subsec:initial_equilibria}). The exact equilibrium is particularly important for a rigorous linear analysis, which is the subject of this paper. 
\item By using the full, 3D, non-hydrostatic Boussinesq equations, we avoid restrictions on the vertical structure of the vortex or background flow that result from the QG or multi-layer models discussed above. Although here we focus on background flows with  stable stratification such that the density decreases linearly with height (i.e., constant Brunt-V\"{a}is\"{a}l\"{a} frequency $\bar{N}$), background flows with more realistic $\bar{N}(z)$ profiles can be easily included in this framework.               
\item The family of Gaussian vortices that is studied here has been shown to fit many types of oceanic and laboratory vortices reasonably well \citep[e.g.,][]{van2009laboratory,Chelton2011} and has been widely-used as a model in various theoretical studies \citep[e.g.,][]{McWilliams1985,Morel1997, Hassanzadeh2012,Negretti2013}. Furthermore in this model, all fields (e.g., velocity, potential vorticity, and density) are continuous and smooth, which eliminate unphysical instabilities that can arise from discontinuities (which are  present, for example, when vortices are modeled with piecewise-constant shells or patches of potential vorticity).
\end{enumerate} 

In this paper we address the linear stability of 3D vortices in rotating stratified flows and discuss the growth rates and most unstable eigenmodes as functions of the Rossby number $Ro$ (for $-0.5<Ro<0.5$), the Burger number $Bu$ (for $0.02<Bu<2.3$), and  $f/\bar{N} = 0.1$ and $0.01$. One of the main purposes of this paper is to extend the linear stability analysis of a specific family of 3D equilibrium vortices beyond some of the approximations or constraints imposed in previous studies and produce the parameter map of stability for 3D non-hydrostatic Boussinesq flows. We also investigate how different modes take over as the most unstable one as the Burger number changes and explore the vertical and horizontal structures of these modes and their critical layers. We discuss how the stability properties found here compare with those reported in other studies using QG or multi-layer equations or using a different vortex model. Furthermore, we show numerically that the linear stability of the family of  3D vortices that we examine is  only weakly dependent on the value of $f/\bar{N}$ for $f/\bar{N} \le 0.1$ and we discuss the reason behind this behavior. 


The results of this paper improve the understanding of the generic stability properties of 3D vortices in rotating stratified flows, and have implications for the dynamics of some of the geophysical and astrophysical vortices. These results are most relevant to the stability of interior (i.e., far from boundaries) oceanic vortices such as Meddies. It is acknowledged that the exclusion of horizontal and vertical background shear, free surface, lateral boundaries, bottom topography, compressible effects, and vertical variation of $\bar{N}$ limit the direct applicability of the current analysis to other oceanic eddies and planetary and astrophysical vortices. However, the numerical framework presented here can be readily adapted to account for the aforementioned boundary conditions/physical processes in future studies, and the results of this paper will be needed to evaluate the influence of these boundary conditions/processes on the stability properties of these vortices.

The remainder of this paper is structured as follows. The equations of motion, numerical method, Gaussian vortex model, and eigenmode solver are discussed in \S\ref{sec:problem_formulation}. The eigenmodes with critical layers are discussed in \S\ref{sec:criticallayers}, and the results of the linear stability analysis and the stability map along with comparison with previous studies are presented in \S\ref{sec:parameter_map_of_stability}. Insensitivity of the most unstable modes to $f/\bar{N}$ is discussed in \S\ref{sec:effect_of_foN} and the radial and vertical structures of the most unstable modes are presented in \S\ref{sec:spatial_distribution}. Discussion and summary are in \S\ref{sec:discussion}.    

\section{Problem formulation} \label{sec:problem_formulation}
\subsection{Equations of motion}  \label{subsec:eqns_of_motion}
The Boussinesq approximation of the equations of motion for 3D rotating, stratified, inviscid flows in the Cartesian coordinates $(x,y,z)$, as observed in a frame rotating with angular velocity $(f/2)\,\hat{\boldsymbol z}$, is \citep{vallis2006atmospheric}
\begin{eqnarray}
\frac{\textrm{D} \boldsymbol{v}}{\textrm{D}t} = -\frac{\nabla p}{\rho_o}+\boldsymbol{v}\times f\hat{\boldsymbol{z}} + b \, \hat{\boldsymbol{z}},~~~~~~\frac{\textrm{D} b}{\textrm{D}t} = - {\bar{N}^2} v_z,~~~~~~\bnabla\bcdot\boldsymbol{v} = 0, \label{eq:1}
\end{eqnarray}
where the operator $\textrm{D}/\textrm{D}t \equiv \partial / \partial t+ \boldsymbol{v}\bcdot\bnabla $ is the material derivative, $t$ denotes time, $\boldsymbol{v}=(v_x,v_y,v_z)$ is the 3D velocity vector, $f$ is the Coriolis frequency (constant in our study), and $g$ is the acceleration of gravity. The total pressure and the total density of the fluid are $p_{tot}\equiv\bar{p}(z)+p(x,y,z,t)$ and $\rho_{tot}\equiv\bar{\rho}(z)+\rho(x,y,z,t)$, where $\bar{\rho}(z=0) = \rho_o$. We define the buoyancy as $b(x, y, z, t) \equiv - g \rho/\rho_o$. Quantities with a bar are properties of the equilibrium background flow (i.e., far from the vortex where $\boldsymbol{v} \rightarrow 0$, $b \rightarrow 0$, $\rho \rightarrow 0$, and $p \rightarrow 0$). The background pressure $\bar{p}$ and density $\bar{\rho}$ are in hydrostatic balance $d \bar{p}/dz = -\bar{\rho}g$. The background Brunt-V\"{a}is\"{a}l\"{a} frequency $\bar{N}\equiv \sqrt{-(g/\rho_o)(d \bar{\rho}/dz)}$ is assumed to be constant, so that $\bar{\rho}(z) = \rho_o(1-\bar{N}^2z/g)$. 

In the above equations, we have ignored  viscosity in the momentum equations and diffusion in the density equation, which are reasonable approximations for atmospheric and oceanic flows. Furthermore, we have dropped the planetary centrifugal term from the momentum equations, assuming that the rotational Froude number $f^2d/g$ is small \citep{Barcilon1967}, where $d$ is the distance between the center of the vortex and the planetary rotation axis. 

\subsection{Numerical method} \label{subsec:numerical_method}
A pseudo-spectral initial-value solver is developed to solve (\ref{eq:1}) in a triply periodic domain with 256 or 512  Fourier modes in each direction. In numerical simulations of strongly rotating stratified flows, resolving the fast inertia-gravity waves can substantially limit the size of the time step $\Delta t$ and thus increase the computational cost. Here we use the  semi-analytic method developed by \citet{Barranco2006} for rotating stratified flows, which enables us to accurately and efficiently deal with large $f\Delta t$ and $\bar{N} \Delta t$. 

A vortex in the middle of a periodic domain interacts with its periodic images. To minimize this interaction and its potential impact on the stability of the vortex (and to simulate having an unbounded flow) the computational domain size is chosen to be large compared to the vortex size: the domain size in the $x$ and $y$ directions, i.e.,
the values of $L_x$ and $L_y$ are 7.5 (or more often 15) times larger than the initial vortex diameter  $(2L)$, and, similarly, the domain size in the $z$ direction $L_z$ is
7.5 (or more often 15) times larger than the initial vortex height $(2H)$.
There are two reasons for sometimes making the domain size very large. First, we wanted to ensure that the periodic boundary conditions had no perceptible effects on the flow dynamics; secondly, in the follow-up paper to this one (see our Discussion \S\ref{sec:discussion}) unstable vortices often fragmented with pieces of the initial vortices becoming widely separated so that the calculations  required a large domain. To help simulate an unbounded flow, we also added
a cylindrical  sponge layer  near the boundaries of the computational domain (see Appendix~\ref{appA}). The sponge layer, implemented as Rayleigh drag and Newtonian cooling in (\ref{eq:1}), damps $\boldsymbol{v}$ and $\rho$ outside a cylindrical surface of diameter $24L$ and height $24H$
(for the large domain calculations) or $12L$ and height $12H$ (for the small domain calculations) around the center of the domain. Another advantage of adding the sponge layer is that it damps the reflection of the outgoing inertia-gravity waves, and occasional detached filaments back to the domain at the periodic boundaries. One more advantage of the axisymmetric sponge layer is that we find that it prevents the (non-axisymmetric) periodic boundary conditions in $x$ and $y$ from adding any significant non-axisymmetric perturbations to the initial vortices. The latter is important when computing the stability of the vortices. One way of determining if the domain size is too small is to compute the ratio of the magnitude of each component of
the velocity and density of a numerically computed eigenmode at a damped location just inside the sponge layer to the maximum value of that component over the entire
domain. With the domain sizes presented here, that ratio is always of order $10^{-4}$ (or smaller), but the ratio increases to values with orders as large as $10^{-2}$ when the computational domain is reduced to
$(10 L)  \times  (10 L)  \times  (10H)$ and a sponge layer with diameter of $8L$ and height $8H$. 

Hyperviscosities and hyperdiffusivities are added to our otherwise inviscid and non-diffusive calculations to stabilize the code. See \citet{Barranco2006} for more details. 

\subsection{Initial equilibria: Gaussian vortices} \label{subsec:initial_equilibria}
In this study we focus on 3D axisymmetric baroclinic vortices that are initially in horizontal cyclo-geostrophic balance and vertical hydrostatic balance, and hence they are in gradient-wind balance \citep{vallis2006atmospheric}. The initial vortex is centered at  $r=0$ and $z=0$, where $r$ denotes the radial coordinate. A widely-used model for oceanic and laboratory vortices is that of an axisymmetric vortex with a Gaussian pressure distribution \citep[e.g.,][]{McWilliams1985}
\begin{eqnarray}
p = p_o \, \chi(r,z), \label{eq:2}
\end{eqnarray}
where $\chi(r,z)\equiv\exp{[-(r/L)^2-(z/H)^2]}$. Using (\ref{eq:2}) and the definitions presented in \S\ref{subsec:eqns_of_motion}, an exact, steady, axisymmetric equilibrium solution to the Boussinesq equations in (\ref{eq:1}) is the vortex
\begin{eqnarray}
v_{\phi}(r, z)  &=& \frac{fr}{2}\left(-1 + \sqrt{1 - \left( 8 p_o \chi(r, z) \right)/\left( \rho_o f^2L^2 \right)}\right),~~~~v_r = v_z = 0, \label{eq:3} \\
b(r, z)  &=& - {\frac{2 p_o z}{\rho_o H^2}} \, \chi(r,z), \label{eq:4} 
\end{eqnarray}
where the cylindrical coordinate is used for convenience ($v_{\phi}$ is the azimuthal velocity). For any vortex, whether or not it is Gaussian, we shall define a quantity written with a subscript ``c'' to mean that the quantity is to be evaluated at the vortex center, so $N_c$ is the Brunt-V\"{a}is\"{a}l\"{a} frequency at the center of a vortex, or $N_c^2 \equiv \bar{N}^2 + (\partial b/\partial z)_c$. For the Gaussian vortex described by (\ref{eq:3})-(\ref{eq:4}),
\begin{eqnarray}
N_c^2 =\bar{N}^2 - 2p_o/(\rho_o\,H^2). 
\label{eq:Nc}
\end{eqnarray}
As discussed in the next section, for some values of $p_o$,  $N_c^2 <0$, which means that the density distribution is locally unstable at the vortex center with heavy fluid over light fluid (i.e., statically unstable). It is convenient to define the Rossby number $Ro$,
which by definition has $Ro>0$ for a cyclone and $Ro<0$ for an anticyclone,  in terms of the maximum (or minimum) value of a vortex's vertical vorticity $\omega_E$, such that $Ro \equiv \omega_E/2f$. For the Gaussian vortices described above $\omega_E = \omega_c$, and 
\begin{equation}
Ro = \omega(r=0, z=0)/2f = - 1/2 + \sqrt{1/4 - 2 p_o/(\rho_o f^2 L^2)}. \label{Rossby}
\end{equation}
Note that the Gaussian vortex has an aspect ratio of 
\begin{equation}
\Biggl({\frac{H}{L}}\Biggr)^2 = \frac{-Ro(1+Ro) f^2}{\bar{N}^2 [1-(N_c/\bar{N})^2]}, \label{eq:slaw}
\end{equation}
in accord with the universal scaling law of \citet{Hassanzadeh2012} and \citet{      Aubert2012}, which is valid for all vortices that are in cyclo-geostrophic and hydrostatic balance. This can be seen by simply replacing $2 p_o/\rho_o$ in (\ref{Rossby}) with $H^2(\bar{N}^2 - N_c^2)$ using (\ref{eq:Nc}), and then solving for $H/L$.  

The three independent dimensional parameters in the governing equations (\ref{eq:1}) are $f$, $\bar{N}$, and $\rho_o$. The sizes of the computational domain
$L_x \times  L_y \times L_z$ have no effect (on the dimensional analysis), due to the fact that the cylindrical sponge layer is far from the vortices, and that the net circulations of the flow  are zero, which makes  the velocity due to the vortices fall off exponentially fast  and be effectively zero at the sponge layer. (See the definition of {\it shielded} below and in the appendices.) The equilibrium Gaussian vortices in (\ref{eq:2})-(\ref{eq:4}) introduce three additional dimensional parameters $H$, $L$, and $p_o$. Thus, there are three independent, dimensionless parameters that describe the dynamics of Gaussian vortices. The choice of these parameters is not unique, but in this paper we choose $Ro$, $f/\bar{N}$, and  
\begin{equation}
Bu \equiv \left(\frac{\bar{N}}{f} \frac{H}{L}\right)^2 = (L_r/L)^2, \label{eq:Burger} 
\end{equation}
where the latter is the Burger number, and $L_r \equiv H \bar{N}/f$ is the deformation radius. It should be noted that whether the vortices studied here are {\it big} or {\it small} depends on the inverse of their Burger number, which is the square of the vortex radius over $L_r$. Big vortices have small $Bu$, and vice versa.

\begin{figure}
\centering
    \subfloat[\small{(a)}]{\includegraphics[trim={6mm 6mm 32mm 2mm},clip,width=0.56\textwidth]{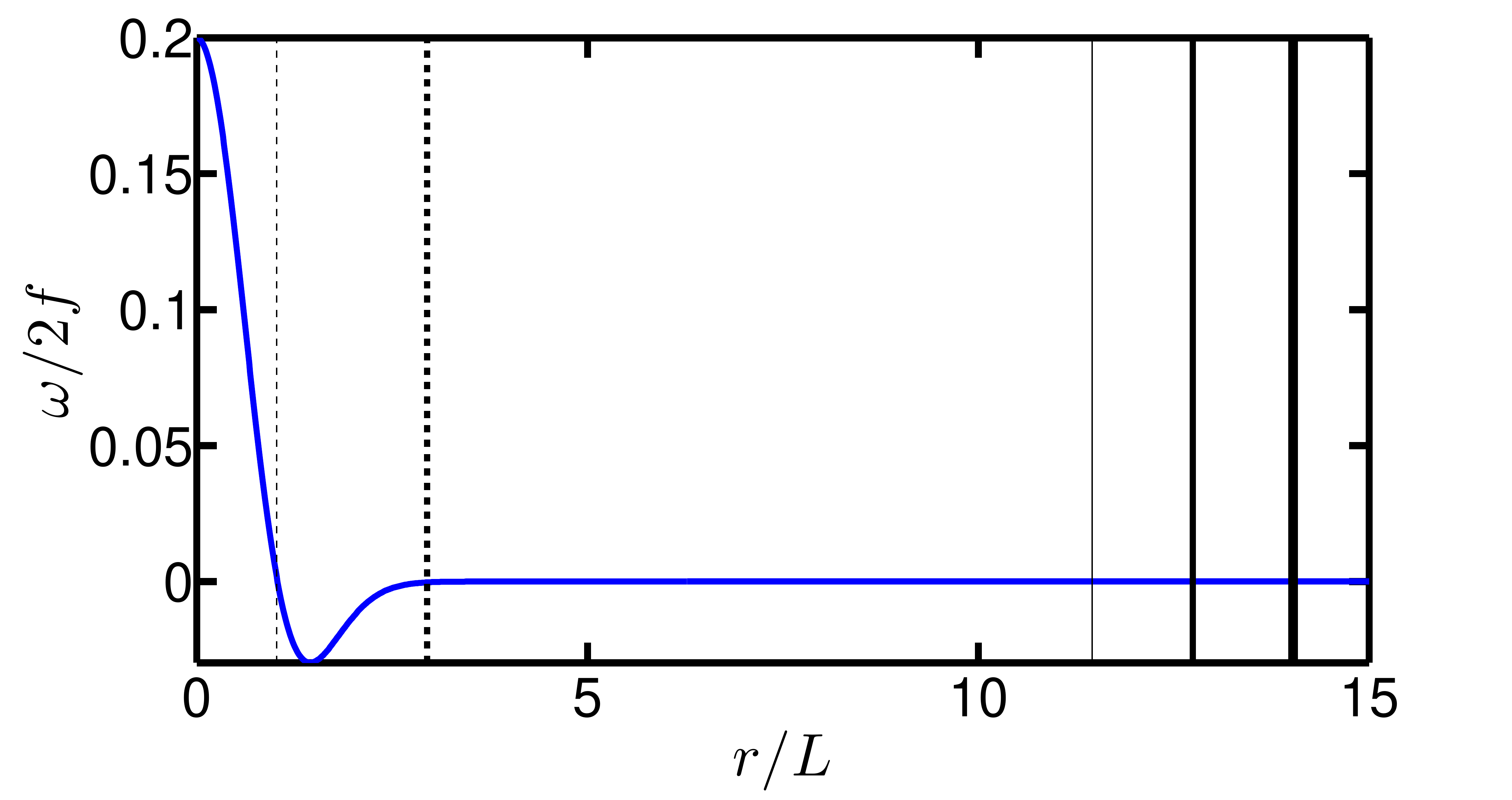}}
    $~~$\subfloat[\small{(b)}]{\includegraphics[trim={96mm 4.5mm 80mm -19mm},clip,width=0.369\textwidth]{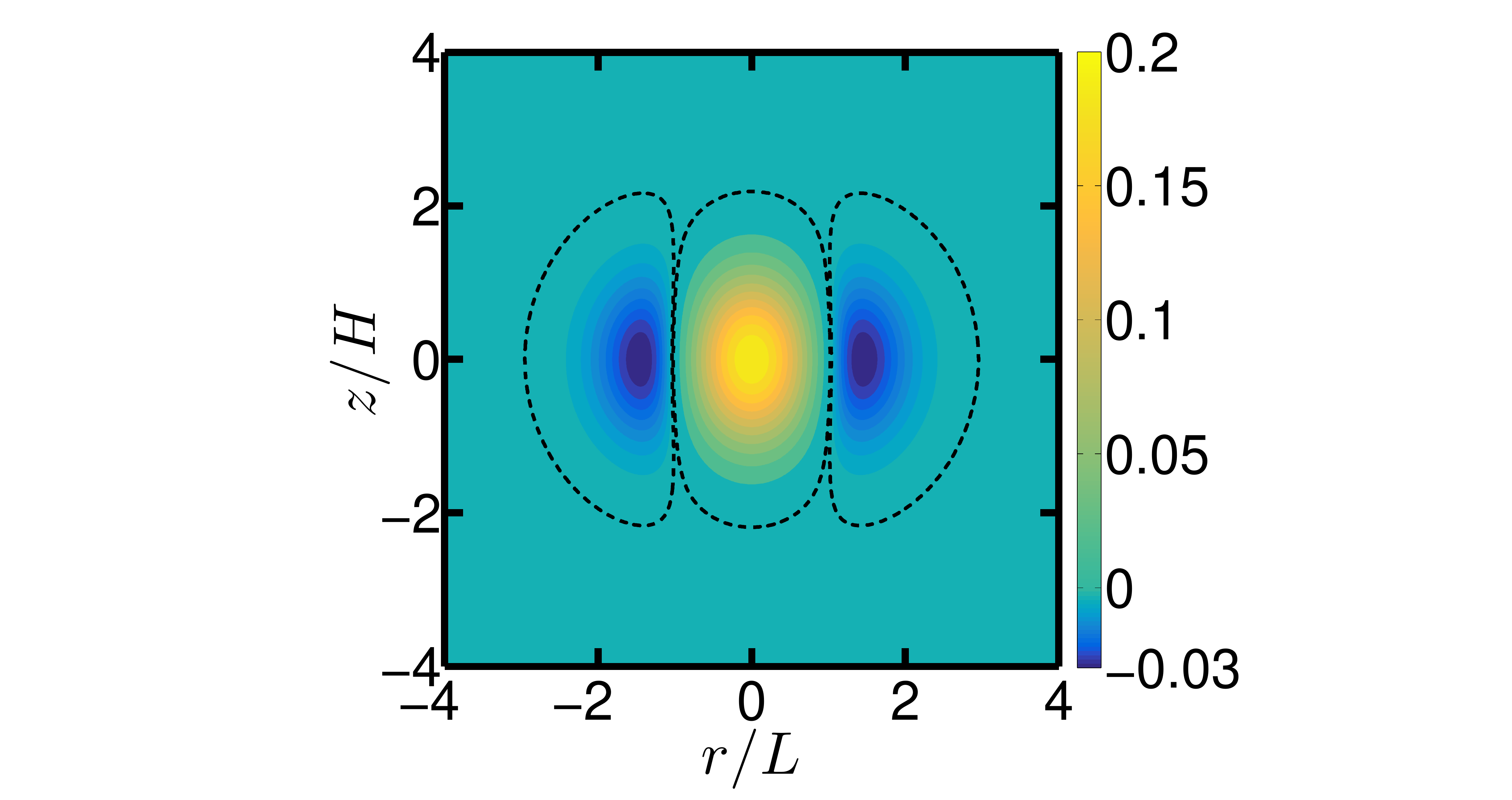}}\\
    \subfloat[\small{(c)}]{\includegraphics[trim={92mm 6mm 68mm  8mm},clip,width=0.34\textwidth]{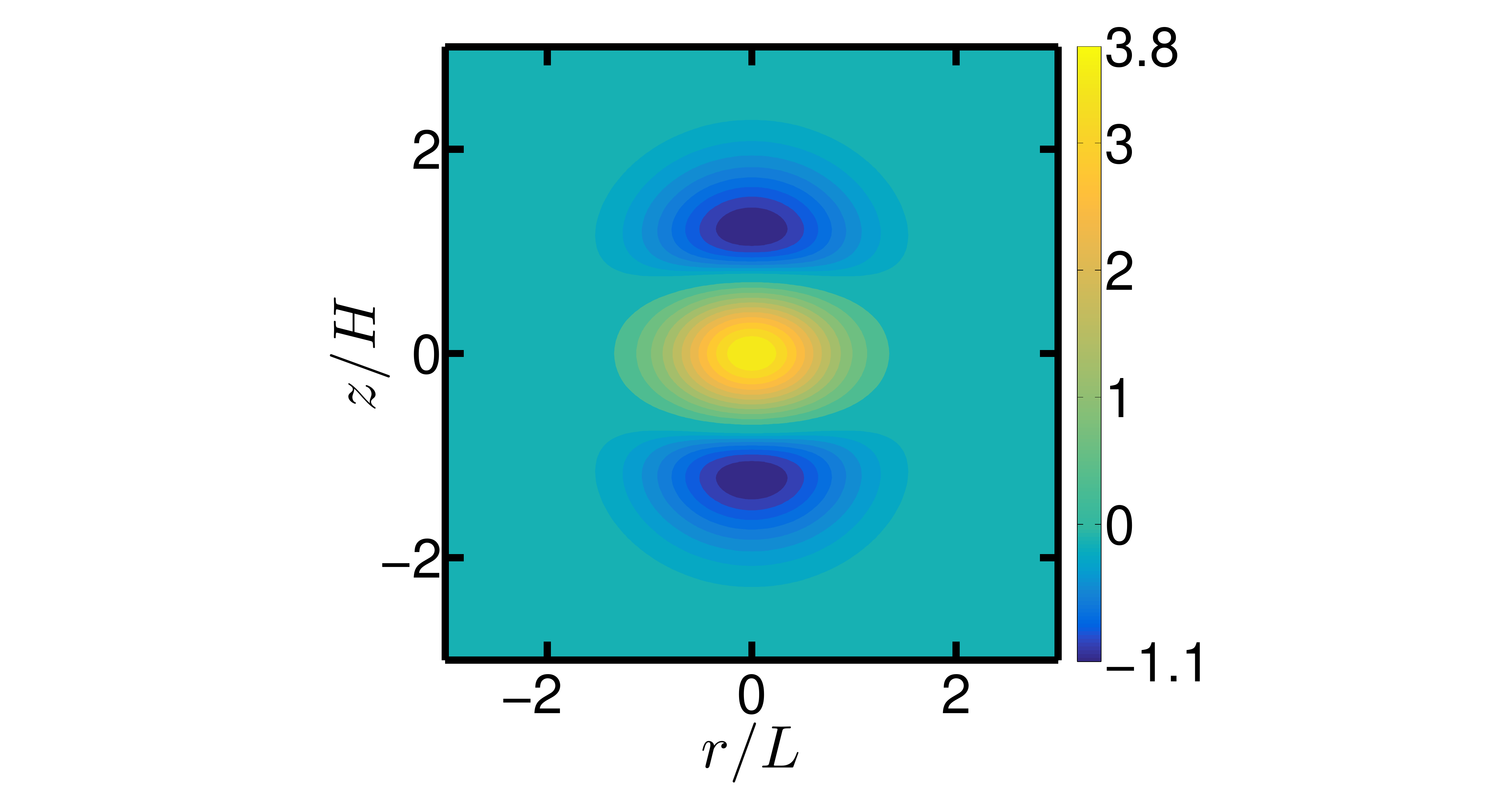}}
    \subfloat[\small{(d)}]{\includegraphics[trim={92mm 6mm 68mm  8mm},clip,width=0.34\textwidth]{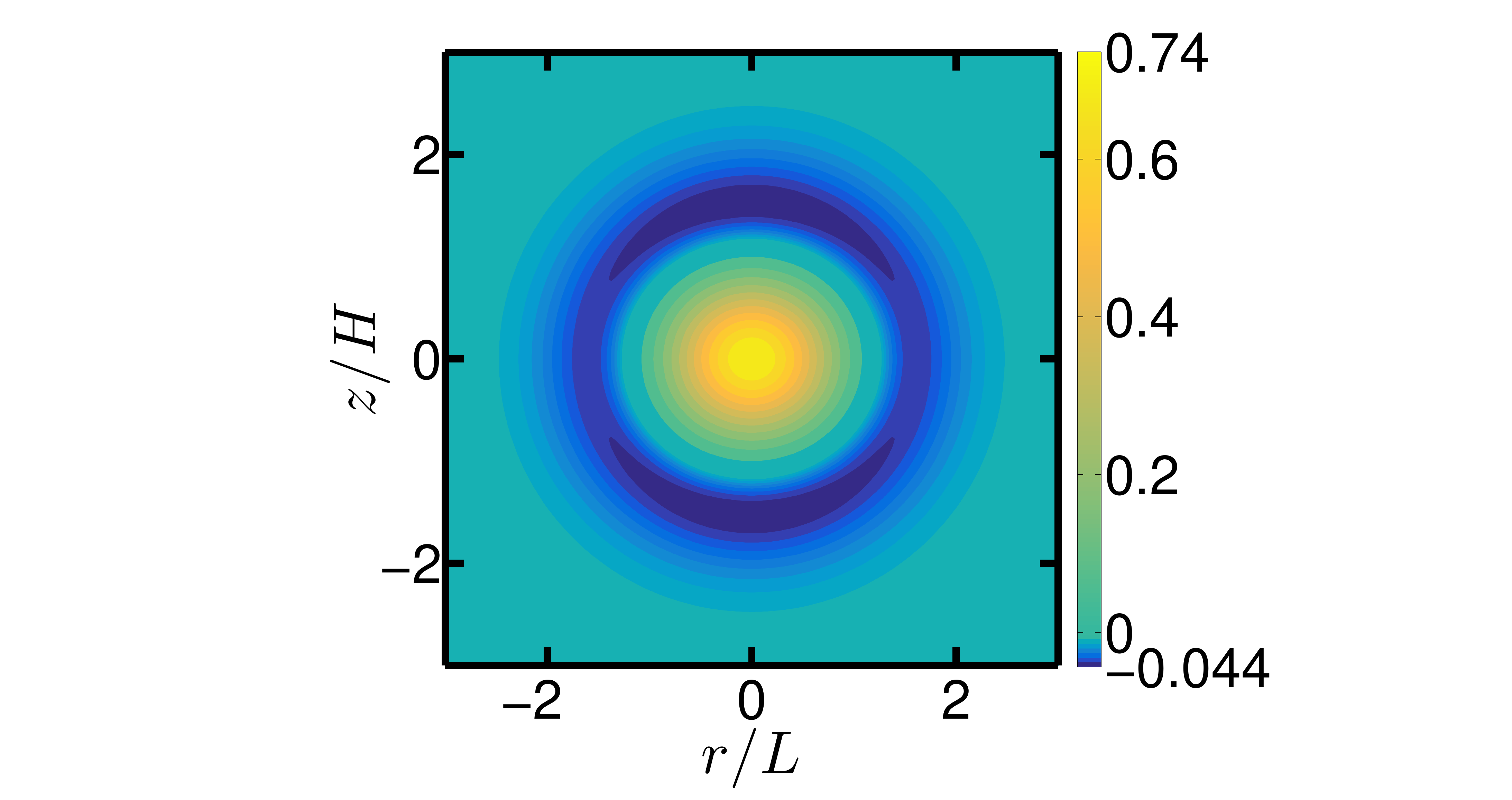}}
    \subfloat[\small{(e)}]{\includegraphics[trim={92mm 6mm 68mm  8mm},clip,width=0.34\textwidth]{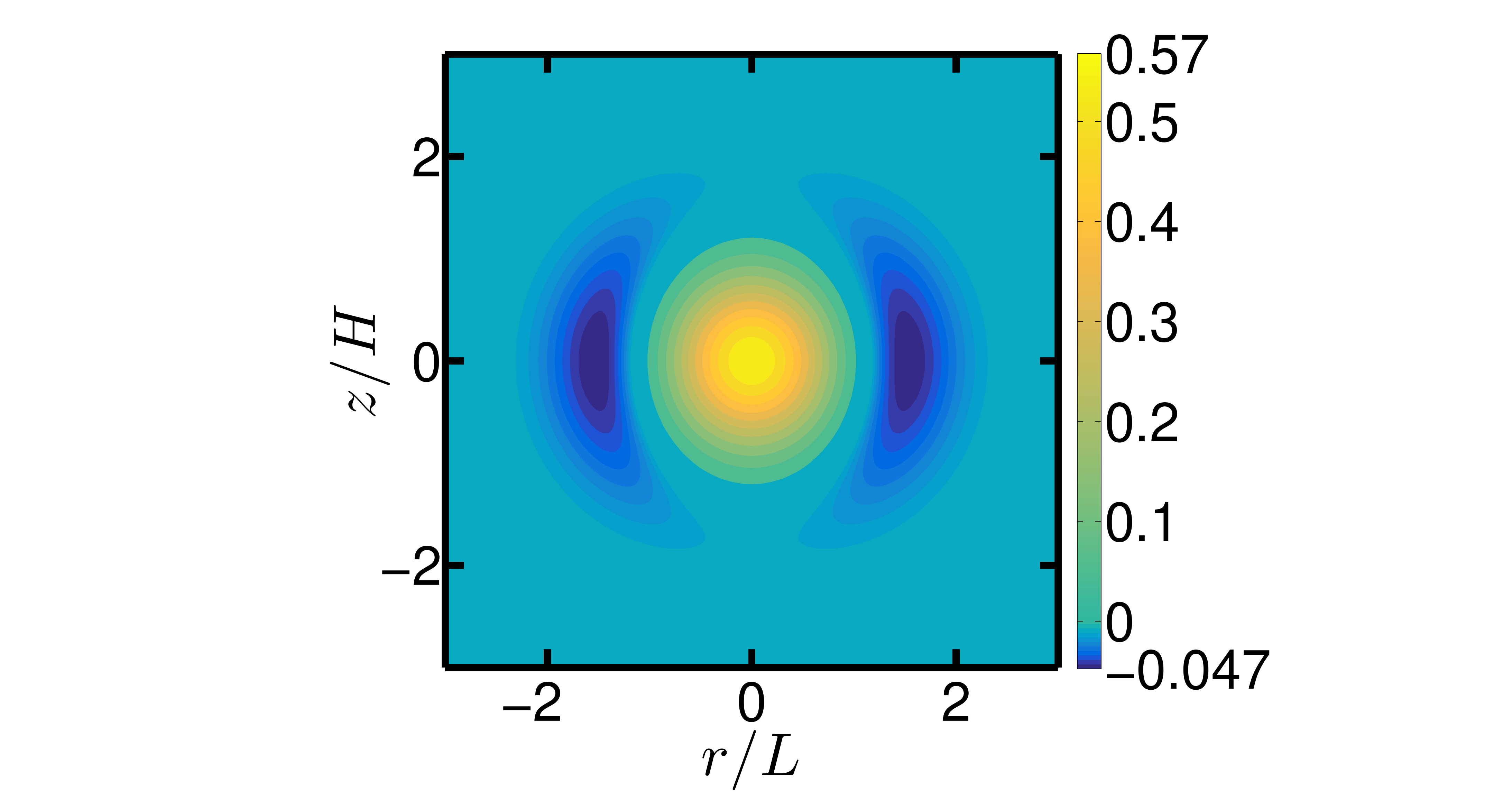}}
    \caption{(Colour online) Vertical vorticity $\omega(r, z)$  and potential vorticity $Q(r, z)$ for Gaussian cyclones defined by (\ref{eq:2})-(\ref{eq:4}). Panel (a) shows $\omega$ at $z=0$ as a function of $r$ as a solid curve (blue, in colour) for $Ro=0.2$, $Bu=0.1$, and $f/\bar{N} = 0.1$. The thin and thick dashed vertical lines show the boundaries of the core and the shield (see Appendix~\ref{appB}). The solid vertical lines at large radii show, with increasing thickness from left to right, where the boundary damping function $f_{bd}$ (see Appendix~\ref{appA}) reaches values of $0.01$, $0.5$ and $0.99$, respectively. Panel (b) shows $\omega(r, z)$ in units of $2f$ in the $r$-$z$ plane of the vortex in panel (a). Dashed lines indicate the boundaries of the core and shield. Panels (c) to (e) show the potential vorticity $Q(r, z)$ for a Gaussian vortex  with $Ro=0.2$ and $f/\bar{N} = 0.1$, for $Bu=0.1$, $Bu=1.0$ and $Bu=2.0$, respectively. For larger values of $Bu$, the distributions of $Q$ and $\omega$ are similar.} 
    \label{f1}
\end{figure}

The Gaussian vortices defined in the above model are {\it shielded}. Here we define a {\it shielded} flow as one in which the circulation computed
with the $z$-component of the vorticity over the entire $(x, y)$-plane for any fixed value of $z$ is zero. In addition,
the circulation computed with the $x$-component of the vorticity over the $(y, z)$-plane for any fixed value of $x$ is zero; and the circulation computed
with the $y$-component of the vorticity over the
$(x, z)$-plane for any fixed value of $y$ is zero. ({\it n.b.} Figure~\ref{f1}(b)  does not violate our definition of {\it shielded} because
the figure shows the vertical component of the
vorticity in an $x$-$z$ plane, not an $x$-$y$ plane.) Our governing equations and boundary conditions show that if the initial flow is shielded, then the flow is shielded for all time. In practical terms, a shielded isolated vortex is one in which the central core of the vortex is
surrounded, or partially surrounded,  by a region (shield) of opposite vorticity and that the circulation quickly vanishes outside the shield. For an arbitrary (i.e., not necessarily Gaussian) cyclonic vortex, the {\it core} of a cyclone is a contiguous region at and near the vortex center where the vertical component of its vorticity $\omega$ is greater than or equal to zero. The {\it shield} is a region around the core (usually looking like a shell or annular ring) located not too far from the core, where 
$\omega < 0$. The precise definitions that we use for {\it core} and
{\it shield} are in  Appendix~B. The  core and shield of an example  Gaussian vortex are illustrated in figures~\ref{f1}(a)~and~(b).
The definitions of the core and shield of an anticyclone are analogous to those of the cyclone.
For Gaussian vortices and many other types of shielded cyclones, outside the shield the amplitude of the vorticity decays exponentially with the radial distance $r$ (or $r^p$ with $p \ge 2$) from the vortex center.  
In our calculations, the circulation due to the vertical component of the vorticity $\int \omega(x, y, z) \, dx  \, dy$ (where the integral is over the entire $x$-$y$ computational domain) at each value of $z$ must remain zero due to the periodic boundary conditions. 

Commonly, in the studies of oceanic and atmospheric vortices, potential vorticity (PV) is used to describe the vortices, instead of vertical vorticity, due to its conservation property \citep{hoskins1985use,Morel1997}. Ertel's PV in figure~\ref{f1} is defined \citep{Ertel1942} as
\begin{equation}
Q \equiv [\boldsymbol{\omega} + f \hat{\boldsymbol{z}}] \cdot \Biggl(\frac{\nabla b + \bar{N}^2 \hat{\boldsymbol{z}}}{f \bar{N}^2}\Biggr) - 1, \label{pv1}
\end{equation}
where
$\boldsymbol{\omega} \equiv \boldsymbol{\nabla} \times \boldsymbol{v}$ is the vorticity vector as observed in the rotating frame. To provide a better sense about the PV structure of the vortices studied here, $Q(r,z)$ for a Gaussian vortex with $Ro = 0.2$ and three values of 
$Bu = 0.1, 1$ and $2$ are depicted in figures~\ref{f1}(c)-(e), showing that the PV structure can significantly change with $Bu$ (see \citet{Morel1997} for a discussion
of potential vorticity of Gaussian vortices). Our purpose for showing the PV of Gaussian vortices is to allow the reader the ability to make comparisons of the vortex model with what is used in some other stability studies such as \citet{Tsang2015} who model the initial vortex with uniform patches of PV.

Finally it should be noted that there is a restriction on the equilibrium of anticyclones in the Gaussian model (\ref{eq:2})-(\ref{eq:4}); there is no equilibrium for anticyclones for $Ro<-0.5$. This is because (\ref{eq:3}) and (\ref{Rossby}) show that $v_{\phi}$ does not have a real solution 
for $Ro<-0.5$, as noted, for example, by \citet{McWilliams1985} and \citet{olson1991rings}. 

\subsection{Eigenmodes} \label{subsec:eigenmodes}
The symmetries of the governing equations in (\ref{eq:1}) linearized around the equilibrium vortex (\ref{eq:2})-(\ref{eq:4}) are presented in dimensionless
form in \S\ref{sec:effect_of_foN} in equations~(\ref{linbegin})-(\ref{linend}). These equations and their 
boundary conditions show that the eigenfunctions are either symmetric or anti-symmetric with respect to the $z=0$ horizontal plane and have an $m$-fold azimuthal symmetry about the $z$-axis.
We use the labels S$m$ or A$m$ for each  eigenmode, to identify it as  Symmetric (or Anti-symmetric) with respect to the $z=0$ horizontal plane and with $m$-fold symmetry. 

The complex eigenvalues $\lambda$ and eigenfunctions 
are of the form
\begin{equation}
e^{\lambda t} \,\, \boldsymbol{g}_{eig}(r,z) \, e^{i m \phi} = e^{\sigma t} \, \boldsymbol{g}_{eig}(r,z) \, e^{i m(\phi-ct)},
\label{eq:6}
\end{equation}
where the eigenvector has 3 velocity components, a density component, and  a pressure component:
\begin{equation}
\boldsymbol{g}_{eig}(r,z) \equiv[v_{r,eig},v_{\phi ,eig},v_{z,eig},\rho_{eig},p_{eig}].
\label{eq:61}
\end{equation}
The three velocity components are with respect to cylindrical coordinates, where $m$ is the integer azimuthal wave number, 
$\sigma$ is a real growth (or decay) rate,  and $c$ is a real azimuthal phase speed.
By taking the complex conjugate of the linearized equation, we can show that if $\lambda$ is an eigenvalue with eigenfunction given by (\ref{eq:61}), then
$\lambda^{\dagger}$ is also an eigenvalue with eigenfunction $\boldsymbol{g}^{\dagger}_{eig}(r,z) \equiv [v^{\dagger}_{r,eig},v^{\dagger}_{\phi ,eig},v_{z,eig}^{\dagger},\rho^{\dagger}_{eig},p^{\dagger}_{eig}]$, with $m$ replaced by $-m$, $c$ unchanged, and    
where the superscript $\dagger$ denotes complex conjugate. Or in other words, the eigenvalues $\lambda$ when plotted in the complex plane are symmetric with respect to the real axis. Because the equations are non-dissipative, replacing $t$ with $-t$ in the 
linearized equations shows that if $\lambda \equiv \sigma - i m c$ is an eigenvalue with eigenfunction given by (\ref{eq:61}), then 
$\lambda' \equiv - \sigma - i m' c'$ is  an eigenvalue that corresponds to 
$\boldsymbol{g}'_{eig}(r,z) \equiv[v_{r,eig}, -v_{\phi ,eig},v_{z,eig}, -\rho_{eig}, -p_{eig}]$, with $m' = -m$ and $c'=c$. 
Or in other words, the eigenvalues $\lambda$ when plotted in the complex plane are symmetric with respect to the imaginary axis, and for each eigenfunction with 
a positive growth rate, there is one with a negative growth rate and vice versa. The flow can never be linearly stable with all of its eigenmodes having 
decay  rates.  The flow can either be unstable or be neutrally stable with  all of its eigenmodes on the imaginary axis with $\sigma =0$ . 
For the Gaussian vortices, the two symmetries of the linearized equations combine and
therefore the eigenvalues appear as quartets of the form $\pm a \pm i b$, with all four possible combinations of the signs, and  where $a$ and $b$ are real functions 
of $m$ and of the parameters of the unperturbed vortex $Ro$, $Bu$, and $f/\bar{N}$. For Hamiltonian systems \citep{de1990hamiltonian}, it can be shown that the quartet of 
eigenvalues is of a more specialized form:
\begin{equation}
\lambda = \pm \sqrt{A} \pm i B, \label{62}
\end{equation}
with all four possible combinations of the signs, and where $A$ and $B$ are real functions of the control parameters of the system. 
For many non-dissipative flows, e.g. unidirectional shears flows with 
vortex sheets and/or vortex layers made up of piecewise-constant vorticity \citep{drazin2004hydrodynamic}, it can be shown that 
the  quartets of the eigenvalues are of the form of (\ref{62}).
Consider a system with eigenvalue quartets such as those in (\ref{62}). When $A>0$, the eigenvalues in the quartets 
are symmetric about the real and imaginary axes, and each quartet has 2 unstable and 2 stable eigenmodes. If a control parameter changes such that $A$ decreases, then eigenvalues
symmetrically approach the imaginary axis and collide when $A=0$. For that parameter value,
there are two pairs of degenerate, neutrally-stable eigenmodes with all 4 eigenvalues on the imaginary axis. If the control parameter is further changed 
such that $A$ continues to decrease and becomes negative, then the eigenvalues are no longer degenerate, but they remain on the imaginary axis and all 4 eigenmodes remain neutrally stable, regardless of how negative $A$ becomes. Although we cannot prove that the eigenvalue quartets of the linear eigenmodes of the Gaussian vortex have the form of (\ref{62}), all of our numerical simulations are consistent with (\ref{62}). (See \S\ref{sec:criticallayers}.)

Note that although we are studying the stability of axisymmetric vortices, we solve (\ref{eq:1}) in the Cartesian coordinates rather than in the cylindrical coordinates. A numerical solver in the Cartesian coordinates avoids the difficulties of handling the singularity at the origin ($r=0$), which requires using special polynomial basis functions \citep{Matsushima1995}. However, our main reason for using  Cartesian coordinates is that future studies can  include 
background shear flows, so that the stability of vortices in planetary atmospheres and protoplanetary disks can be examined, as discussed in the Introduction. To minimize the effect of the square computational domain, we have used a circular sponge layer as described in \S\ref{subsec:numerical_method}. 
In order to find the eigenmodes with various classes of azimuthal (and vertical) symmetry in the Cartesian coordinates, we use our initial-value solver as an eigenvector/eigenvalue solver and additionally use a spatial symmetrizer (see Appendix~\ref{appC} for details). Using the spatial symmetrizer, the eigenmodes can be restricted to be symmetric or anti-symmetric in the vertical direction, while in the azimuthal direction we can enforce one of the following classes of symmetry: $m$ odd; $m$ even not divisible-by-$4$; and $m$ even and divisible-by-$4$. We use these specific symmetry groups to apply the azimuthal symmetry directly in the Cartesian coordinates, which greatly speeds up the convergence of the calculations, and also avoids introducing additional errors due to transformation between Cartesian and cylindrical coordinates (see Appendix~\ref{appC}).

\section{Critical layers} \label{sec:criticallayers}

Eigenmodes of unidirectional  equilibrium flows such as the Gaussian vortices studied here can have critical layers, i.e., singularities at locations where the azimuthal
phase speed $c$ is equal to the azimuthal velocity $v_\phi(r, z)$ of the unperturbed vortex \citep{maslowe1986critical,benilov2003instability}.\footnote{In stratified unidirectional flows, critical layers can appear
  at other locations as well \citep{marcus2013three,marcuszombie}.} Here we show examples of eigenmodes with critical layers and discuss, for a few cases, how different modes take over as the fastest-growing mode as $Bu$ changes, which will be used later to interpret the results of \S\ref{sec:parameter_map_of_stability}. It should be noted that despite the peculiar nature of critical layers, it is not difficult to accurately compute them using high-resolution numerical simulations. For example, \citet{Nguyen2012} and \citet{Yim2016} have simulated critical layers in 3D QG and Boussinesq vortices, respectively. Recently, we have numerically computed critical layers, with and without dissipation, in stratified, rotating, unidirectional flows and found that with sufficient spatial resolution the locations, widths and other analytically-known properties of the critical layers can be quantitatively reproduced
\citep{marcus2013three,marcuszombie}. In the results presented here, the location of the critical layers and the phase speed of the eigenmode containing the critical layer are insensitive to the numerical resolution and remain the same when the resolution is increased by a factor of $4$ by halving the domain size in each direction to $(15L) \times (15L) \times (15H)$ and increasing the Fourier modes from $256^3$ to $512^3$ (the figures showing the structure of the eigenmodes in this section are from the higher resolution). 


The singularity in the eigenmode occurs where the coefficient $[v_{\phi}(r, z)/r - c - i \sigma/m]$ in front of the highest-order derivative terms in the governing equations of the eigenmode
becomes zero. Unless the growth rate $\sigma$ is zero and the eigenmode is neutrally stable,  the eigenmode is no longer formally singular. 
However,  the amplitudes of the
eigenmodes remain large at locations where $v_{\phi}(r, z)/r = c$
for parameter values where $\sigma > 0$ and the mode is {\it weakly} growing. For parameter values where the analytically computed eigenmode has $\sigma =0$, but the eigenmode is computed
numerically with a modified initial-value code (as done here) with weak hyperdissipation, the computed eigenmode has large amplitude at $v_{\phi}(r, z)/r = c$, and the magnitude of the
numerically computed growth rates $\sigma$ are typically less than or equal to $0.002$ in inverse units of the vortex turnaround time $\tau \equiv 4\pi/\omega_c$, where $\omega_c$ is the absolute value of the vertical vorticity at the center of the vortex.


We argued in \S\ref{subsec:eigenmodes} that as a parameter value, such as the Burger number, is changed such that a growing/decaying pair of eigenmodes has its eigenvalues $\lambda$ collide on the imaginary axis, the  eigenmodes become  neutrally stable and degenerate. As the parameter value further changes, the eigenvalues remain neutrally stable and their phase speeds become distinct from each other. Here we demonstrate in detail that this scenario of eigenvalue collision, in which the families of eigenmodes {\it{continue}} after the collision rather than ceasing to exist due to the singularity of the critical layer, is correct by illustrating the collision for  three distinct families of eigenmodes with critical layers. In particular, we show that as the $Bu$ changes and the eigenmode goes from unstable to neutrally stable, the family containing that eigenmode  continues to exist and remains neutrally stable as
the $Bu$ is further changed. We need these three demonstrations to not only show that our numerical computations of eigenmodes are accurate, but also to highlight the physics of the collisions.

%
%

\begin{figure}
  \centering
  \subfloat[\large{(a)}]{\includegraphics[trim={4mm 18mm -10mm 2mm},clip,width=0.93\textwidth]{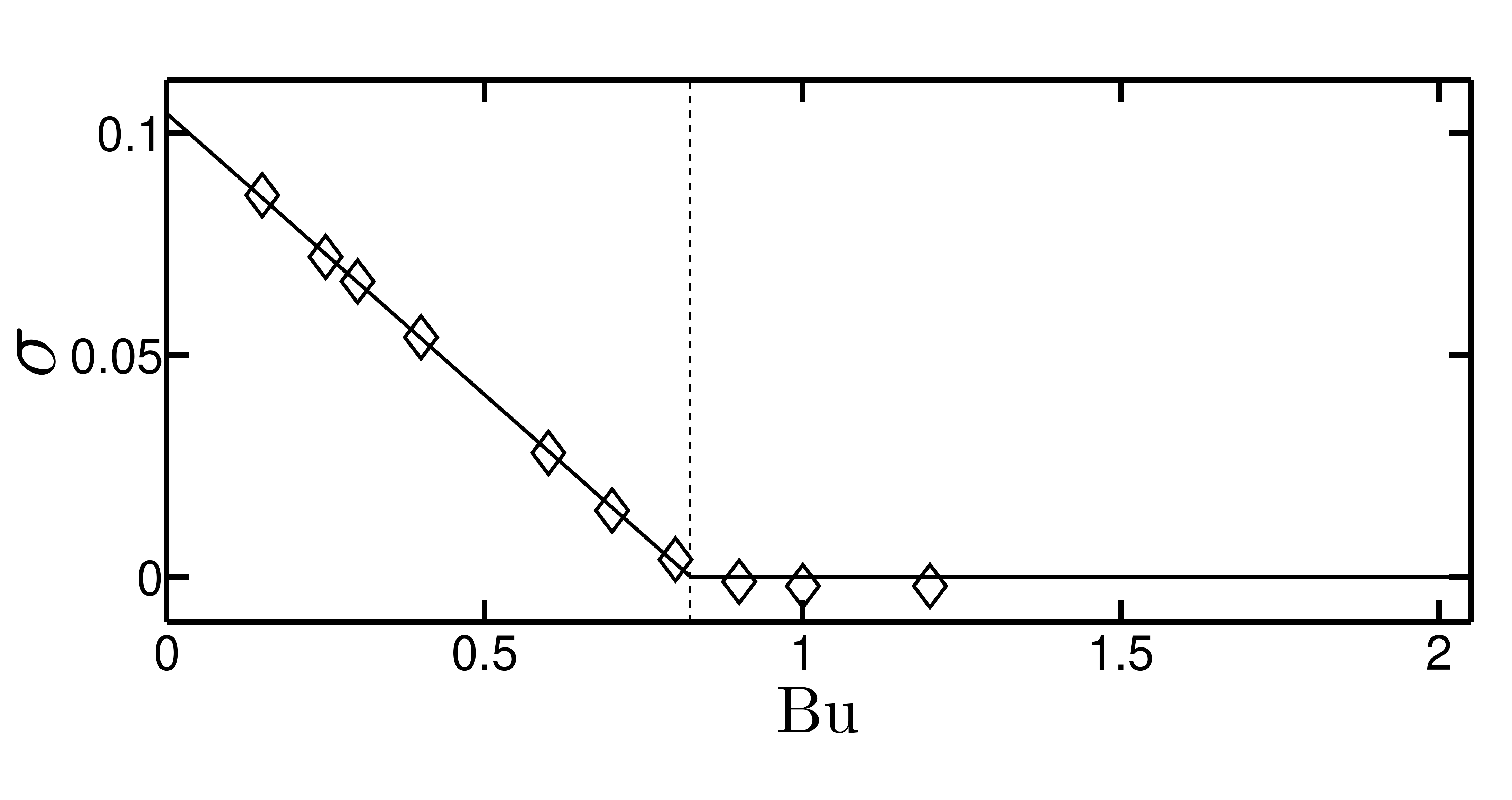}}\\
  \subfloat[\large{(b)}]{\includegraphics[trim={4mm 18mm -10mm 22mm},clip,width=0.93\textwidth]{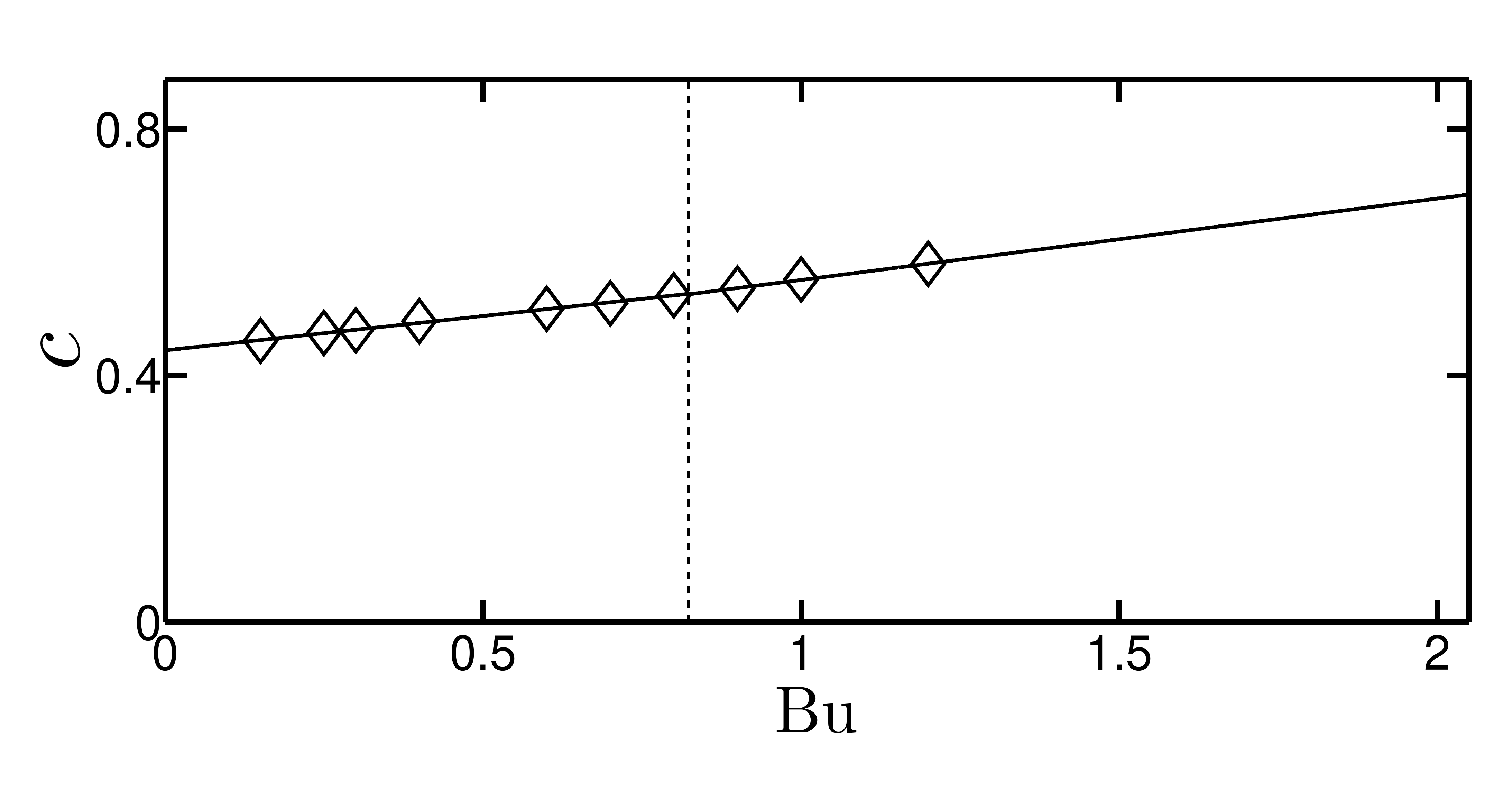}}\\
  \caption{(a) Growth rates $\sigma$ (in units of $\tau^{-1}$) of the eigenmodes with S2 symmetries as functions of $Bu$ for $Ro=0.05$ and $f/\bar{N}=0.1$. The lines connecting the  symbols are to ``guide the eye''. The eigenmodes with S2 symmetry are unstable in the range $Bu \lesssim 0.823$ (they are the fastest-growing for $0.2 \lesssim Bu \lesssim  0.823$). As $Bu$ increases, the eigenmode changes from unstable to neutrally stable at $Bu \simeq 0.823$
(shown with the  vertical broken line), but the family of eigenmodes
does not terminate there. (b) The phase speed $c$  (in units of $\tau^{-1}$) corresponding to the growth rates shown in panel~(a).
The lines connecting the  symbols are to ``guide the eye''. The phase speed is continuous when it passes through the vertical broken line, which is a necessary condition for the unstable and neutrally-stable eigenmodes to belong to the same family. Note that because our computation uses a small hyperdissipation, the ``neutral'' modes in panel (a) have a slight decay rate of $\sim 0.002 \tau^{-1}$; however, as the value of the hyperdissipation decreases (with a corresponding increase in spatial resolution to prevent an accumulation of energy and enstrophy at the smallest resolvable length scales), so does the decay rate,  suggesting that a dissipationless calculation would show that family of eigenmodes with $Bu > 0.823$ are truly neutral.
}
  \label{f3_new_after_1}
\end{figure}

\begin{figure}
  \centering
  \subfloat[\large{~~~~~~(a)}]{\includegraphics[trim={90mm 3mm 125mm -8mm},clip,width=0.40\textwidth]{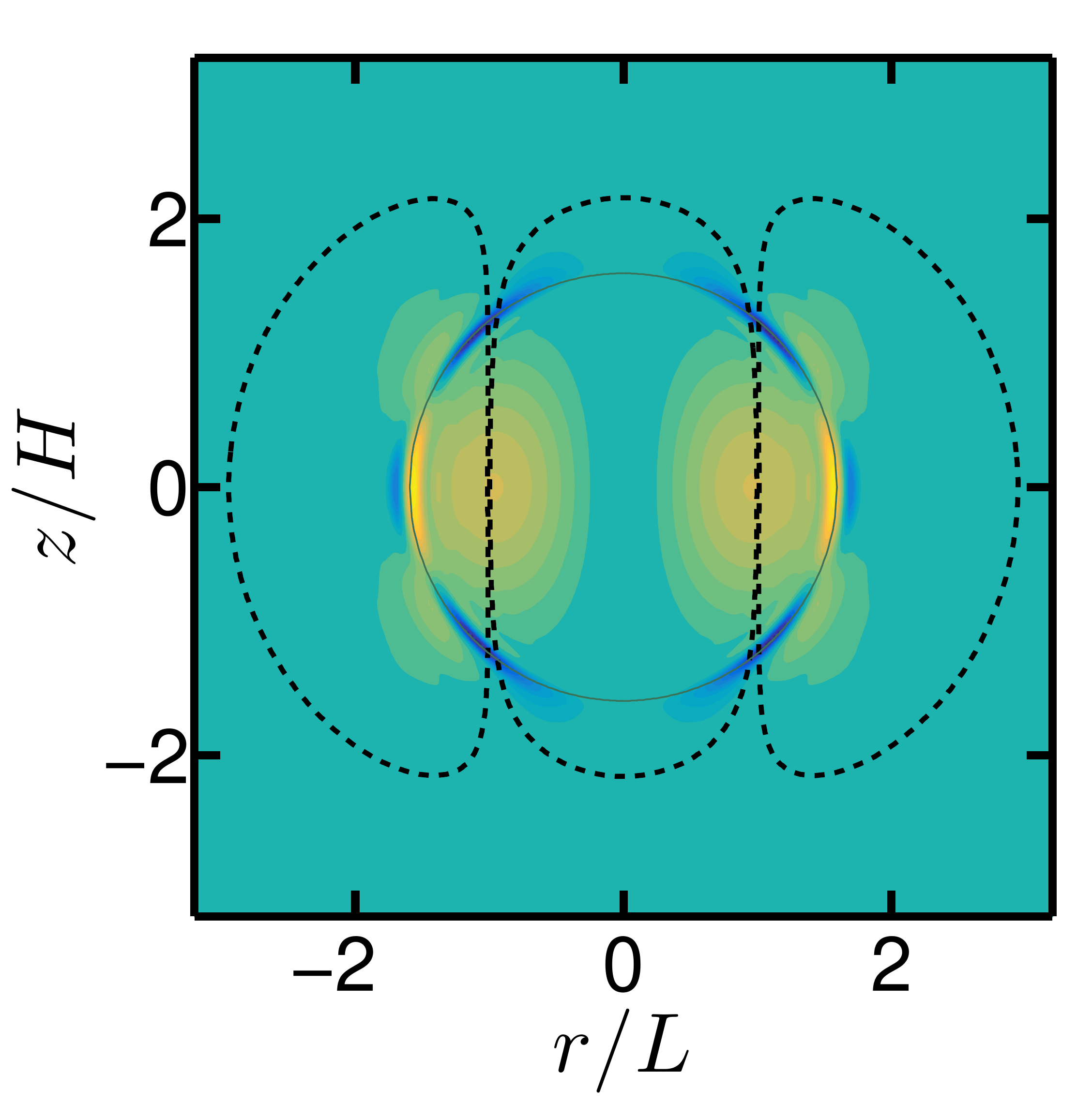}}
  $~~$\subfloat[~~~~~~\large{(b)}]{\includegraphics[trim={90mm 3mm 125mm -8mm},clip,width=0.40\textwidth]{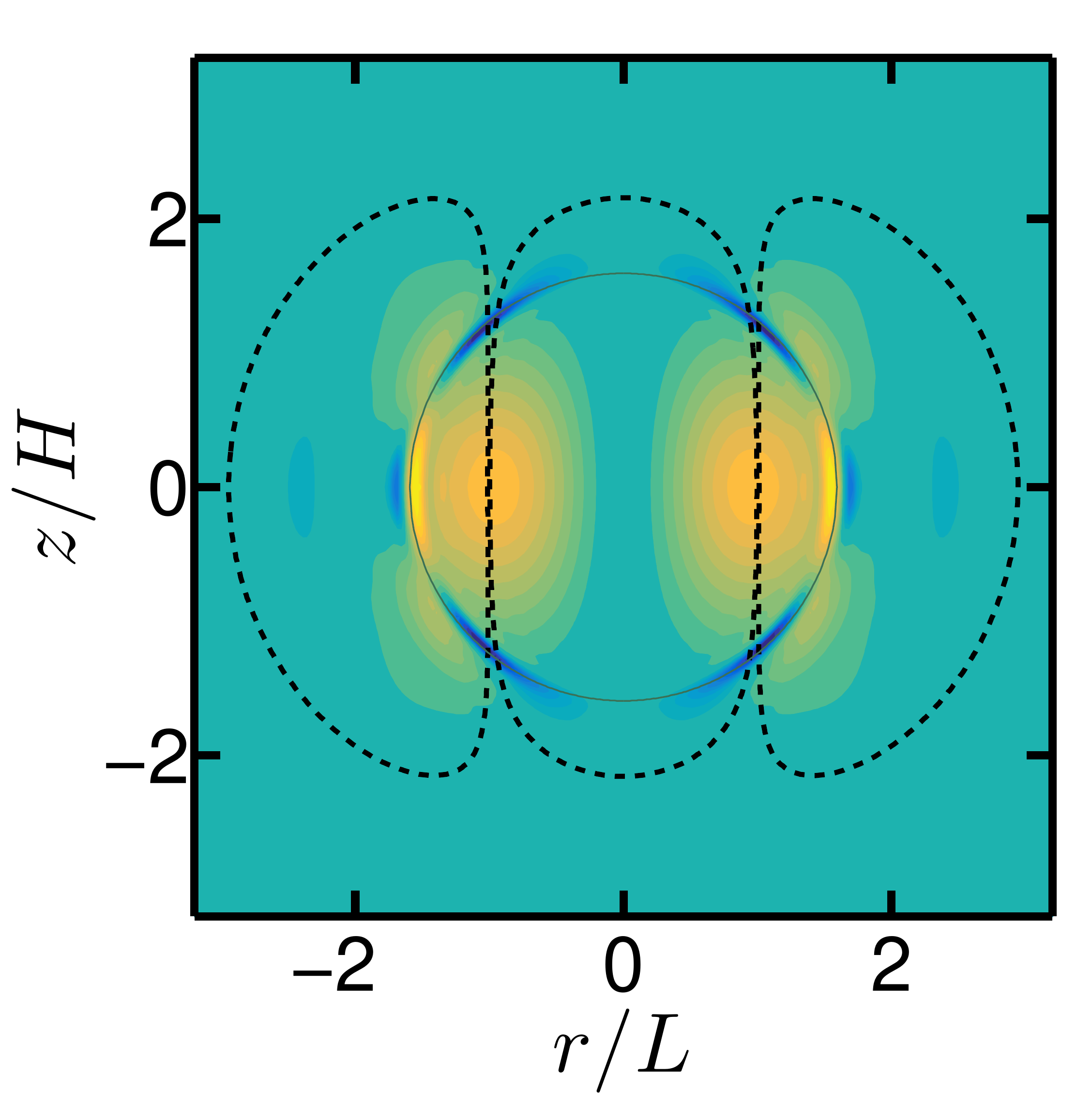}}
  \subfloat{\includegraphics[trim={195mm -8mm 189mm -8mm},clip,height=0.283\textheight]{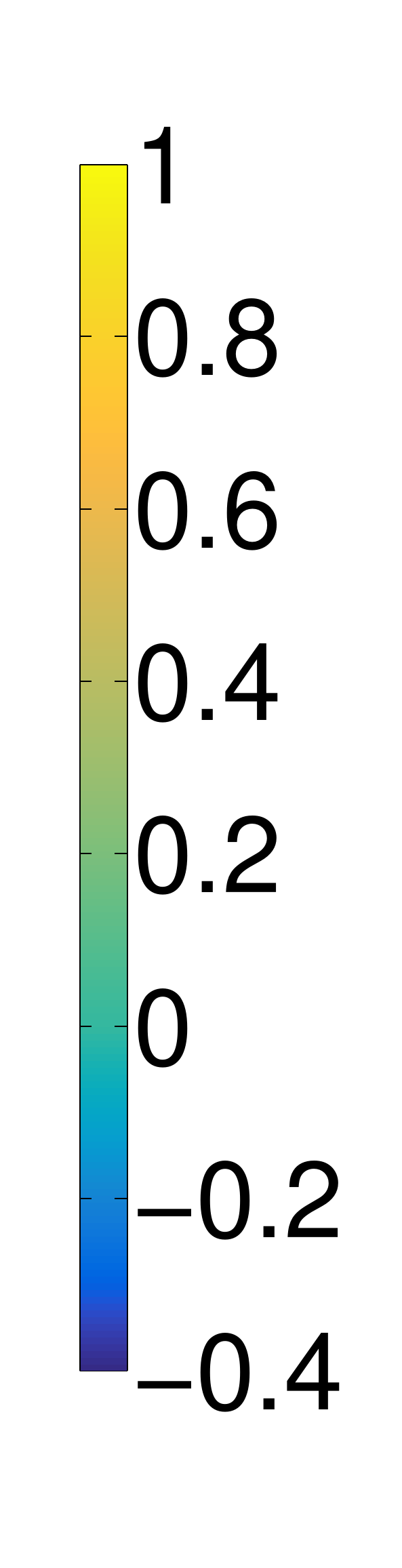}}
  \caption{(Colour online) Vertical vorticity in the $(r-z)$ plane of two of the eigenmodes shown in figure~\ref{f3_new_after_1}, with medium shade being zero (cyan, in colour), light shade being the most cyclonic (yellow, in colour), and dark shade the most anticyclonic (blue, in colour). The center of each panel corresponds to the center of the unperturbed Gaussian vortex. The azimuthal angle of each panel was  chosen so that the critical layer is prominent. The theoretical location of each critical layer is indicated by the continuous, nearly circular curve (dark green, in colour), which is where the phase speed $c$ is equal to the azimuthal velocity of the unperturbed vortex. Both eigenmodes have S2 symmetry.
    (a) For the unstable eigenmode at $Bu = 0.7$. (b) For the neutrally-stable eigenmode at $Bu = 0.9$. The similarity of the
    radial structure of the unstable and neutrally-stable eigenmodes indicate that they are part of the same family and that the
    family does not terminate when the growth rate changes from positive to zero.
}
  \label{criticallayer_1}
\end{figure}

Figure~\ref{f3_new_after_1} shows the growth rate $\sigma$ and phase speed $c$ of the fastest-growing eigenmode with S2 symmetry for $Ro = 0.05$ and
$0 \le Bu \le 2.1$. As $Bu$  increases,  the growth rate in figure~\ref{f3_new_after_1}(a) changes from positive (unstable) to zero (neutrally stable)
at $Bu \simeq 0.823$. Note that we have computed three  neutrally-stable
eigenmodes in this family. There can be  multiple  neutrally-stable S2 eigenmodes for the same $Ro$, $f/\bar{N}$,  and $Bu$ so
it is necessary to show that the eigenvalues with $Bu \lesssim 0.823$ and $Bu \gtrsim 0.823$ belong to eigenmodes in the same family.
We do this in two ways.  Figure~\ref{f3_new_after_1}(b) shows the phase speeds $c$ for the eigenmodes illustrated in
figure~\ref{f3_new_after_1}(a). According to (\ref{62}), a necessary condition that the eigenmodes belong to the same family is that  there is
no discontinuity in $c$ at the value of $Bu$ where $\sigma$ changes from positive to zero.\footnote{Note that the slope of $c$ can be discontinuous at   the $Bu$ where $\sigma$ changes from positive to zero.} Figure~\ref{f3_new_after_1}(b) shows that this condition is met.
Figure~\ref{criticallayer_1} shows the vertical vorticity of the eigenmodes whose eigenvalues are shown in figure~\ref{f3_new_after_1}
with $Bu= 0.7$ (where the eigenmode is unstable) and $Bu = 0.9$ (where the eigenmode is neutrally stable). The eigenmodes clearly have similar radial structures and are therefore part of the same family. The continuous, nearly-circular curve (dark green, in colour) is the locus in the $r-z$ plane where $v_{\phi}(r, z)/r = c$ and indicates the theoretical location of the critical layer. The large vorticity that is nearly coincident with the continuous curve is the critical layer.

\begin{figure}
  \centering
 \subfloat[\large{(a)}]{\includegraphics[trim={4mm 18mm -10mm 2mm},clip,width=0.93\textwidth]{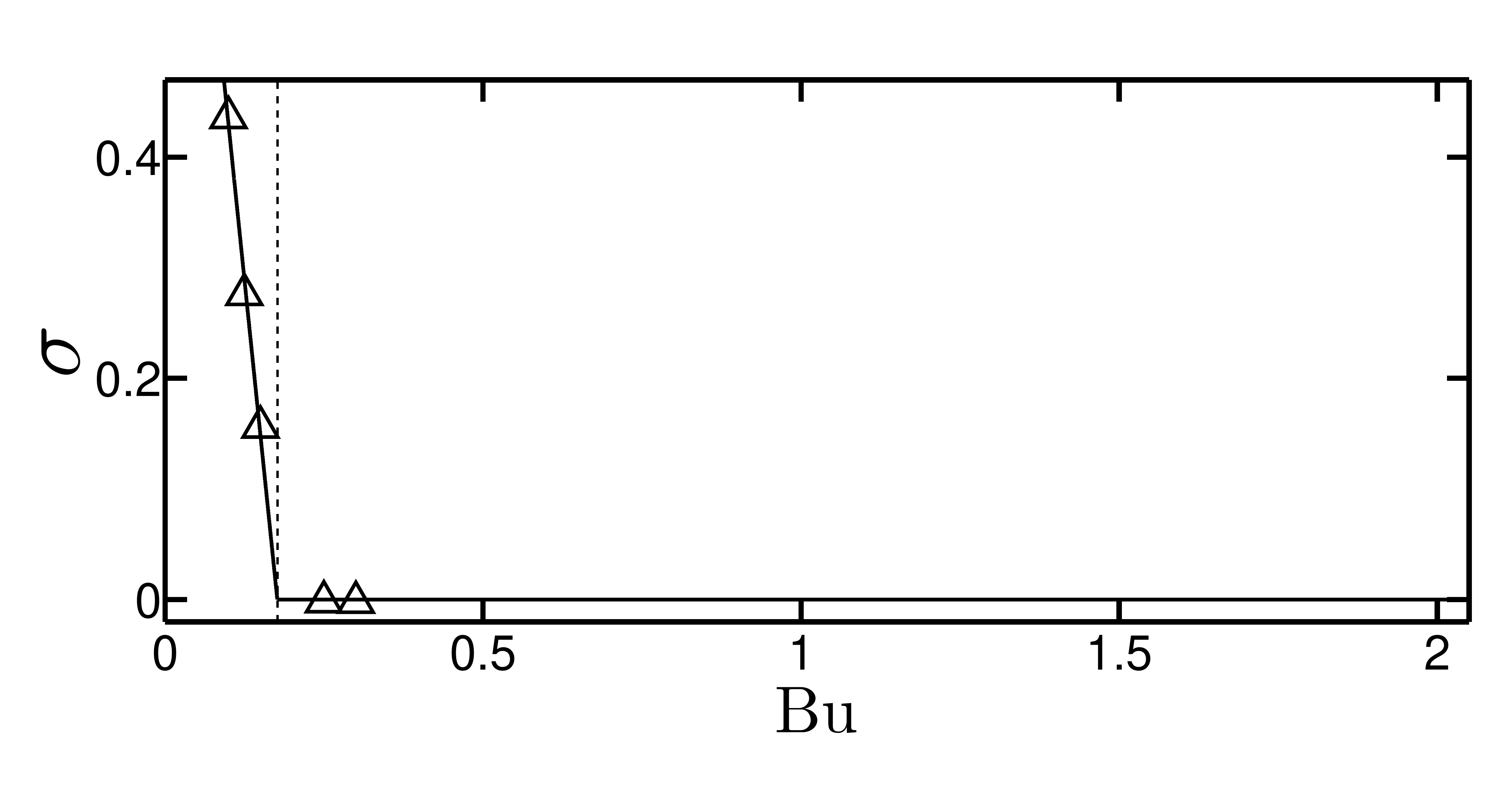}}\\
  \subfloat[\large{(b)}]{\includegraphics[trim={4mm 18mm -10mm 22mm},clip,width=0.93\textwidth]{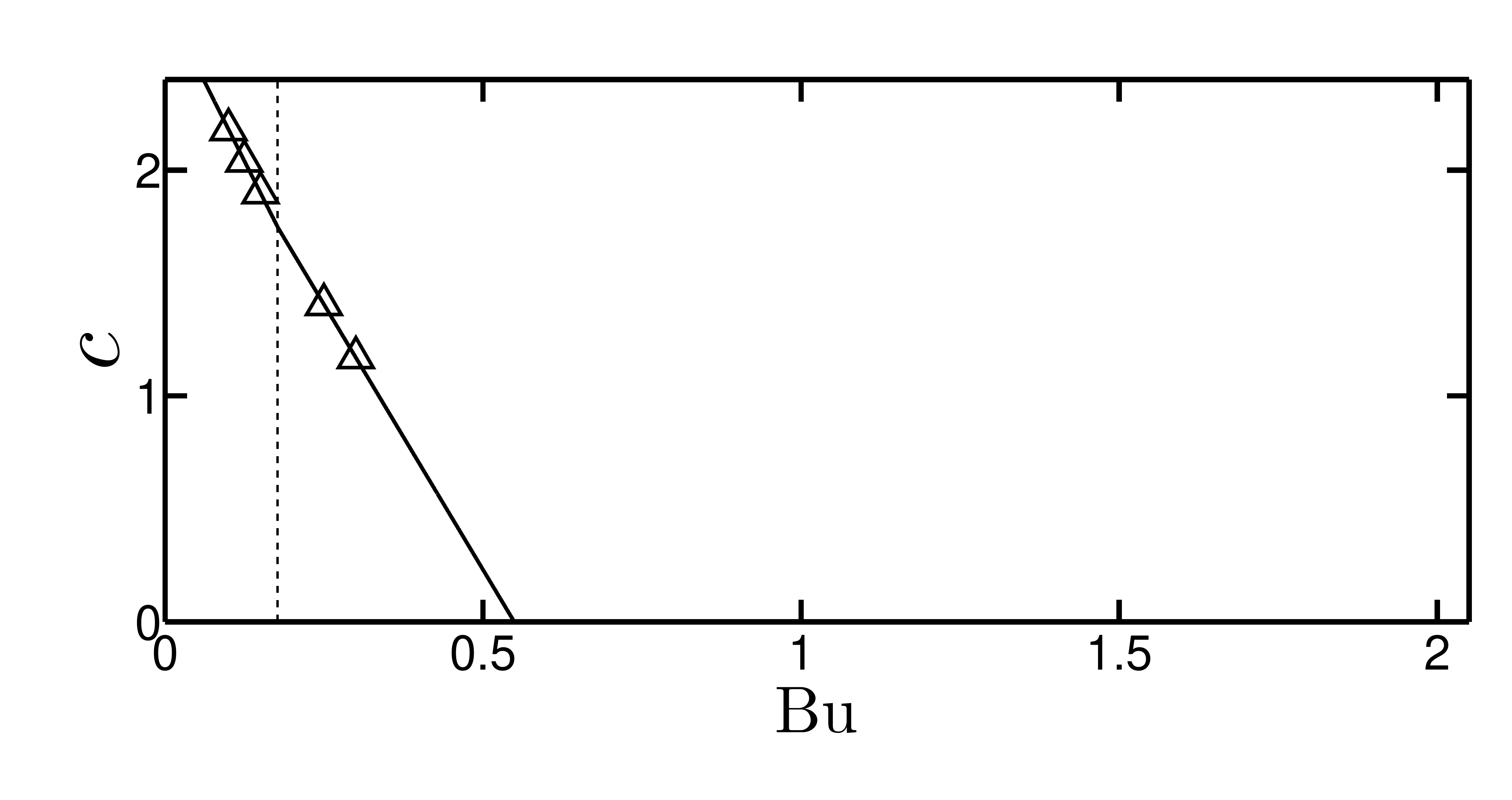}}\\
  \caption{As in figure~\ref{f3_new_after_1} but for the family of A1 eigenmodes that are the fastest growing for
$Ro=0.05$, $f/\bar{N}=0.1$ and in the range  $Bu \lesssim 0.2$.
    Triangles indicate the numerically computed values of $\sigma$ and $c$.  The eigenmode goes from unstable to neutrally stable at $Bu \simeq 0.177$, indicated by the vertical broken line.
}
  \label{f3_new_after_2}
\end{figure}

\begin{figure}
  \centering
  \subfloat[\large{~~~~~~(a)}]{\includegraphics[trim={90mm 3mm 125mm -8mm},clip,width=0.40\textwidth]{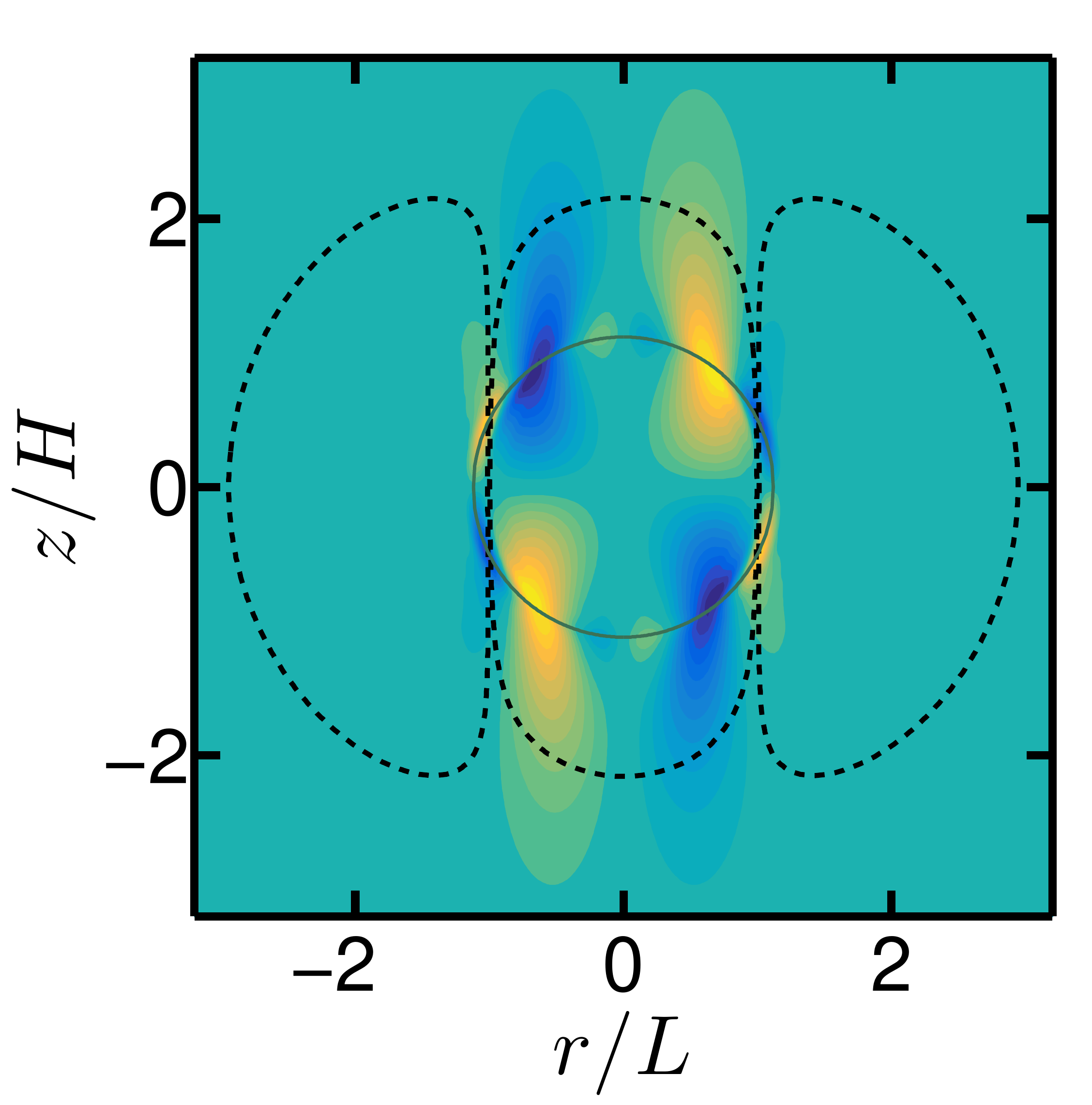}}
  $~~$\subfloat[~~~~~~\large{(b)}]{\includegraphics[trim={90mm 3mm 125mm -8mm},clip,width=0.40\textwidth]{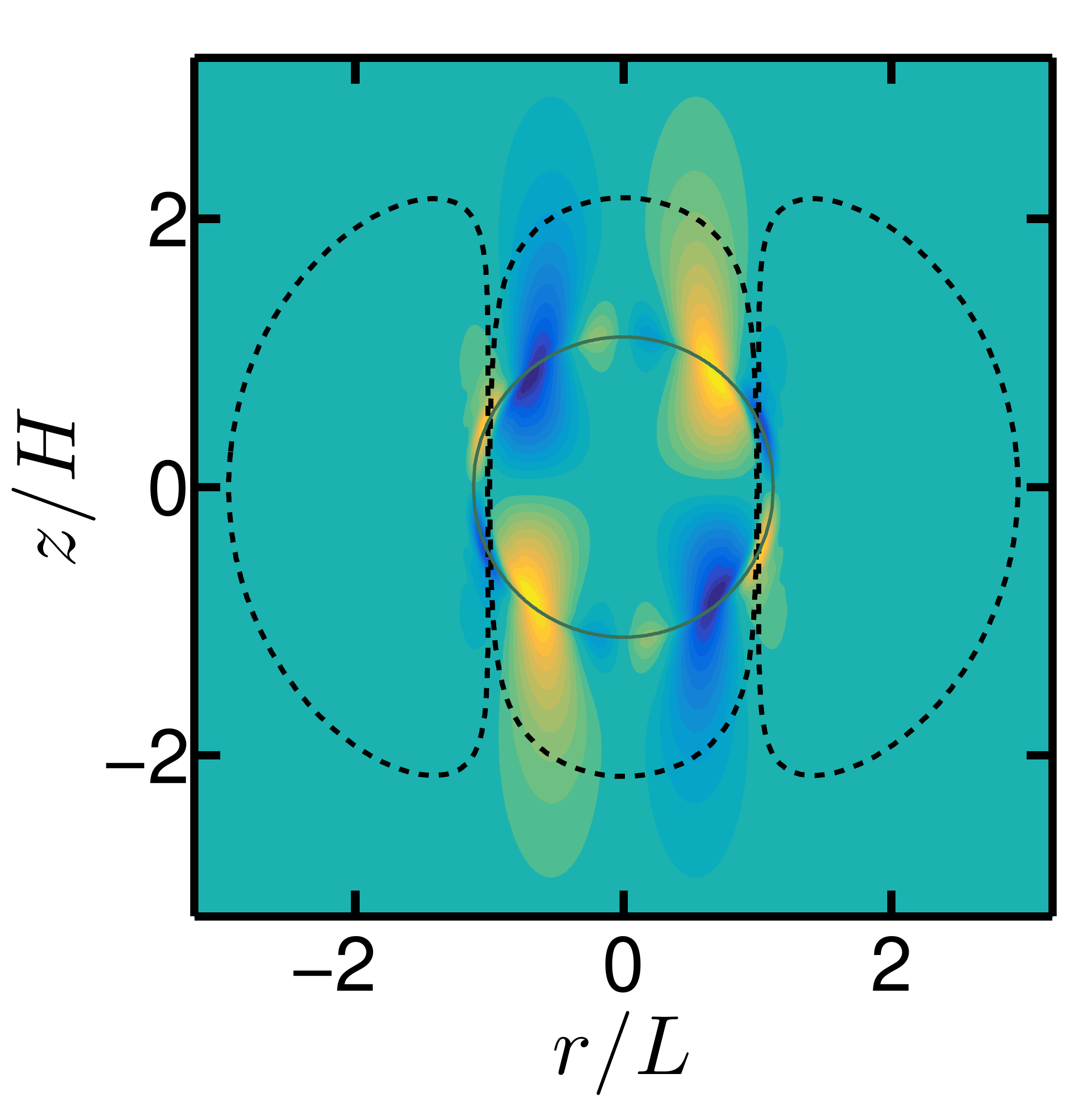}}
  \subfloat{\includegraphics[trim={195mm -8mm 189mm -8mm},clip,height=0.283\textheight]{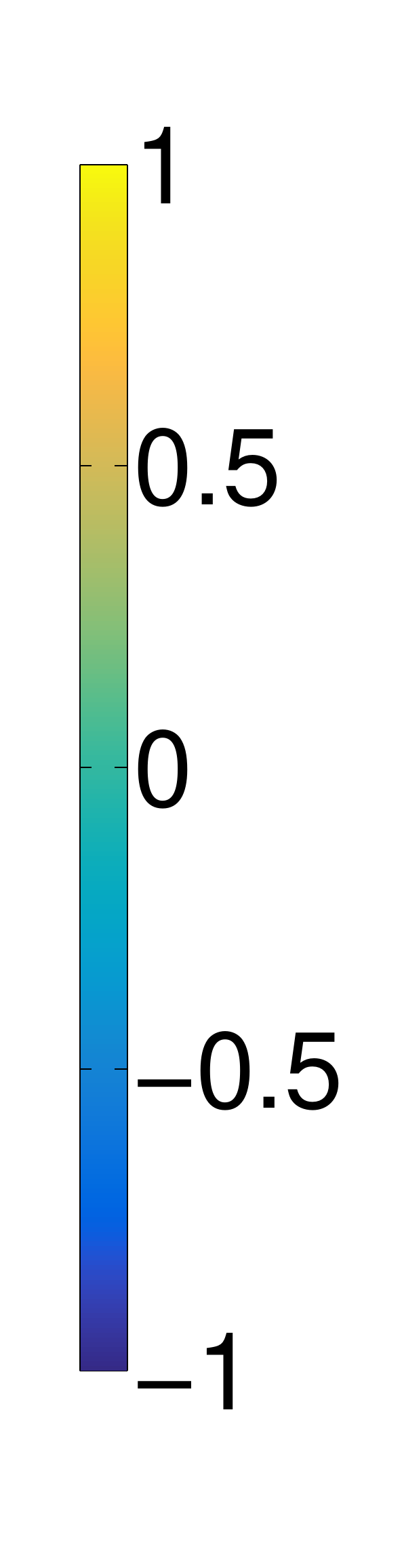}}
\caption{(Colour online) As in figure~\ref{criticallayer_1} but for two of the eigenmodes shown in figure~\ref{f3_new_after_2} with  A1 symmetry. (a) For the unstable eigenmode at $Bu = 0.15$. (b) For the neutrally-stable eigenmode at $Bu = 0.25$. 
}
  \label{criticallayer_2}
\end{figure}

Figures~\ref{f3_new_after_2}~and~\ref{criticallayer_2} show the growth rates, phase speeds, and the vertical vorticity of another family of eigenmodes with  critical layers for $Ro = 0.05$ and $0 \le Bu \le 2.1$. These eigenmodes have A1 symmetry and are the fastest growing eigenmodes when $Bu \lesssim 0.2$. As $Bu$  increases,  the growth rate changes from positive (unstable) to zero (neutrally stable)
at $Bu \simeq 0.177$. The continuity of $c$ and the similarity of the vorticity distributions for the unstable and neutrally-stable
eigenmodes indicate that the unstable and neutrally-stable eigenmodes belong to the same family and that the family does not
end abruptly at the value of  $Bu$ where the eigenmodes pass from unstable to neutrally stable.

\begin{figure}
  \centering
  \subfloat[\large{(a)}]{\includegraphics[trim={4mm 18mm -10mm 2mm},clip,width=0.93\textwidth]{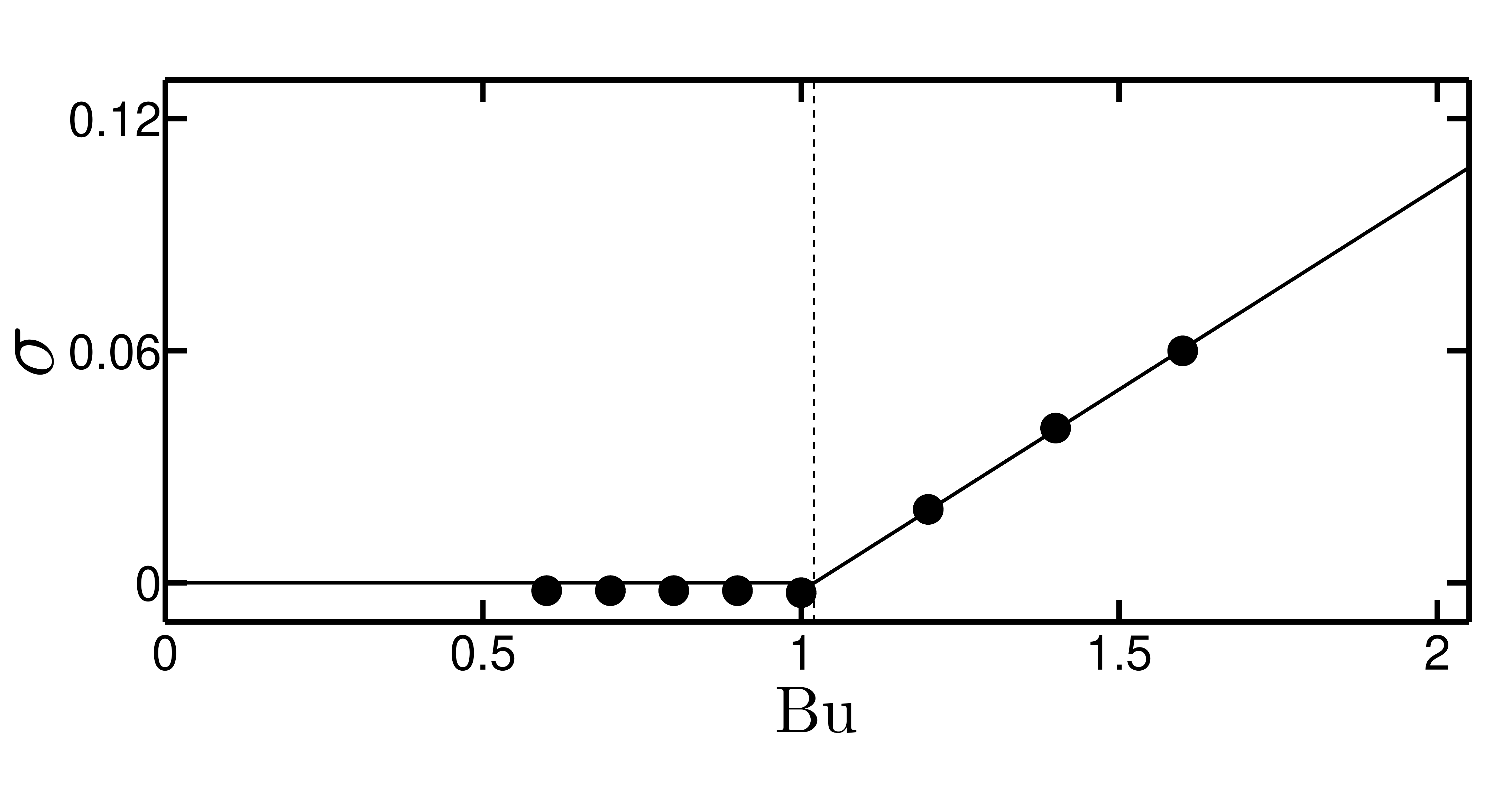}}\\
  \subfloat[\large{(b)}]{\includegraphics[trim={4mm 18mm -10mm 22mm},clip,width=0.93\textwidth]{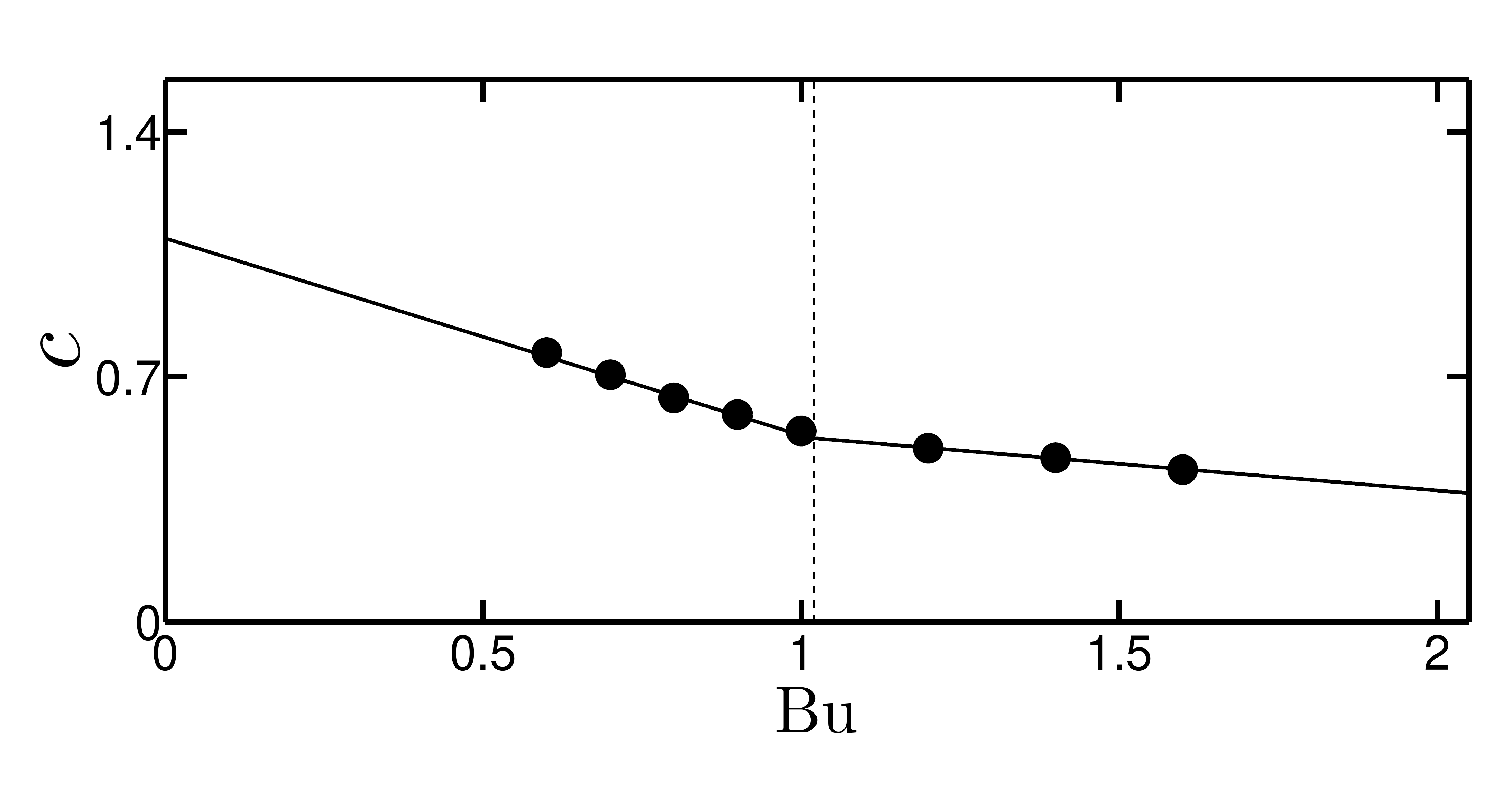}}\\
  \caption{As in figure~\ref{f3_new_after_1} but for the family of A1 eigenmodes that are the fastest growing for
$Ro=0.05$, $f/\bar{N}=0.1$, and  $Bu \gtrsim 1$.  Solid circles indicate the numerically computed values of $\sigma$ and $c$. The eigenmode goes from neutrally stable to unstable at $Bu \simeq 1.02$ indicated by the vertical broken line. Note that the families illustrated here and in figure~\ref{f3_new_after_2} both have A1 symmetry, but they are different families. 
}
  \label{f3_new_after_3}
\end{figure}

\begin{figure}
  \centering
  \subfloat[\large{~~~~~~(a)}]{\includegraphics[trim={90mm 3mm 125mm -8mm},clip,width=0.40\textwidth]{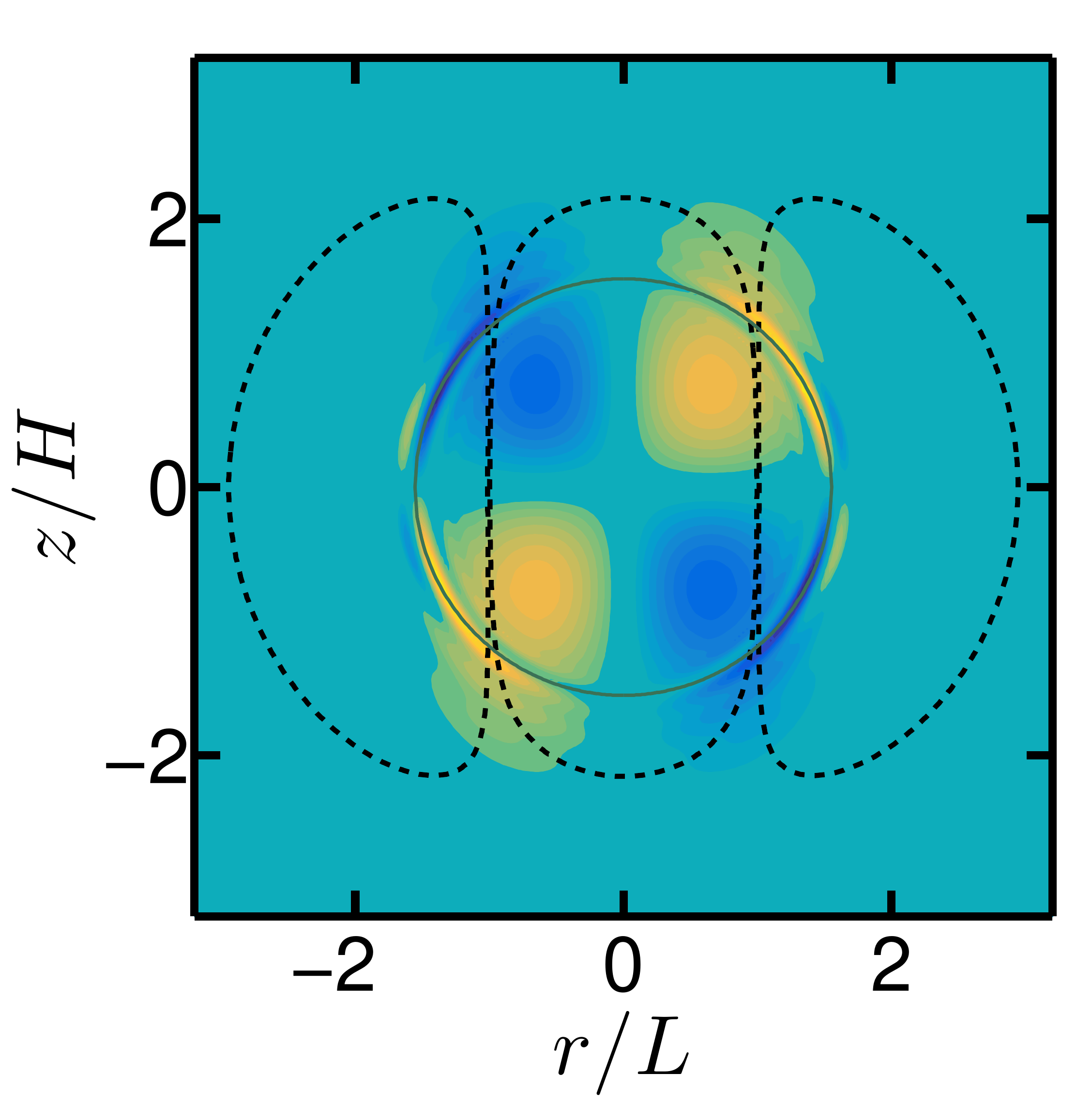}}
  $~~$\subfloat[~~~~~~\large{(b)}]{\includegraphics[trim={90mm 3mm 125mm -8mm},clip,width=0.40\textwidth]{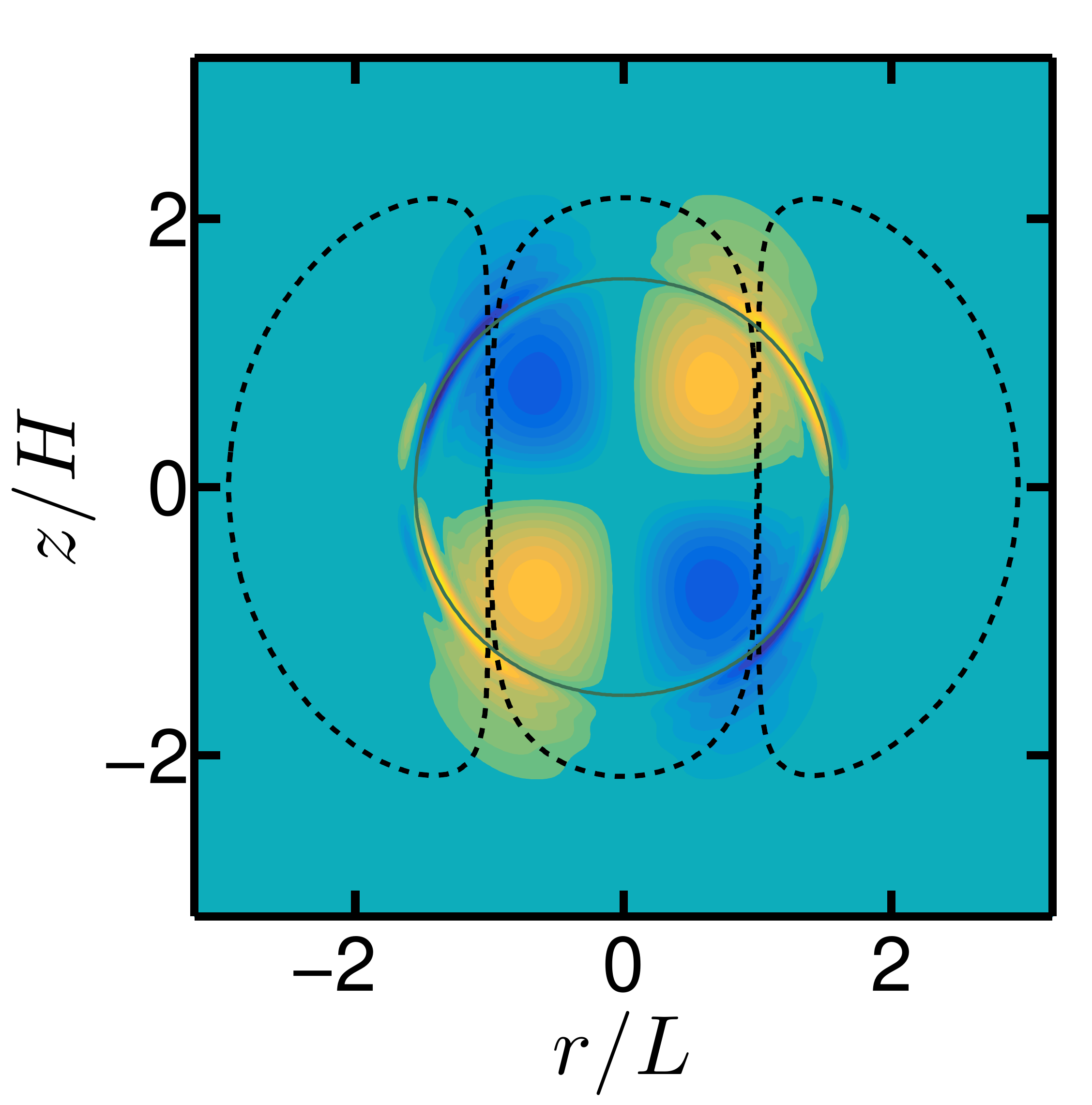}}
  \subfloat{\includegraphics[trim={195mm -8mm 189mm -8mm},clip,height=0.283\textheight]{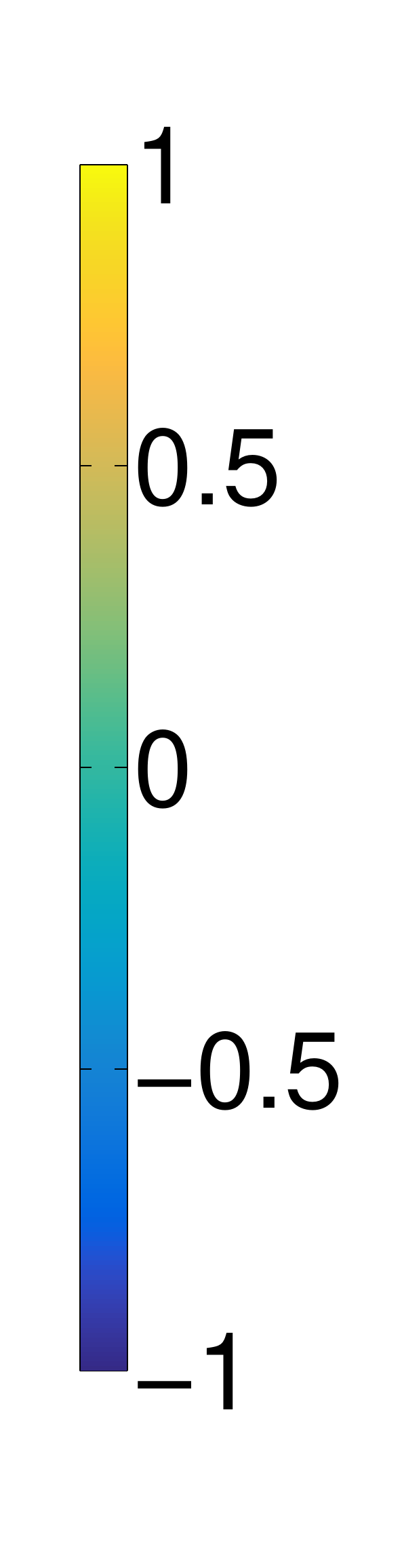}}
  \caption{(Colour online) As in figure~\ref{criticallayer_1} but for two of the eigenmodes shown in figure~\ref{f3_new_after_3}
    with  A1 symmetry.
    (a) For the unstable eigenmode at $Bu = 1.2$. (b) For the neutrally-stable eigenmode at $Bu = 0.9$.
        Note that the eigenmodes illustrated here and in figure~\ref{criticallayer_2} both have A1 symmetry, but they are different eigenmodes. 
}
  \label{criticallayer_3}
\end{figure}

Figures~\ref{f3_new_after_3}~and~\ref{criticallayer_3} also show the growth rates and phase speeds and the vertical vorticity of a
different  family of eigenmodes with  critical layers with A1 symmetry for $Ro = 0.05$ and $0 \le Bu \le 2.1$. For this family as
$Bu$  decreases,  the growth rate changes from positive (unstable) to zero (neutrally stable)
at $Bu \simeq 1.02$. Again, the continuity of $c$ and the similarity of the vorticity distributions for the unstable and neutrally stable
eigenmodes indicate that the unstable and neutrally-stable eigenmodes belong to the same family and that the family does not
end abruptly at the value of $Bu$ where the eigenmodes pass from unstable to neutrally stable. Note that although the set of
figures~\ref{f3_new_after_2}~and~\ref{criticallayer_2} and the set of figures~\ref{f3_new_after_3}~and~\ref{criticallayer_3} both illustrate A1 eigenmodes, they are different families of eigenmodes. The distinction is easily seen because the radial structures of the eigenmodes differ and because the
phase speeds differ. We have illustrated these two different families of A1 eigenmodes to emphasize the fact that we can
easily determine when  two families of eigenmodes are distinct and when they are not. These results demonstrate that the unstable and neutrally-stable eigenmodes
in figure~\ref{f3_new_after_1} (or in  figure~\ref{f3_new_after_2} or in figure~\ref{f3_new_after_3}) are  part of the same family and confirm that when a pair of eigenvalues of eigenmodes of the vortices studied here collide on the imaginary axis, the families of eigenmodes do not terminate. This finding  will be used later to interpret the results of \S\ref{sec:parameter_map_of_stability} (specifically, figure~\ref{f3_new}).

Finally, it should be mentioned that for the cases examined here ($Ro=0.05$, $0.1 \lesssim Bu \lesssim 1.6$), the peripheral location of critical layers is found to be generic (figures~\ref{criticallayer_1},~\ref{criticallayer_2},~\ref{criticallayer_3},~and~\ref{f8}(f)), which is consistent with the QG analysis of \citet{Nguyen2012}.

\section{Parameter map of stability} \label{sec:parameter_map_of_stability}
Here, we explore the stability and linear growth rates of Gaussian vortices as functions of $Ro$ and $Bu$ for $f/\bar{N}=0.1$. Like many other studies, for most cases we have used $f/\bar{N} = 0.1$, rather than $f/\bar{N} = 0.01$ (which is a better representative of the mid-latitude oceans, see \citet{Chelton1998,Lelong2005}), because small values of $f/\bar{N}$ are computationally expensive to tackle \citep[see, e.g.,][]{Suzuki2012,Tsang2015}. However in this paper, we  use the semi-analytic method of \citet{Barranco2006}, which allows us to compute flows efficiently for a wide range of $f/\bar{N}$, including the more physically relevant value of $0.01$. Some cases are repeated with $f/\bar{N}=0.01$ and discussed in \S\ref{sec:effect_of_foN}. The results presented in this section are all obtained using the computational domain of $(30L) \times (30L) \times (30H)$ and resolution of $256^3$.

For each of the vortices we examined, we computed the eigenvalues and eigenvectors [as given by (\ref{eq:6})] of the fastest-growing eigenmode and also for the fastest-growing eigenmodes of each of the six symmetry classes that could be computed by the simultaneous application of the spatial symmetrizer in $z$ (which forced the eigenmode to be symmetric or anti-symmetric in $z$) and the azimuthal symmetrizer (which forced the eigenmode to have an odd azimuthal wave number $m$, or to have an even $m$ that was not divisible by $4$, or to have an even $m$ that was divisible by $4$). For some cases, the fastest-growing eigenmodes were also computed without a spatial symmetrizer, which were found to  be identical (up to 3 significant digits) to the fastest-growing eigenmode of the six eigenmodes that were computed with one of the enforced symmetries.

The results are compared and contrasted with the most relevant published results obtained from analyzing the QG, shallow-water, and full Boussinesq equations in \S\ref{sec:comp}.   

\subsection{Spatial symmetries and growth rates of the eigenmodes} \label{subsec:symmetries_and_growth_rates}
\begin{figure}
    \centering
    \subfloat[\small{(a)}]{\includegraphics[trim={0 1.5mm 0 -8mm},clip,width=0.92\textwidth]{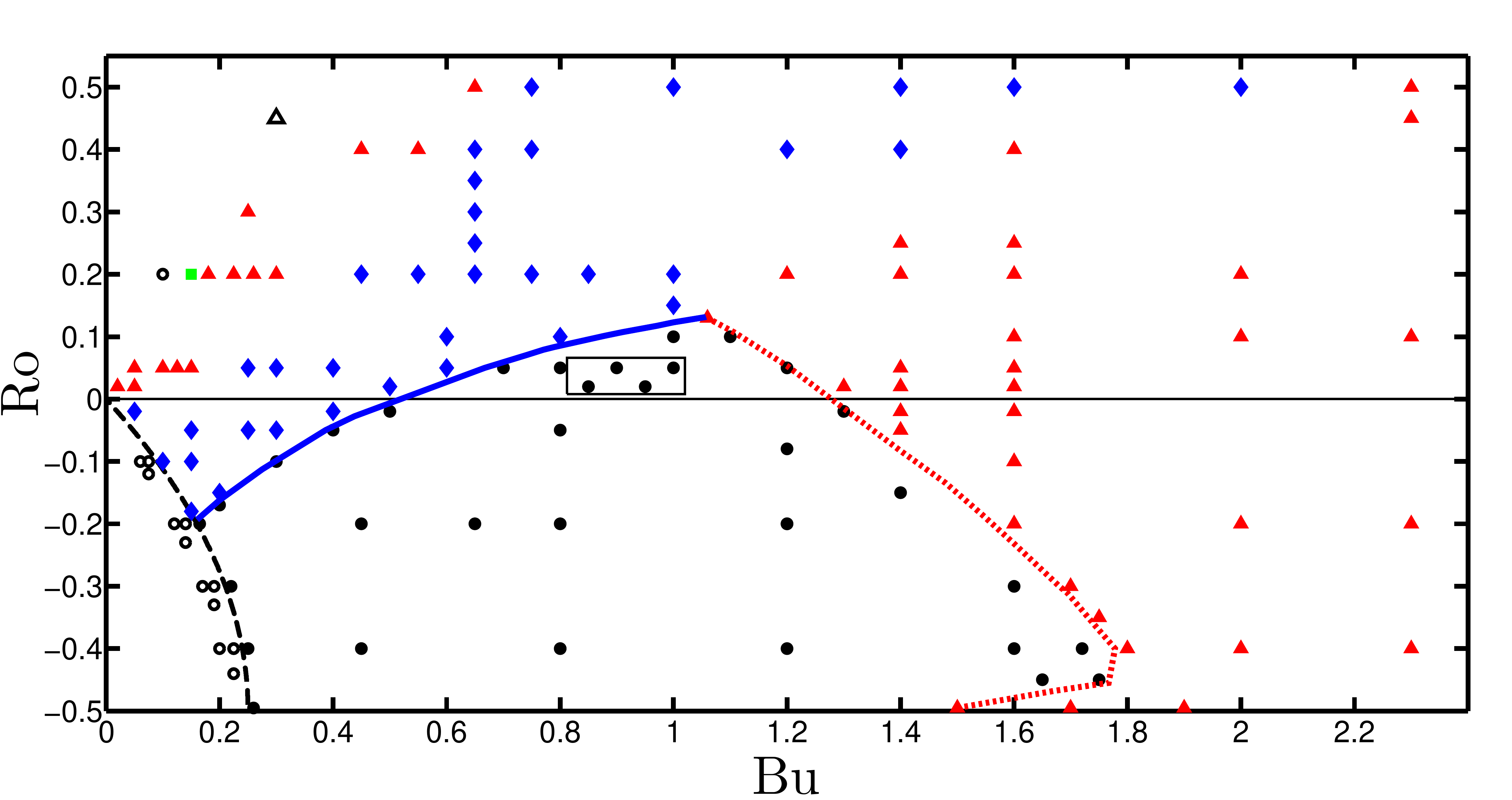}}\\[1mm]
    \subfloat[\small{(b)}]{\includegraphics[trim={0 1.5mm 0 7.5mm},clip,width=0.92\textwidth]{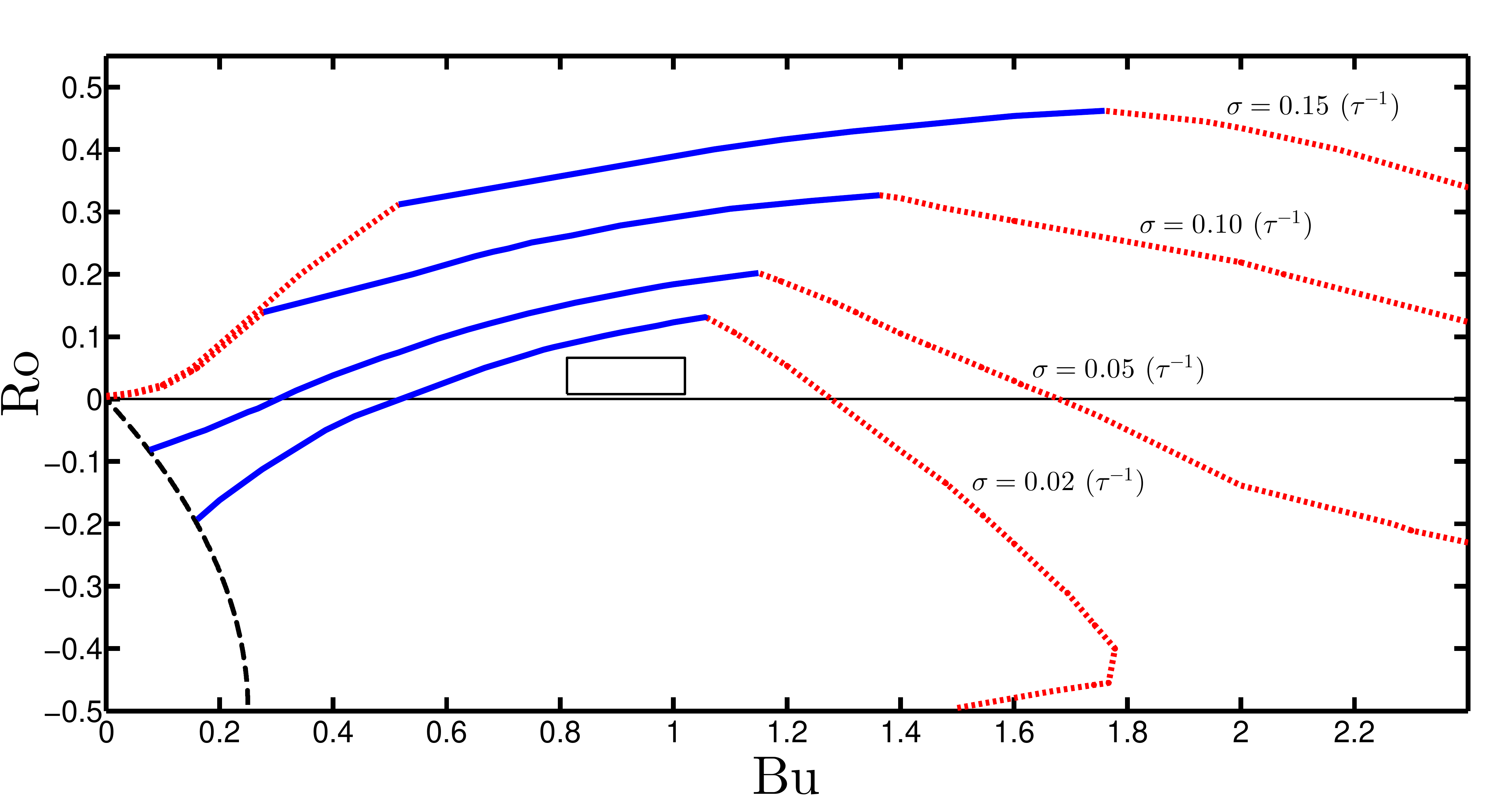}}
    \caption{(Colour online) Parameter map of the stability of Gaussian vortices in the $Ro-Bu$ space. No equilibrium Gaussian vortices exist with $Ro<-0.5$. The thick black dashed line in the lower left corner indicates the locus over which $N_c^2=0$, i.e., $Bu=-Ro(1+Ro)$, with $N_c^2<0$ for vortices with smaller $Ro$ or $Bu$. Panel (a): The thick solid (blue, in colour) and thick dotted (red, in colour) lines indicate the iso-surface where $\sigma$ of the fastest-growing eigenmode is  $0.02\:\tau^{-1}$. The region bounded by this iso-surface, the thick black dashed curve (but see the caveat in the text describing figure~\ref{f4}), and the bottom of the figure has  $\sigma<0.02\:\tau^{-1}$ (the iso-contour is to ``guide the eye'' and is approximated by interpolating among the growth rates calculated at the locations of the discrete symbols). The symbols denote the spatial symmetry of the fastest-growing eigenmode, with  diamonds (blue, in colour) as S2, solid triangles (red, in colour) as A1, squares (green, in colour) as A2, hollow triangles as A3, and hollow circles as A4. Black solid circles correspond to vortices for which the most unstable eigenmodes have growth rates slower than $0.02\:\tau^{-1}$. Panel (b): Four iso-contours [approximated as in (a)] of growth rate $\sigma$ of the fastest-growing eigenmode. Each contour consists of one solid curve (blue, in colour) and one or two dotted curves (red, in colour). The fastest-growing eigenmodes along the dotted curves (red, in colour) have A1 symmetry and along the solid curves (blue, in colour) have S2 symmetry. The small rectangular box near $Bu=1$ is to guide the eye and shows the approximate, very small, region where all of the eigenmodes of the cyclones are neutrally stable. The $\sigma$ and the symmetries of the most unstable eigenmodes with $\sigma>0.02\:\tau^{-1}$ for vortices with $N_c^2> 0$ in panel (a) are given in Appendix~\ref{appD}.} 
    \label{f2}
\end{figure}

The parameter map of stability in the $Ro-Bu$ space is shown in figure~\ref{f2}(a). Gaussian anticyclones do not exist
with $Ro<-0.5$ (see \S\ref{subsec:initial_equilibria}). The region to the lower left of the thick dashed black curve corresponds to equilibrium Gaussian vortices for which $N_c^2<0$ [or $Bu<-Ro(1+Ro)$ according to (\ref{eq:slaw})]. These vortices are not unphysical, but near their cores they have heavy fluid above light fluid (i.e., $\partial \rho/\partial z>0$ at the vortex center).  

As shown in figure~\ref{f2}(a), the most unstable eigenmodes (i.e, those with the largest growth rates) of the vortices generally have either S2 or A1 symmetries. A few points in the figure correspond to vortices for which the fastest-growing eigenmode is A2, A3 or A4. We found that no vortex had a fastest-growing eigenmode with a symmetry different from those just listed. To our surprise, only $4$ out of the 130 vortices that we examined were neutrally stable. All the neutrally-stable vortices were cyclones with $0.02 \lesssim Ro \lesssim 0.05$ and $0.8 \lesssim Bu \lesssim 1$.  The neutrally-stable eigenmodes are denoted in figure~\ref{f2}(a) as solid  circles in the region circumscribed by a small rectangle. The rectangle is to ``guide the eye'' and is  used to denote the approximate boundary of the region of neutral stability. Computing the actual boundary between the regions where vortices are all neutrally stable and where they are unstable  would be expensive and rather pointless given how small the neutrally-stable region is. Anticyclones have linear growth rates that are slow and would not destroy
a vortex in less than 50 vortex turnaround times if $0.5 \lesssim Bu \lesssim 1.3$.
For  nearly geostrophic cyclones with $|Ro| < 0.05$, linear growth rates are slow and would not destroy
a vortex in less than 50 vortex turnaround times if $0.7 \lesssim Bu \lesssim 1.2$. As $Ro$ increases, the growth rates of large-diameter cyclones
(i.e., with $Bu \lesssim 1.05$ or $L \gtrsim 0.98 L_r$) becomes faster.

Considering the smallness of the region of neutral stability, clearly, linear stability cannot be used to explain the differences between the numbers of observed cyclones and anticyclones in the oceans or in planetary atmospheres. On the other hand, ocean vortices can survive for more than $50$ of their own turn-around times, $\tau$. So, one plausible explanation of the cyclonic/anticyclonic asymmetry in the frequency of observation of mesoscale oceanic eddies and of planetary vortices might depend on the differences of the {\it growth rates} of the linear instabilities, rather than just the fact that some vortices are not linearly unstable and others are. For example, if there are physical processes (such as turbulence, interactions with other vortices or currents or boundaries) that are likely to destroy a vortex after $50\:\tau$, which is more than $\sim1/2$ year for ocean Meddies  \citep{McWilliams1985,Armi1989,Hebert1990,Pingree1993,D'Asaro1994,Prater1994,Paillet2002}, then a vortex need not be neutrally stable to be observed, it needs only have growth rates less than $\sim1/50\:\tau^{-1}$. 
So, it is plausible that the asymmetry between the numbers of observed cyclones and anticyclones depends upon the relative amount of area in $Ro-Bu$ parameter space for which the fastest-growing eigenmodes grow slower than $\sim1/50\:\tau^{-1}$, or some other critical growth rate. For Gaussian vortices, the region in $Ro-Bu$ parameter space where the growth rate of the fastest-growing eigenmode is less than $1/50\:\tau^{-1}$ (i.e., the ``slow growth region'' for linear instability) is the region bounded above by the solid (blue, in colour) and dotted (red, in colour) curves in figure~\ref{f2}(a) and to the lower left  by the thick dashed curve. Along the solid curve (blue, in colour), the fastest-growing eigenmode has S2 symmetry, whereas along the dotted curve (red, in colour) it is A1. The solid and dotted curves are drawn to ``guide the eye'', and the vortices corresponding to the black solid circles  have $\sigma<1/50\:\tau^{-1}$. In general, for large $Bu$, the fastest-growing eigenmodes have A1 symmetry, while for smaller $Bu$, they have S2 symmetry. However, for cyclones with $Bu \lesssim 0.4$, some of the fastest-growing eigenmodes also have A1 symmetry, or even A2, A3 or  A4 symmetry, and the growth rates are often faster than $1\:\tau^{-1}$. There are two regions in the $Ro-Bu$ parameter space where the fastest-growing eigenmodes of the cyclones have A1 symmetry. In the region with higher $Bu$, the growth rate  of the fastest-growing modes is smaller than that in the lower $Bu$ region, and, as discussed previously and elaborated on in \S\ref{sec:spatial_distribution}, the radial structures of the fastest-growing A1 eigenmodes in the large and small $Bu$ regions differ as well. 

Of course, our choice of $50\:\tau$ to define the ``slow growth region'' for linear instability is arbitrary, so figure~\ref{f2}(b) shows how the ``slow growth'' region changes when we change our choice from $50\:\tau$ to $20\:\tau$, $10\:\tau$, or $6.67\:\tau$. That is, the two sets of (solid/broken) curves are iso-surfaces in $Ro-Bu$ parameter space where $\sigma$ is 0.02, 0.05, 0.1, and 0.15 in units of $\tau^{-1}$. For the iso-surface for the growth rate of $0.15\:\tau^{-1}$ in figure~\ref{f2}(b), the fastest-growing eigenmode has S2 symmetry for $0.5 \lesssim Bu \lesssim 1.8$, otherwise the fastest-growing eigenmode  has A1 symmetry. Note that the iso-surfaces for the growth rates of $0.10\:\tau^{-1}$ and $0.15\:\tau^{-1}$ are very close to each other for $Bu \lesssim  0.3$.  Most of the fastest-growing eigenmodes in the $Ro-Bu$ parameter space shown in figure~\ref{f2} have $\sigma<0.2\:\tau^{-1}$. However, cyclones in the upper left corner of figure~\ref{f2} can have  $\sigma$ of order one $\tau^{-1}$. As shown below, anticyclones to the lower left of the thick dashed curve in the lower left side of figure~\ref{f2} (with $N_c^2 < 0$) can have much larger $\sigma$.

\begin{figure}
  \centering
  \subfloat[\normalsize{(a)}]{\includegraphics[trim={4mm 18mm -10mm 2mm},clip,width=0.93\textwidth]{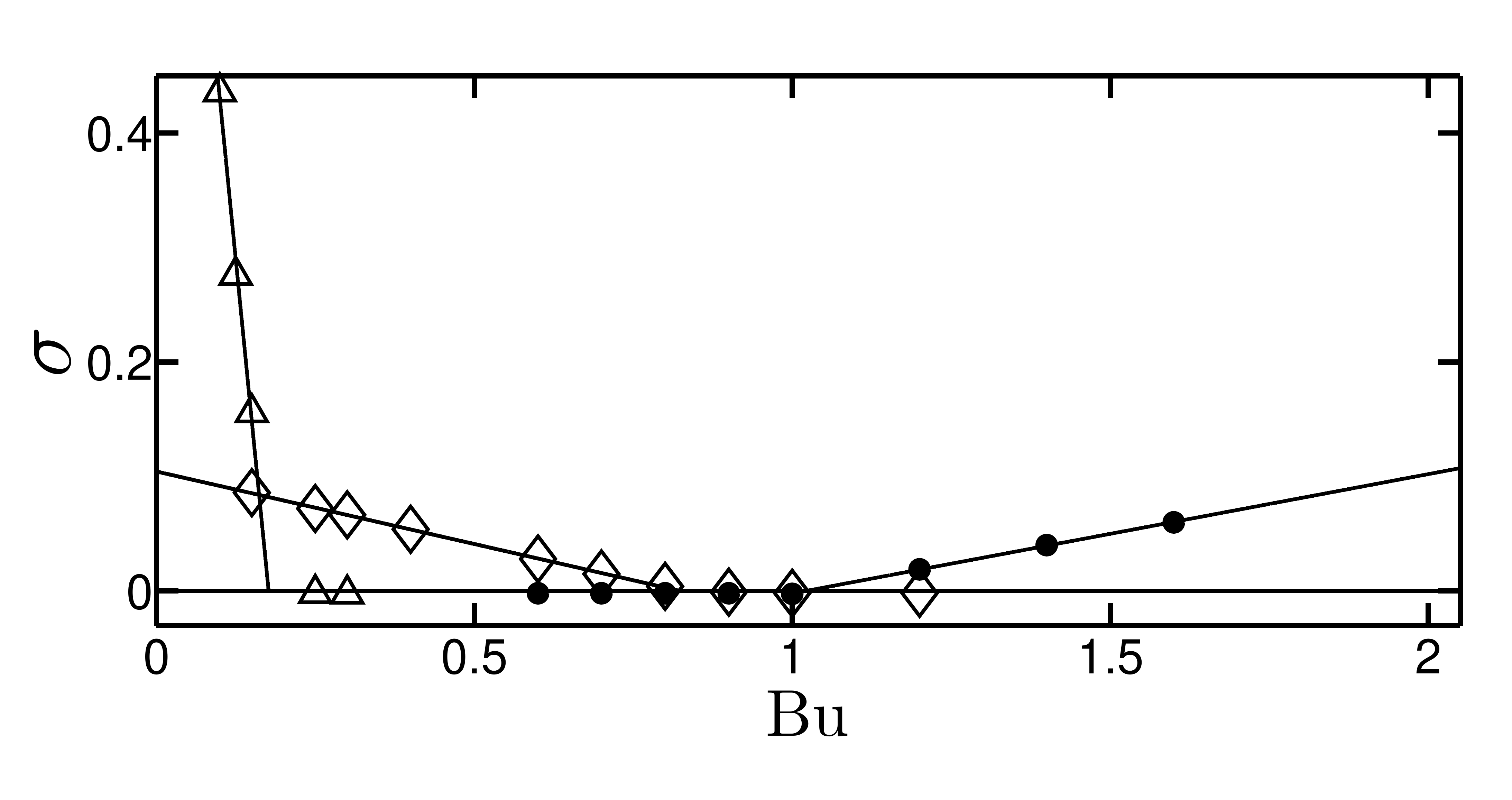}}\\
  \subfloat[\normalsize{(b)}]{\includegraphics[trim={4mm 18mm -10mm 22mm},clip,width=0.93\textwidth]{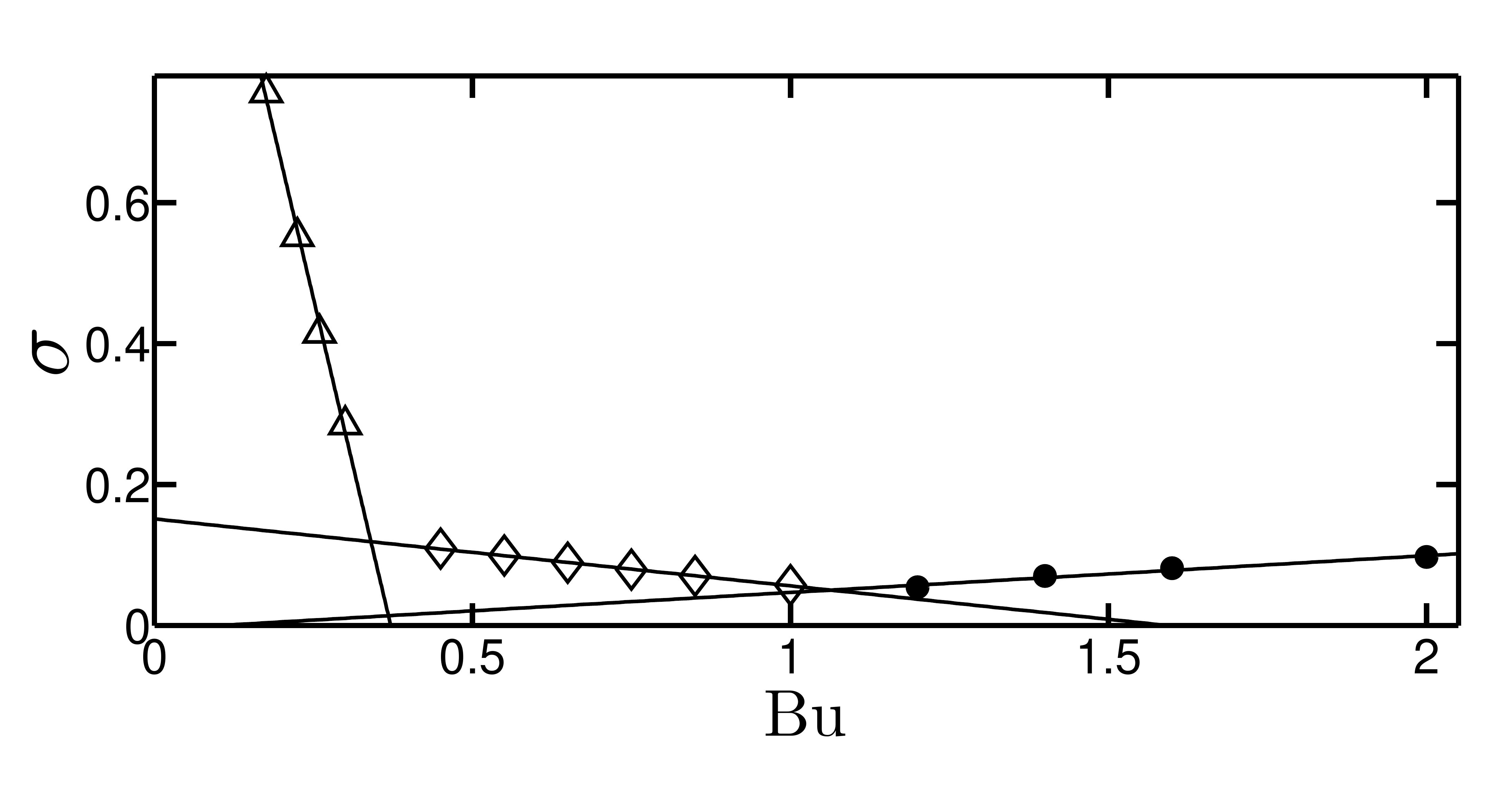}}
  \caption{Growth rates $\sigma$  (in units of $\tau^{-1}$) of the three fastest-growing modes as functions of $Bu$ for fixed $Ro$. $f/\bar{N}=0.1$. Triangles, filled circles, and diamonds, respectively  indicate the fastest growing eigenmodes at low $Bu$ (which have A1 symmetry), the fastest growing eigenmodes at high $Bu$ (which also have A1 symmetry), and  the fastest growing eigenmodes for intermediate $Bu$ (which have S2 symmetry). The three lines connecting the three sets of symbols are to ``guide the eye'' to show the three families of eigenmodes. (a) $Ro=0.05$; In this case as $Bu$ increases, the fastest-growing mode changes from A1 to S2; then all modes are linearly neutrally stable; then the fastest-growing mode is A1. (b) $Ro=0.2$; the fastest-growing mode changes from A1 to S2 and again to A1 as $Bu$ increases.}
  \label{f3_new}
\end{figure}

The growth rates of the three fastest-growing eigenmodes for $Ro = 0.05$ as functions of $Bu$ are plotted in figure~\ref{f3_new}(a) (combining figures~\ref{f3_new_after_1},~\ref{f3_new_after_2},~and~\ref{f3_new_after_3}) showing that the fastest growing 
eigenmode is A1 for $Bu \lesssim 0.2$; is S2 for $0.2 \lesssim Bu \lesssim 0.8$; and is A1 for
$1 \lesssim Bu \lesssim 2.1$. However, for $0.8 \lesssim Bu \lesssim 1$, the eigenmodes
are all neutrally stable. This region of neutral stability is consistent with the neutrally stable region shown in figure~\ref{f2}. The change in the spatial symmetry from A1 to S2 back to A1 of the fastest growing eigenmode as $Bu$ increases was discussed in \S\ref{sec:criticallayers} and it was shown that i) the family of eigenmodes continues to exist even after the eigenmodes become neutrally stable, and ii) the A1 modes at small and large $Bu$ belong to two different families of eigenmodes. Similar changes in the symmetries of the most unstable mode are observed at $Ro = 0.2$ (figure~\ref{f3_new}(b)); however, at $Ro = 0.2$ there is not a region where the vortex is neutrally stable to all eigenmodes. Similar to $Ro=0.05$, the two families of A1 eigenmodes shown in figure~\ref{f3_new}(b) with triangles and with filled circles are distinct families with different radial structures. How these results, particularly at the small $Ro$ of $0.05$, compare with those obtained from analyzing the QG equations is discussed in \S\ref{sec:comp}.  

\begin{figure}
\centering
  \subfloat[]{\includegraphics[trim={8mm 0mm 10mm -8mm},clip,width=1.0\textwidth,height=0.36\textheight]{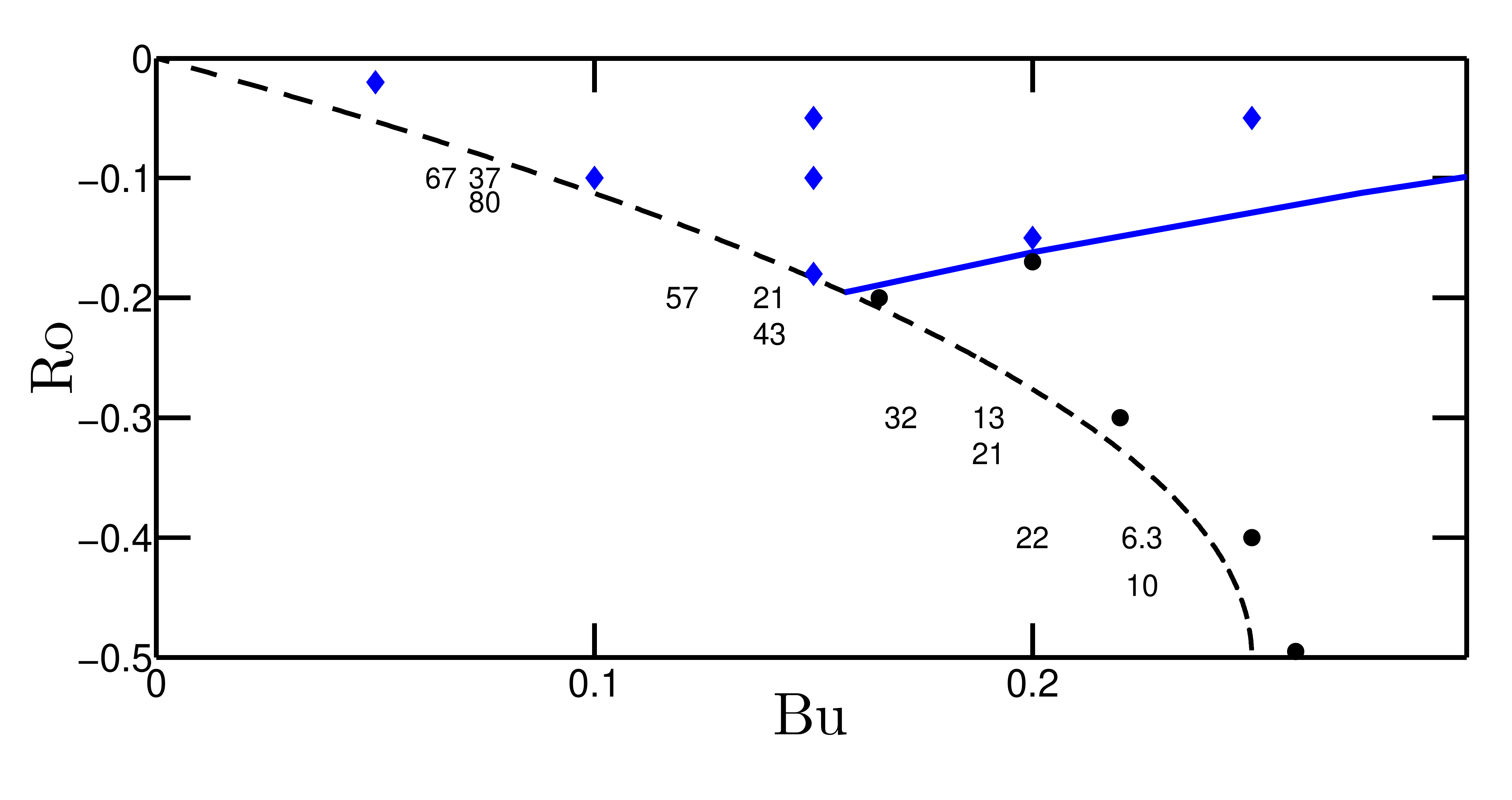}}
  \caption{(Colour online) Blow up of the lower left corner of figure~\ref{f2}(a), showing details of the eigenvalues in the region where the Gaussian vortices have $N^2_c<0$. The axes of the figure, line styles, and symbols have the same meaning as they do in figure~\ref{f2}(a). In the lower left region, below the broken line, numbers rather than symbols are used to indicate where in parameter space we have carried out linear stability calculations. The numbers are the values $\sigma$ (in units of $\tau^{-1}$) of the fastest-growing eigenmode (which in all cases has an A4 symmetry).}
  \label{f4}
  \end{figure}

\begin{figure}
    \centering
    \subfloat[\normalsize{(a)}]{\includegraphics[trim={0 6mm 0 -2mm},clip,width=0.94\textwidth]{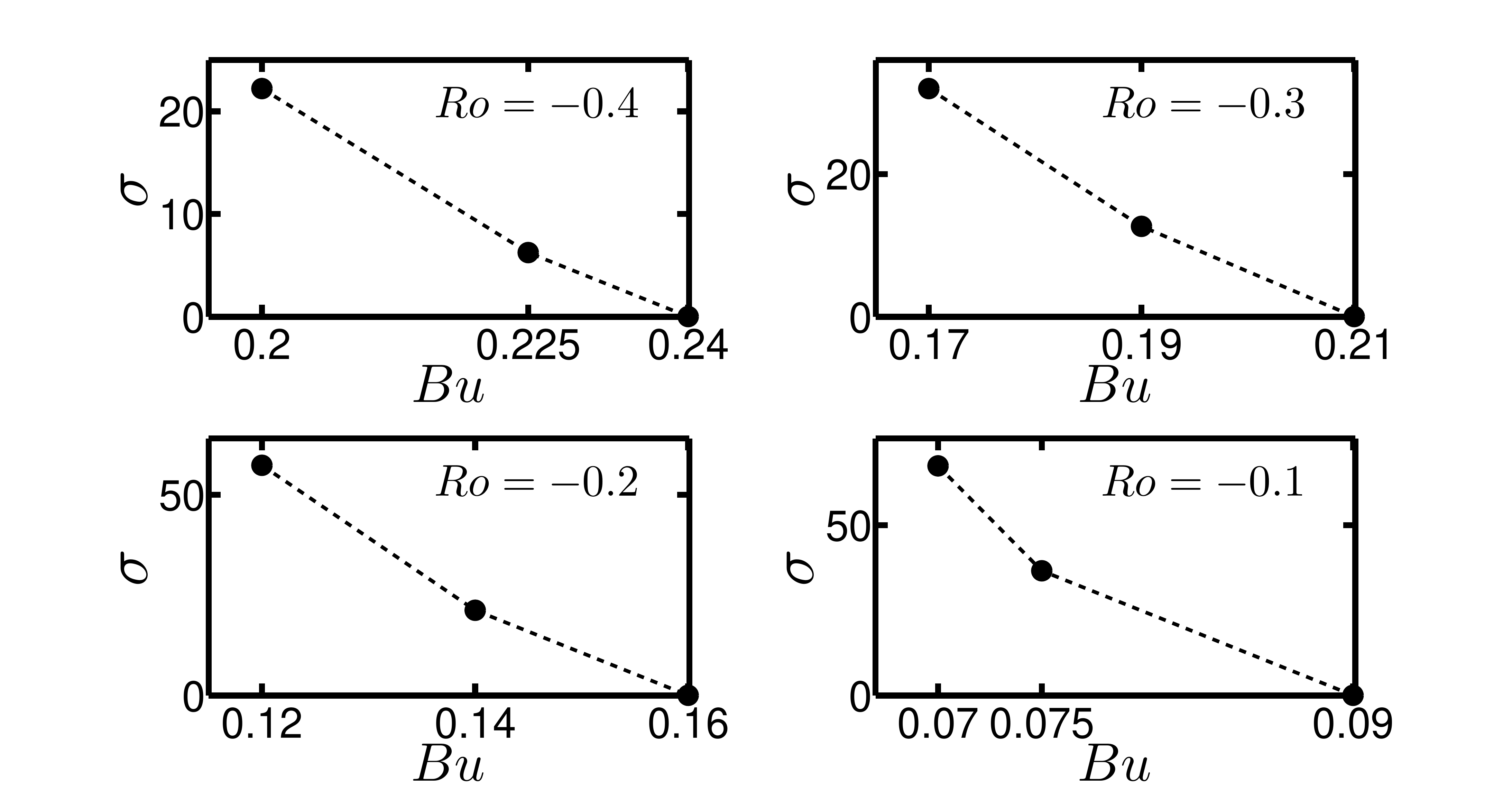}}\\
    \subfloat[\normalsize{(b)}]{\includegraphics[trim={0 6mm 0 12mm},clip,width=0.94\textwidth]{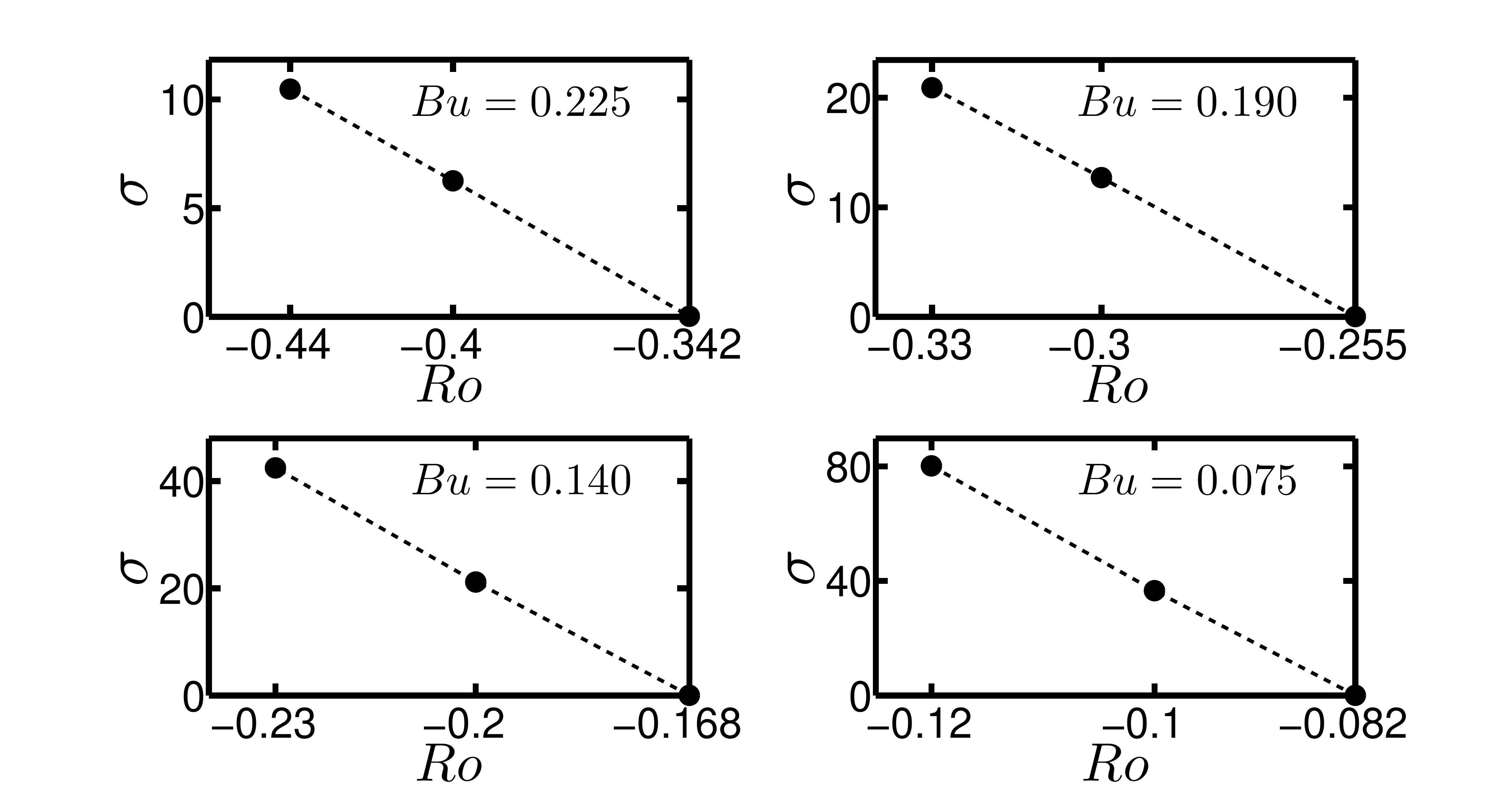}}
    \caption{Growth rates  (in units of $\tau^{-1}$) for the most unstable eigenmode of vortices with $N_c^2<0$ for fixed $Ro$ and $Bu$. $f/\bar{N}=0.1$. For all vortices examined in this region, the fastest-growing eigenmode has A4 symmetry; Panel (a): $\sigma$ as a function of  $Bu$ for $Ro=-0.4,\,-0.3,\,-0.2$ and $-0.1$; Panel (b): $\sigma$ as a function of  $Ro$ for $Bu=0.225,\,0.19,\,0.14$ and $0.075$. In each panel, the value of the horizontal coordinate axis on the  right side of the panel corresponds to a vortex with $N_c^2=0$. The dotted lines are to ``guide the eye''.}
    \label{f5}
\end{figure}

The growth rates for region with statically unstable vortex cores, i.e., with $N_c^2 < 0$, are shown in figure~\ref{f4}. Eigenmodes for this region have A4 symmetry and the growth rates can be as large as  $\sim100\:\tau^{-1}$. The $\sigma$ as a function of $Bu$ (for fixed $Ro$), and as a function of $Ro$ (for fixed $Bu$) for vortices with $N_c^2 < 0$ are shown in figures~\ref{f5}(a)~and~(b), respectively. In each of the eight panels, the value of the horizontal coordinate axis on the right side of the panel corresponds to a vortex with $N_c^2=0$ (i.e., a point on the thick dashed curve in figure~\ref{f2} or in the broken curve in figure~\ref{f4}).  The figure shows that $\sigma$ increases rapidly as a function of distance from the $N_c^2=0$ boundary. Due to this rapid growth in $\sigma$, for all practical purposes we can consider the thick dashed line at $N_c^2=0$ to be the left boundary of the region in figure~\ref{f2}(b) in $Ro-Bu$ for which $\sigma<0.02\:\tau^{-1}$, and also the boundary for the region $0.02\:\tau^{-1}\le\sigma<0.05\:\tau^{-1}$, and for the region $0.05\:\tau^{-1}\le\sigma<0.10\:\tau^{-1}$.

\subsection{Comparison with previous studies}\label{sec:comp}
As discussed in \S\ref{sec:introduction}, this paper extends the analyses of previous studies by using the full 3D non-hydrostatic Boussinesq equations and by employing the 3D Gaussian vortex model, which has continuous velocity and density (and PV) fields and is initially in exact equilibrium. The latter is necessary for a rigorous linear stability analysis. A comparison of our results with those of many previous studies is not straightforward because various different vortex models and flow models have been used. Below we compare our parameter map of stability with the results of the most relevant study in the QG limit \citep{Nguyen2012} and with the results of several relevant studies using multi-layer models. We also discuss the results of \citet{Yim2016}, who used the full Boussinesq equations but studied a different family of vortices.

In the limit of vanishing $Ro$, the most relevant study to ours is that of \citet{Nguyen2012}, who numerically calculated the unstable modes of a Gaussian vortex using the QG equations. They found that the fastest-growing mode changes from S2 to A1 around $Bu=1$, which along with the general dependence of the growth rate of the fastest-growing mode on $Bu$ in their figure~1(a) agrees overall with the results of current study (see figure~\ref{f3_new}(a) which is for $Ro=0.05$). However, they also found that for $Bu$ as small as $0.05$, modes with higher $m$ dominate. In our results, for anticyclones, as $Bu$ decreases, the most unstable mode changes from S2 to A4 once the vortex becomes statically unstable (this instability is not considered in the QG framework used by \citet{Nguyen2012}). For cyclones, as $Bu$ decreases, the most unstable mode changes from S2 to A1 for small $Ro$ and to A2 or A3 for moderate $Ro$ (see figure~\ref{f2}).   

There are a number of studies which have used the shallow-water equations with the Gaussian vortex model and are relevant to current work. Consistent with the results of our analysis, these studies find that anticyclones become more stable as the absolute value of the Rossby number increases, whereas for cyclones the growth rates decrease with decreasing the Rossby number \citep{Stegner2000,Baey2002,benilov2008effect}. (In this section note that our results are only for vortices with stably-stratified interiors.)

How the growth rates in these studies vary with the Burger number, however, shows a strong dependence on the vertical structure of the vortex and the background flow. \citet{Stegner2000} studied the stability of isolated Gaussian vortices using a $1-1/2$~layer model and found that for vortices with small Rossby numbers, the growth rate decreases with decreasing the Burger number. This is consistent with our results only for $Bu \gtrsim 1$. \citet{benilov2008effect} used a two-layer model to examine the stability of the ``compensated" (i.e., $\boldsymbol{v}=\boldsymbol{0}$ in the bottom layer) Gaussian vortices, and also Gaussian vortices with uniform PV in the lower layer. They found that compensated vortices are neutrally stable for intermediate Burger numbers, while vortices with uniform PV in the lower layer are neutrally stable for Burger numbers smaller than a critical value of order 1. \cite{Baey2002} studied two-layer Gaussian vortices and found, in contrast to the previous results and those of ours, that the growth rate decreases with Burger number for both cyclones and anticyclones and the eigenmodes are stable for Burger numbers larger than a critical value. It is apparent that identifying a unique stability behavior with Burger number in these studies is difficult and the behavior is highly dependent on the vertical structure of the flow/vortex. An example of such dependence is given by \cite{Sutyrin2015}, who examined two and three layer compensated shallow water vortices and showed that the addition of a third middle layer with uniform PV weakens the coupling between the upper and lower layers and enhances the stability of vortices. Considering these results, comparing the Burger number dependence of the stability behavior of 3D vortices in continuously-stratified Boussinesq flows and vortices in shallow water and layer models is not particularly useful. 

Only few studies have used the full Boussinesq equations, and even those have focused on very different vortex models such as barotropic Taylor columns \citep{Smyth1998}, evolving (out-of-equilibrium) 3D vortices interacting with large-scale internal waves \citep{Suzuki2012}, out-of-equilibrium, ellipsoidal 3D vortices with discontinuous PV profiles \citep{Tsang2015}, and 3D equilibrium vortices with Gaussian angular velocity \citep{Yim2016}. Here we focus on the latter, because the main difference between our analysis and that of \citet{Yim2016} is in the vortex model: Gaussian pressure anomaly in the current study versus their Gaussian angular velocity (also note that the flow in their study is not inviscid). Such comparison provides some understanding of how the stability properties depend on the vortex profile.          

\citet{Yim2016} conducted a linear stability analysis of 3D equilibrium vortices with Gaussian angular velocity in unbounded, rotating, stratified flows for a wide range of Rossby number, $|Ro| \le 20$. Here we only focus on their results for $|Ro| \le 0.5$ and inviscid and non-diffusive flows, which are relevant to the present study. Consistent with our results, for $Bu \gtrsim 1$, they found A1 as the most unstable mode for both cyclones and anticyclones (their figures~39(d)~and~(f)), which they attributed to the instability mechanism of \citet{Gent1986} (this is also consistent with the results of \citet{Smyth1998} for Taylor columns). For $Bu \lesssim 1$, \citet{Yim2016} found anticyclones neutrally stable for $0.5 \lesssim Bu \lesssim 1$ (while we found them weakly unstable), and they found S2 as the most unstable mode for anticyclones between the statically-unstable region and $Bu \sim 0.4-0.5$ (depending on $Ro$), which is consistent with our results. For cyclones with $Bu \lesssim 1$, \citet{Yim2016} found a neutrally-stable region between $0.5 \lesssim Bu \lesssim 1$ (variable with $Ro$), which is much larger than (and encompasses) the neutrally-stable region we found; they also found that as $Bu$ decreases from one, modes with $m=2$ become the most unstable ones before modes with $m=1$ also becoming unstable at lower $Bu$, which is overall consistent with our results. At $Bu$ as low as $0.3$, the family of vortices studied by \citet{Yim2016} can have statically-unstable cyclones, while cyclones in the family of vortices we studied are always statically stable. The comparison of the results of the current study and those of \citet{Yim2016}, as summarized above, suggests that for these two vortex families, while the linear stability properties are not sensitive to the vortex profile for $Bu \gtrsim 1$, the stability properties strongly depend on the vortex profile for $Bu \lesssim 1$. Whether this behavior is generic or not requires further studies with other vortex families  

\section{Effect of $f/\bar{N}$ on linear stability} \label{sec:effect_of_foN}

Despite the fact that $f/\bar{N}$ is of order  $0.01$  in the mid-latitude oceans \citep{Chelton1998, Sundermeyer2005}, $f/\bar{N}\sim0.1$ is commonly used in studies of the oceanic vortices to reduce the computational cost; small values of $f/\bar{N}$ in explicit codes makes the equations of motion
numerically ``stiff'', which means they must be computed with  small time steps.
In this paper calculations are done with $f/\bar{N}=0.1$ for the purpose of sweeping a large region of the $Ro-Bu$ parameter space and comparing our results with those of others who have used
this value.

Several other studies \citep{Smyth1998,Sundermeyer2005,Suzuki2012,Dritschel2014,Tsang2015} have shown numerically that the stability properties and some aspects of the dynamics of vortices in rotating, stratified flows are not very sensitive to the specific value of $f/\bar{N}$ as long as this value is small. Here, we show numerically that the eigenvectors and eigenvalues of Gaussian vortices (with $N^2_c > 0$),
when properly scaled, are nearly independent of $f/\bar{N}$ for small $f/\bar{N}$. Furthermore, by properly non-dimensionalizing the linearized equations of motion, we explain the insensitivity of the eigenvalues and eigenvector structures of the fastest-growing modes to the value of $f/\bar{N}$.

\begin{figure}
\centering
\subfloat{\includegraphics[trim={90mm 1mm 115mm -8mm},clip,width=0.217\textwidth]{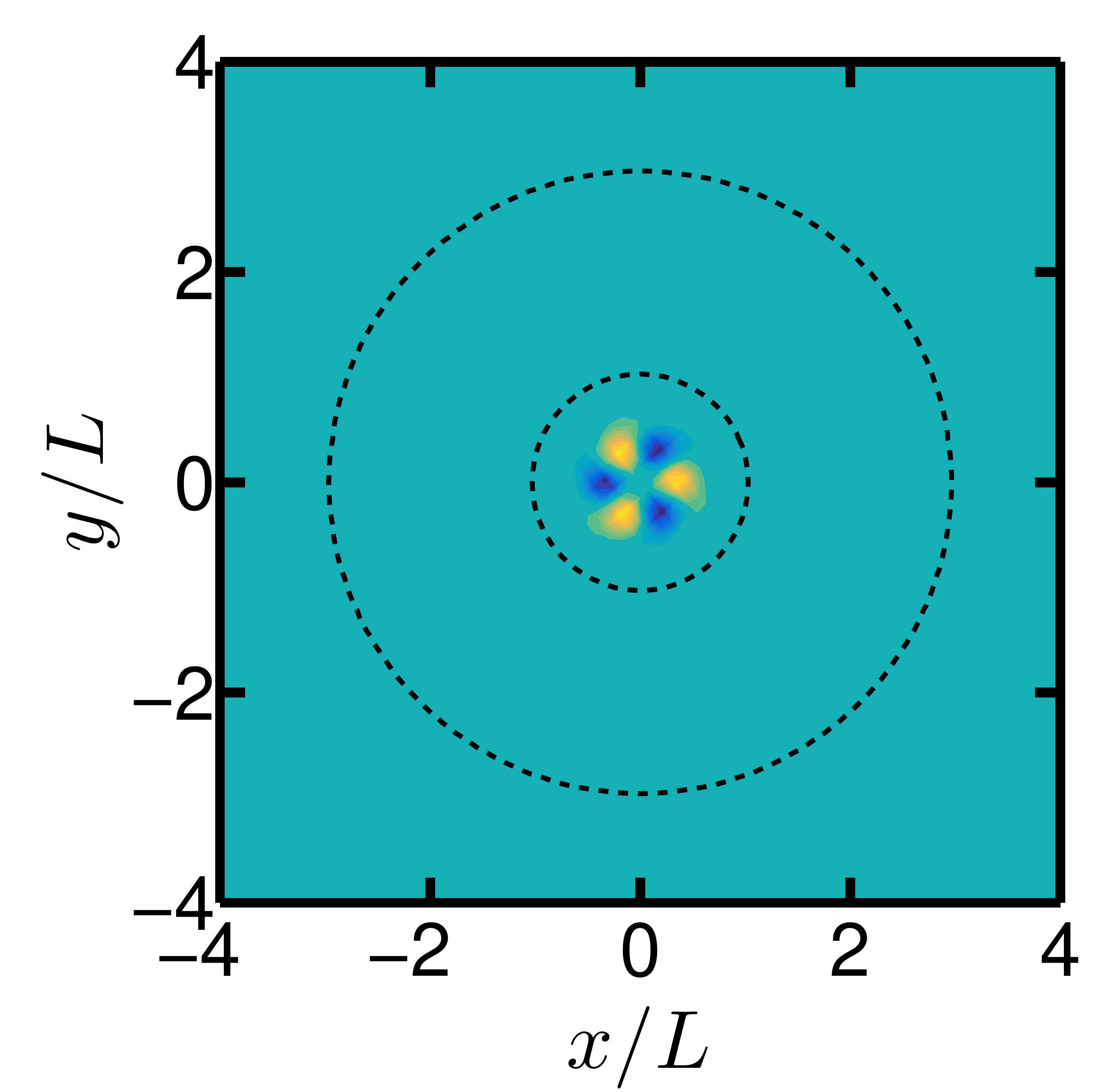}}
\subfloat{\includegraphics[trim={90mm 1mm 115mm -8mm},clip,width=0.217\textwidth]{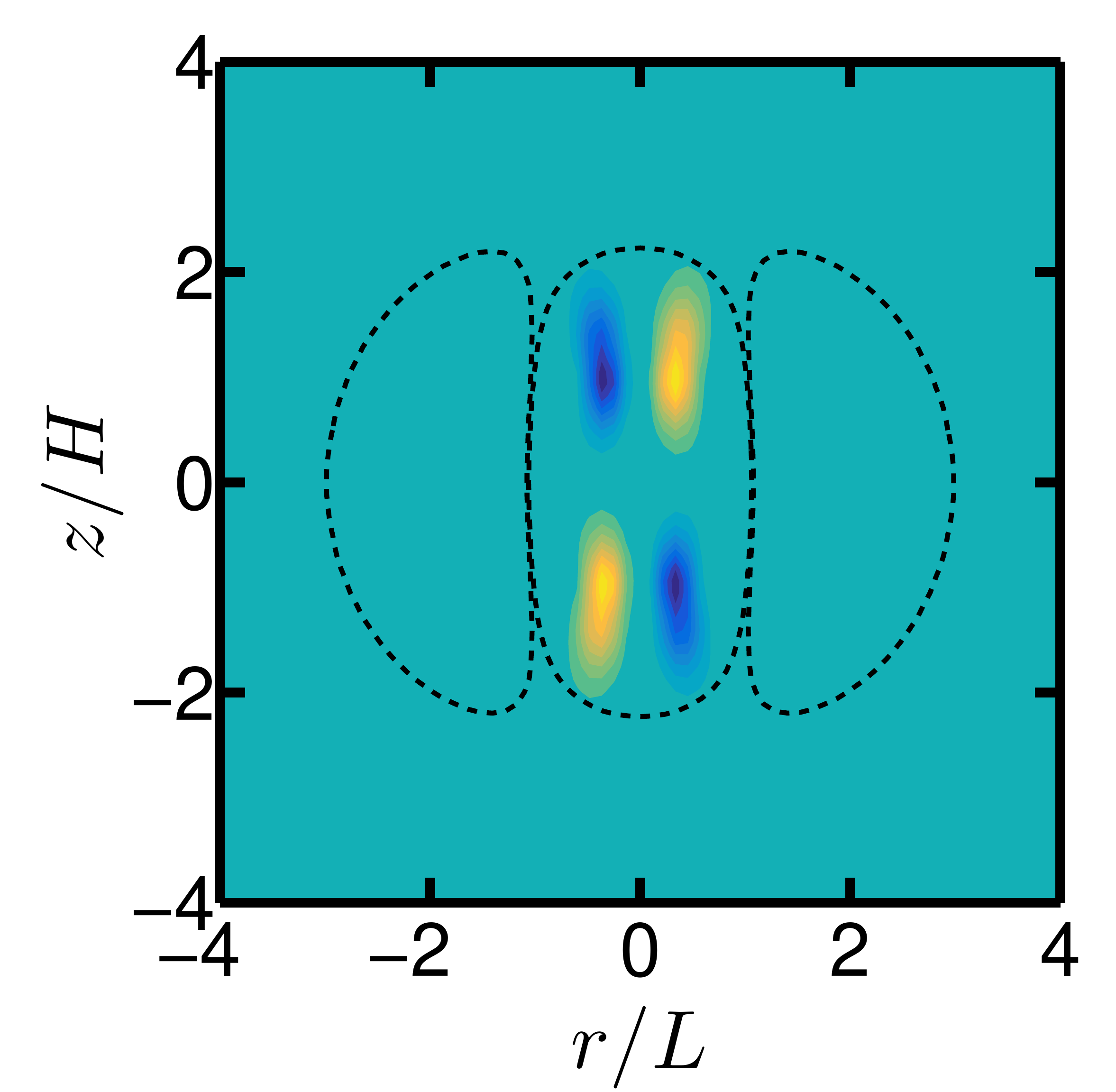}}
$~~~~~$\subfloat{\includegraphics[trim={90mm 1mm 115mm -8mm},clip,width=0.217\textwidth]{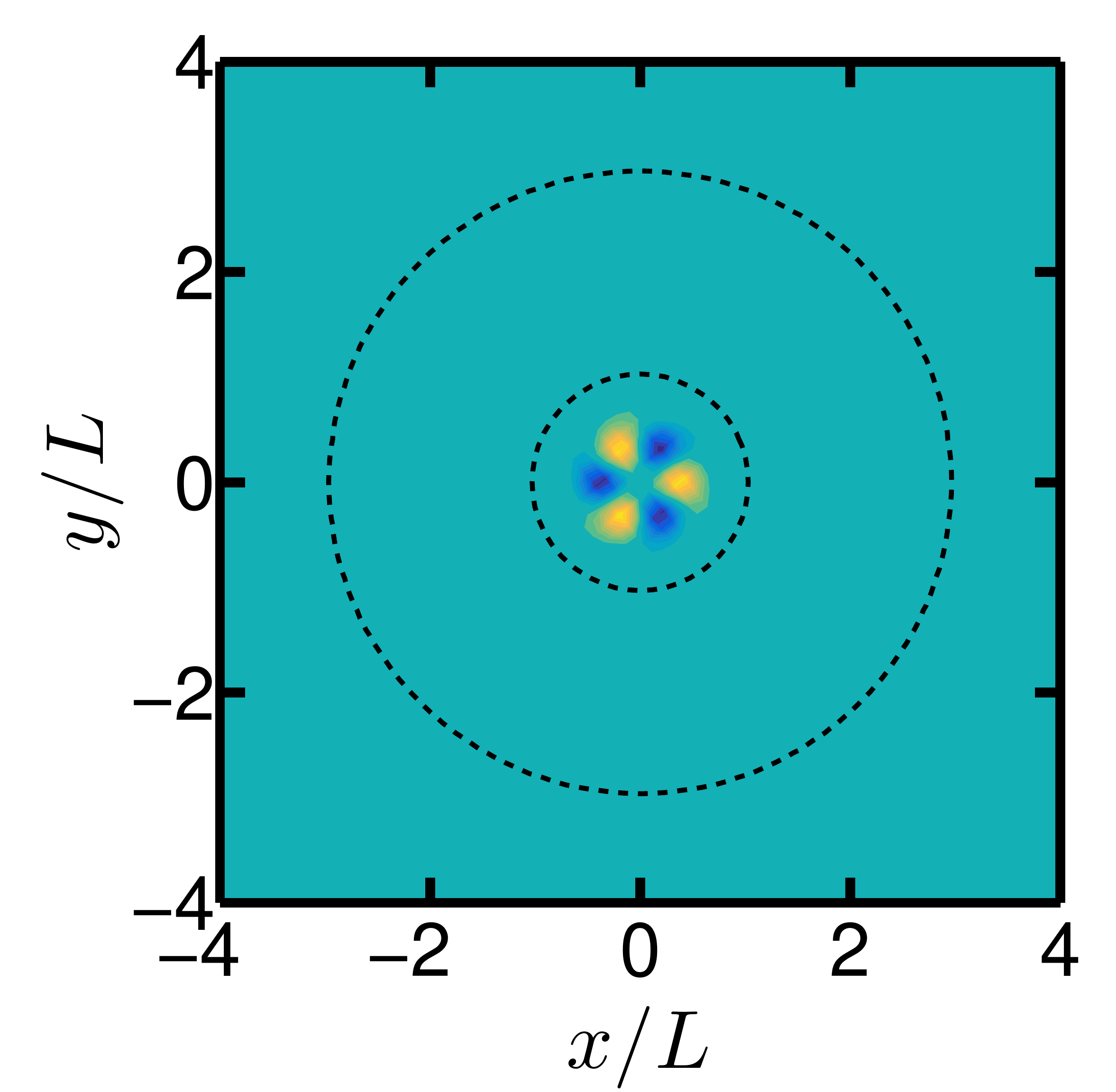}}
\subfloat{\includegraphics[trim={90mm 1mm 115mm -8mm},clip,width=0.217\textwidth]{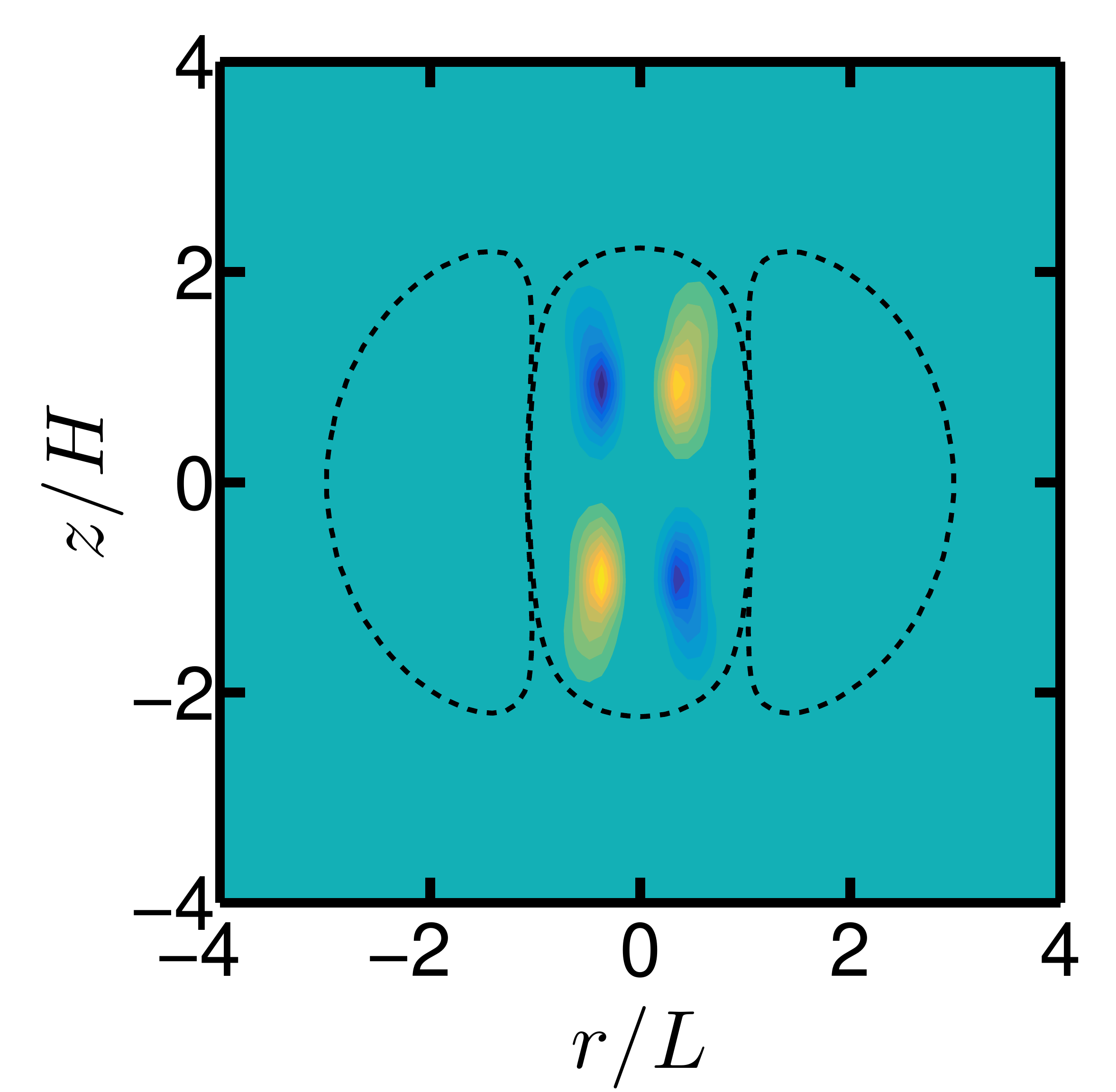}}\\
\subfloat{\includegraphics[trim={90mm 1mm 115mm 7mm},clip,width=0.217\textwidth]{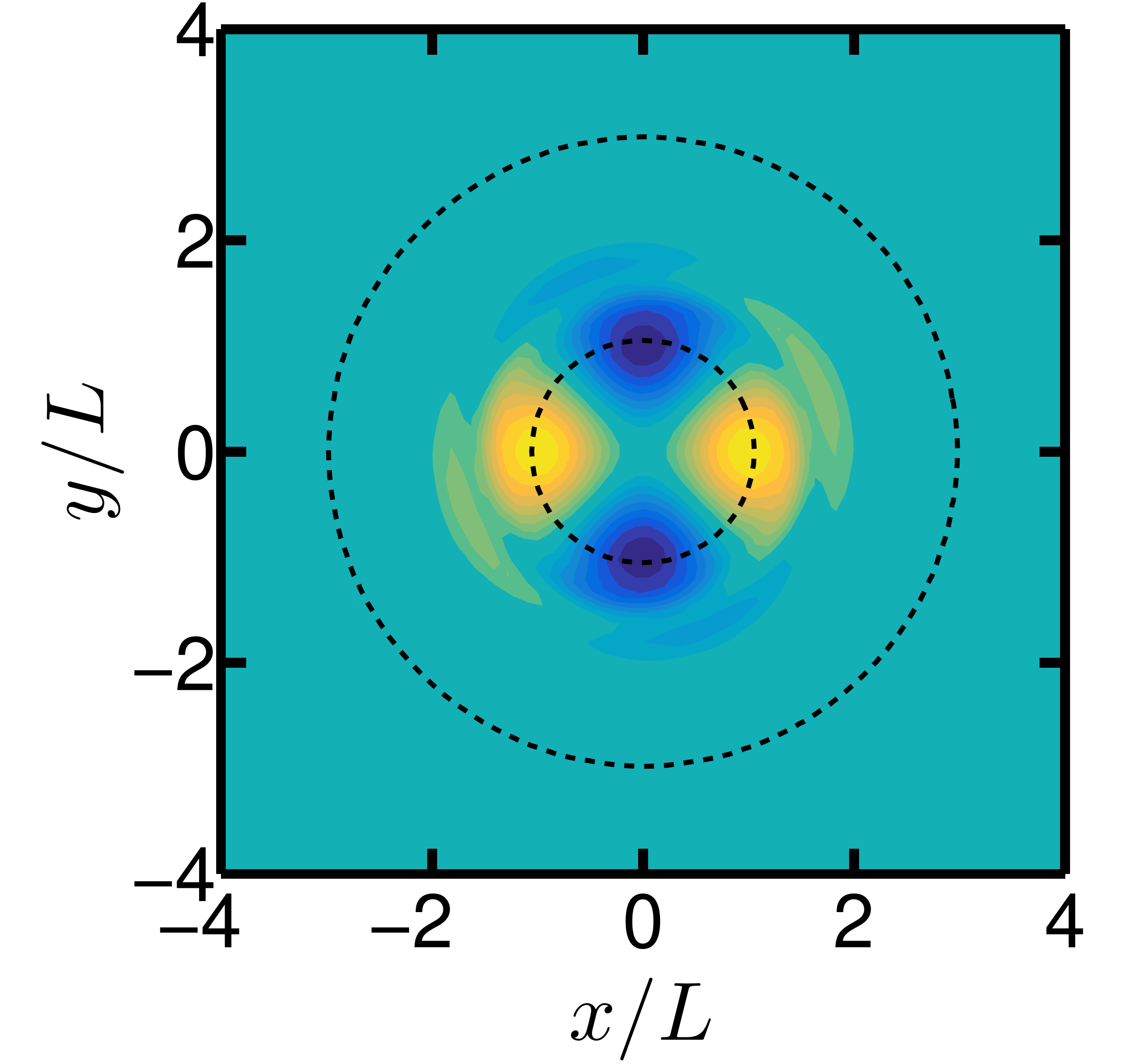}}
\subfloat{\includegraphics[trim={90mm 1mm 115mm 7mm},clip,width=0.217\textwidth]{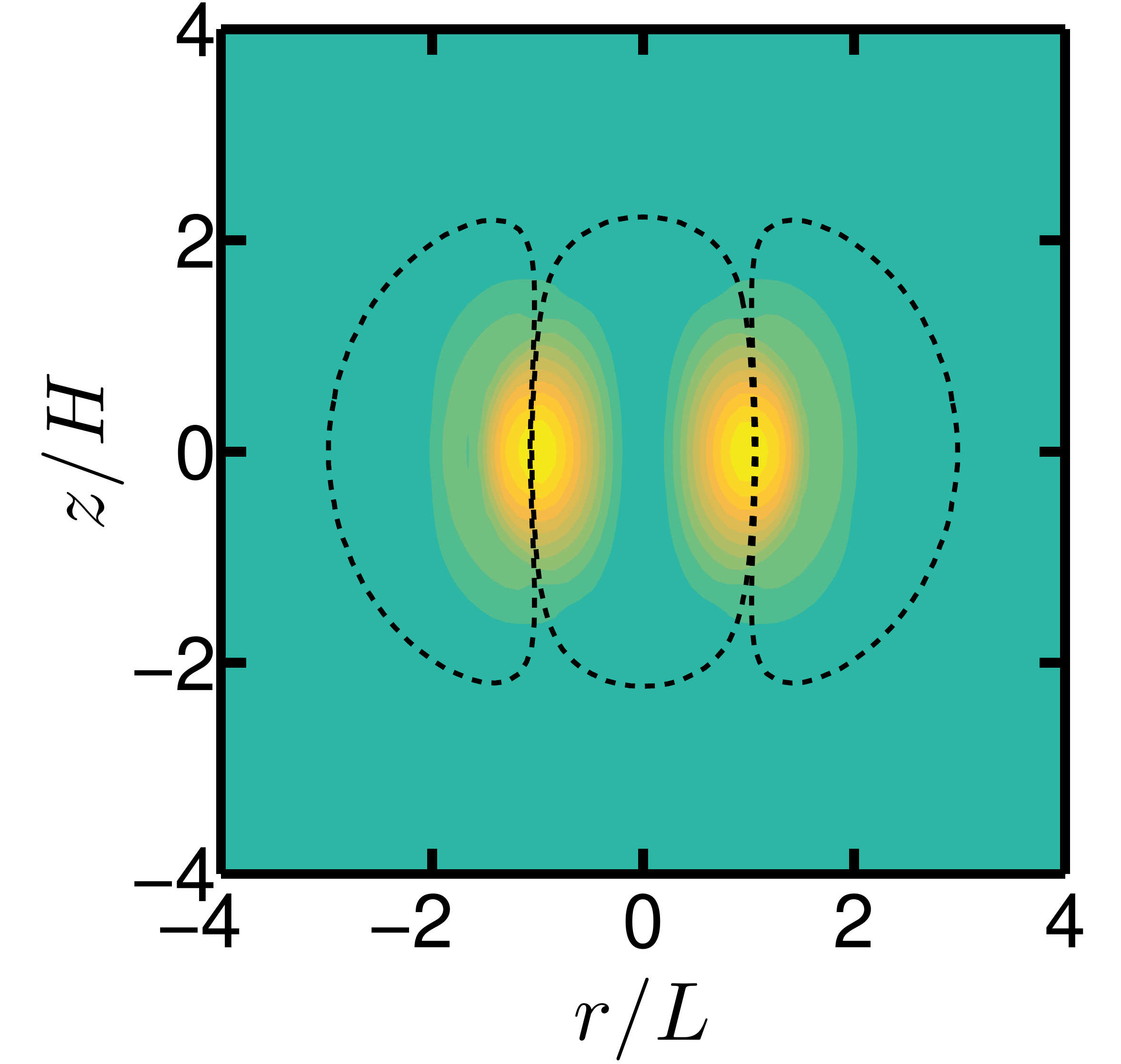}}
$~~~~~$\subfloat{\includegraphics[trim={90mm 1mm 115mm 7mm},clip,width=0.217\textwidth]{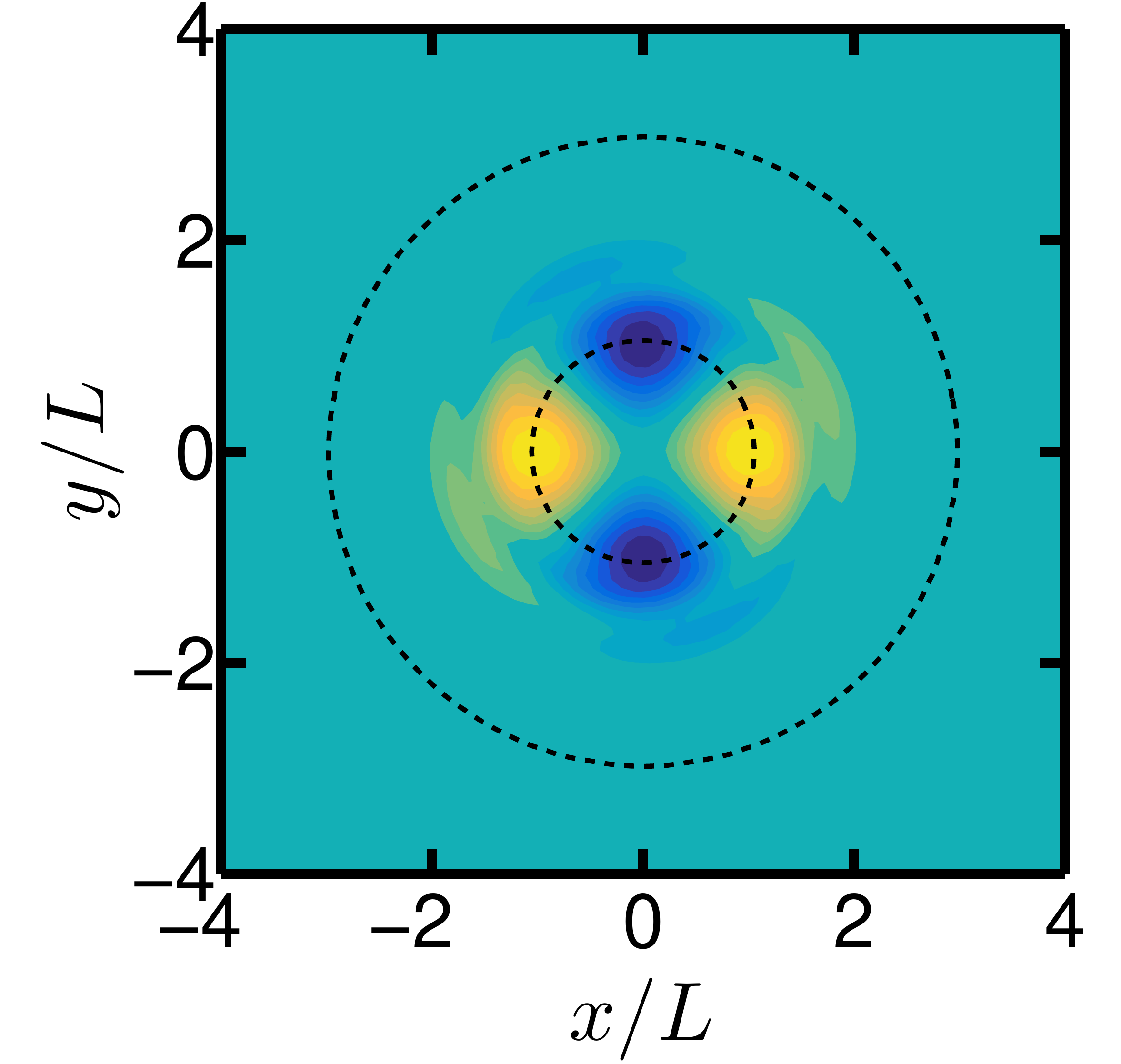}}
\subfloat{\includegraphics[trim={90mm 1mm 115mm 7mm},clip,width=0.217\textwidth]{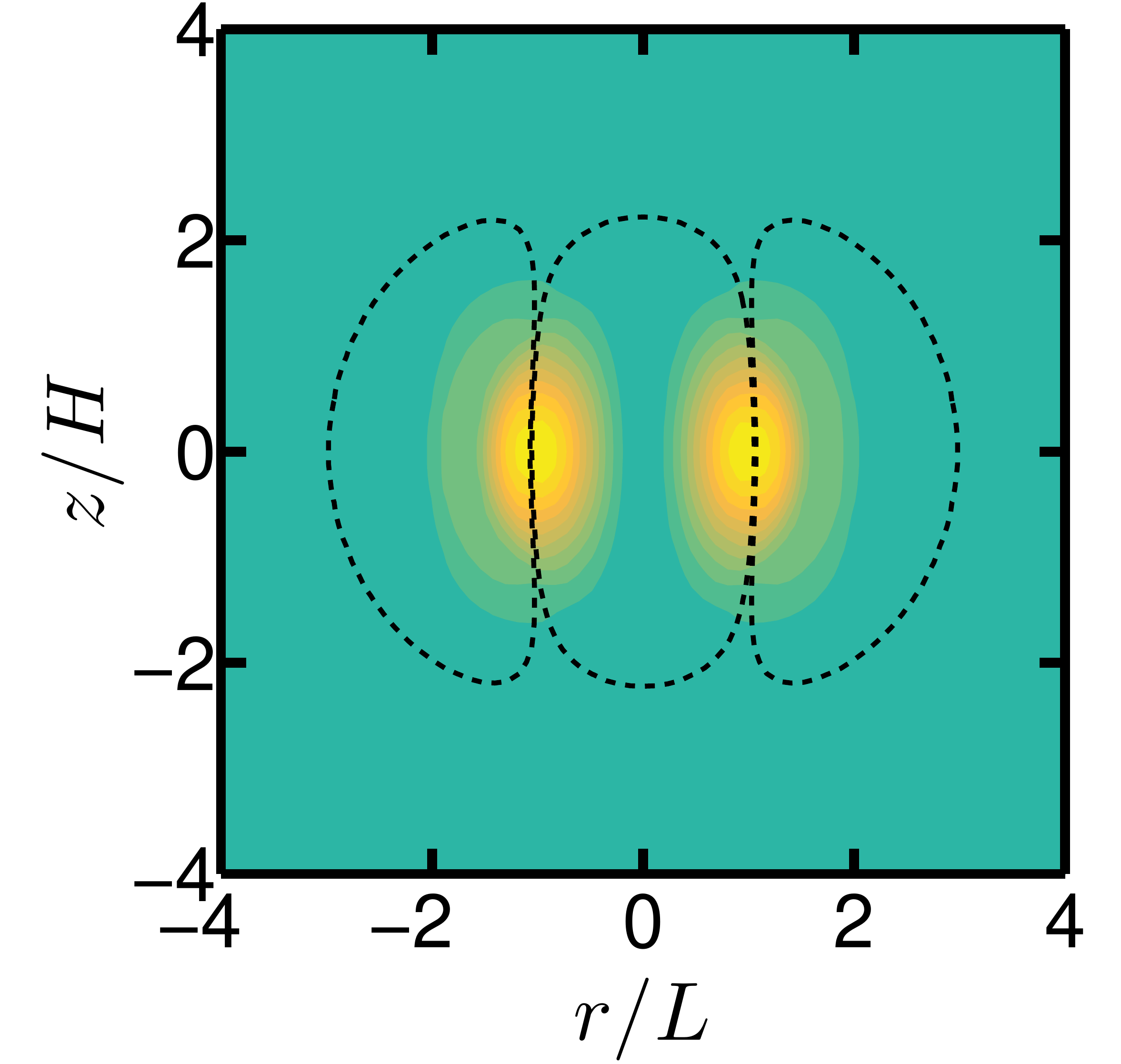}}\\
\subfloat{\includegraphics[trim={90mm 1mm 115mm 7mm},clip,width=0.217\textwidth]{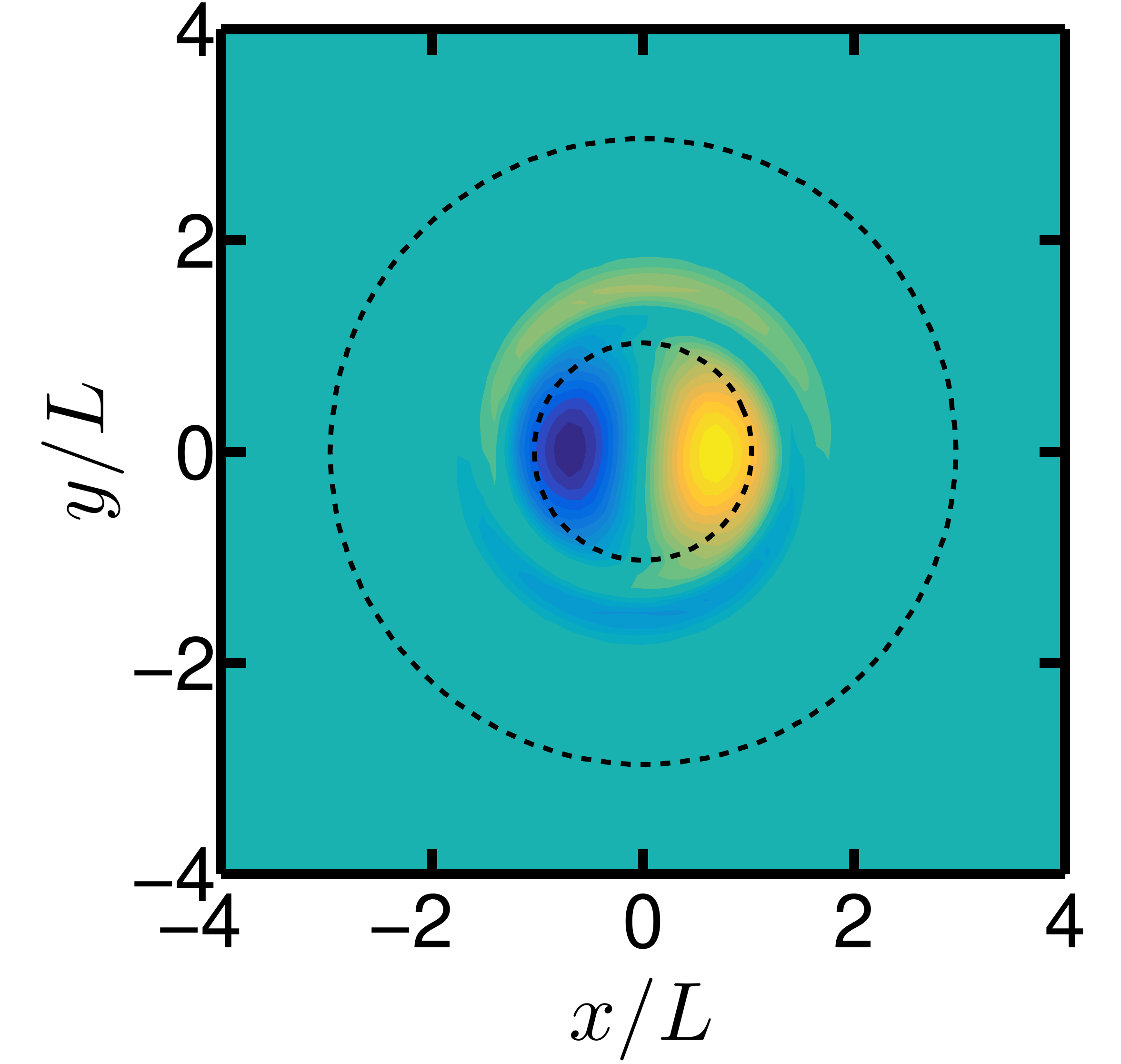}}
\subfloat{\includegraphics[trim={90mm 1mm 115mm 7mm},clip,width=0.217\textwidth]{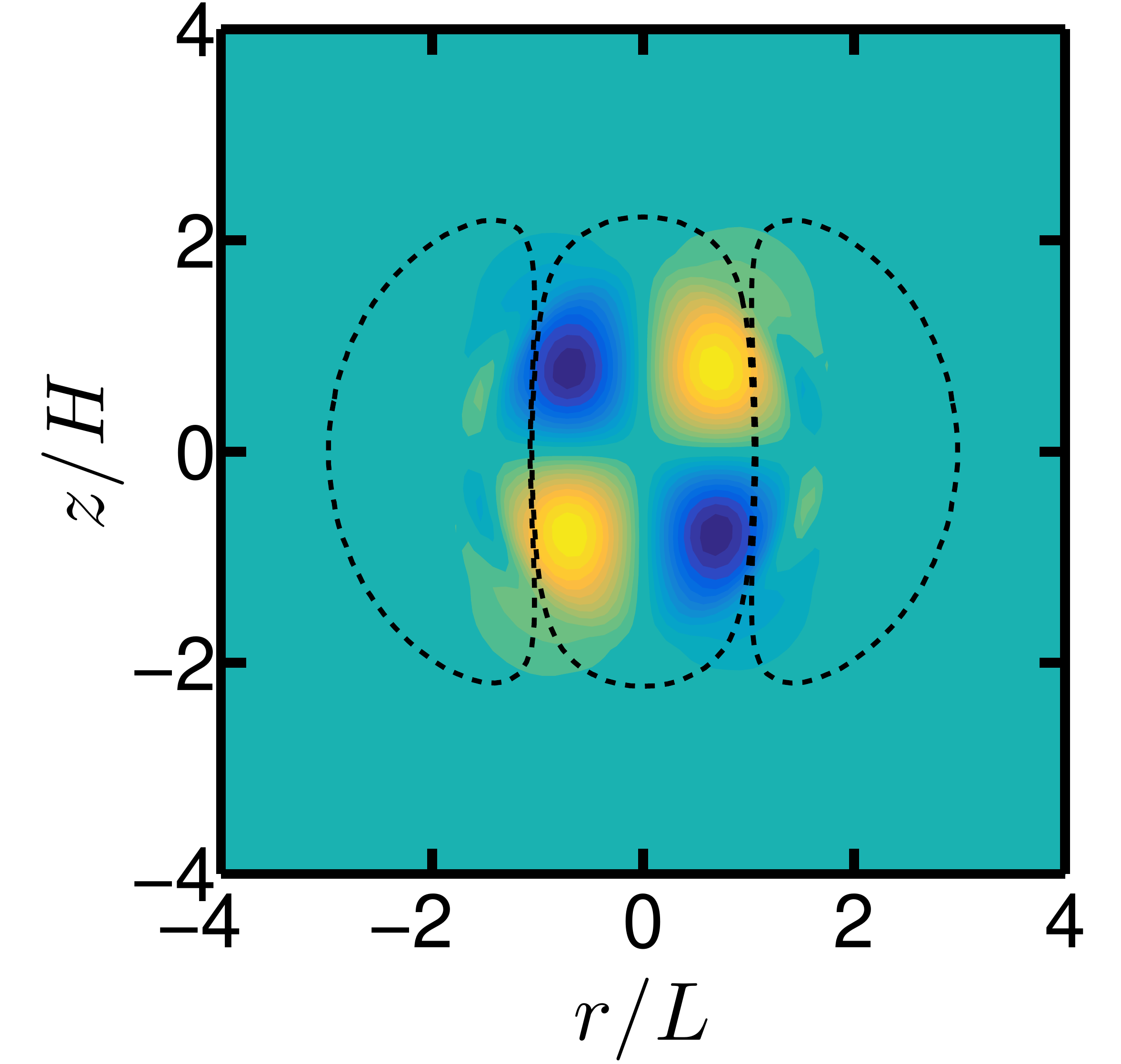}}
$~~~~~$\subfloat{\includegraphics[trim={90mm 1mm 115mm 7mm},clip,width=0.217\textwidth]{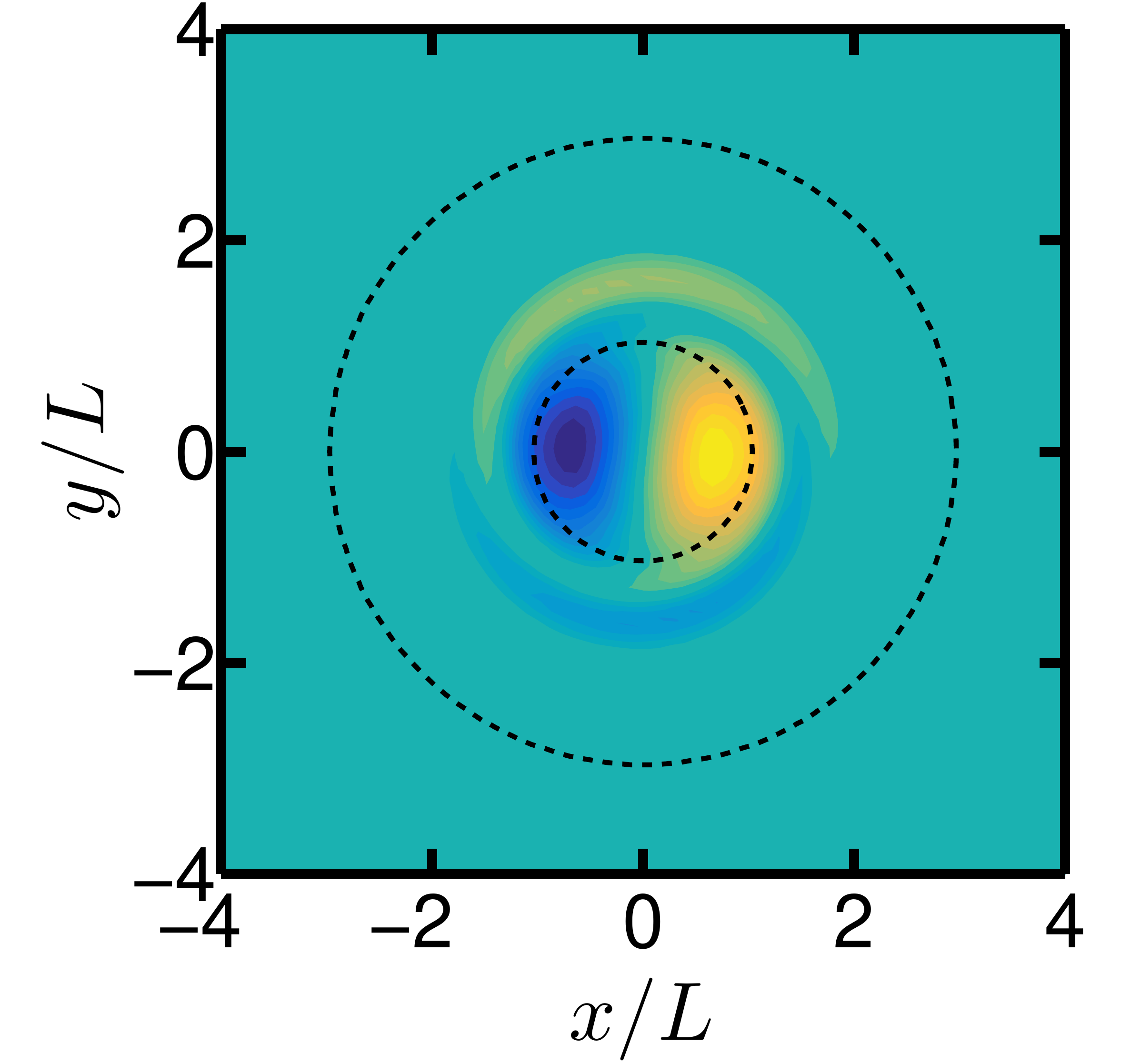}}
\subfloat{\includegraphics[trim={90mm 1mm 115mm 7mm},clip,width=0.217\textwidth]{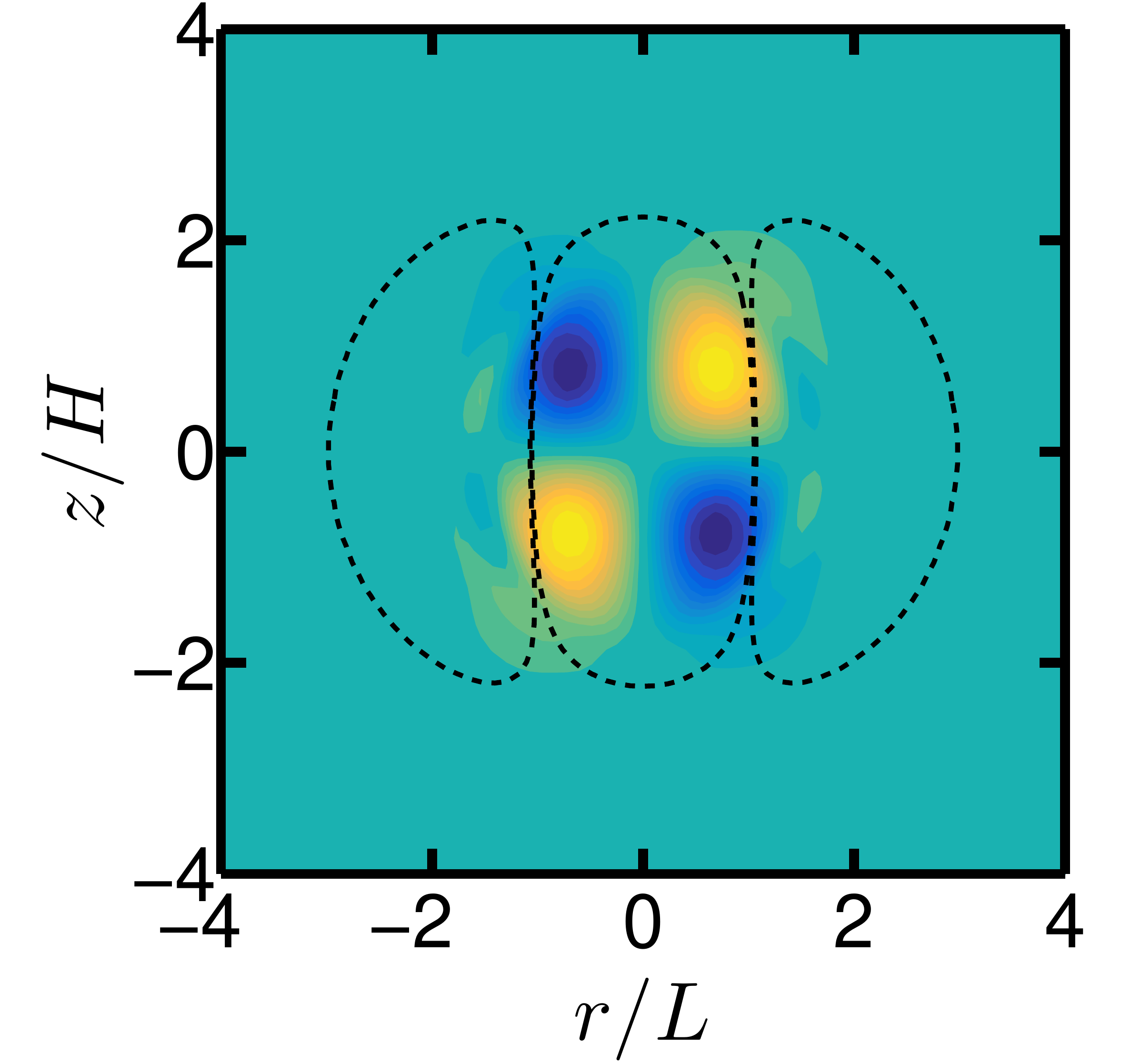}}\\
\subfloat{\includegraphics[trim={90mm 1mm 115mm 7mm},clip,width=0.217\textwidth]{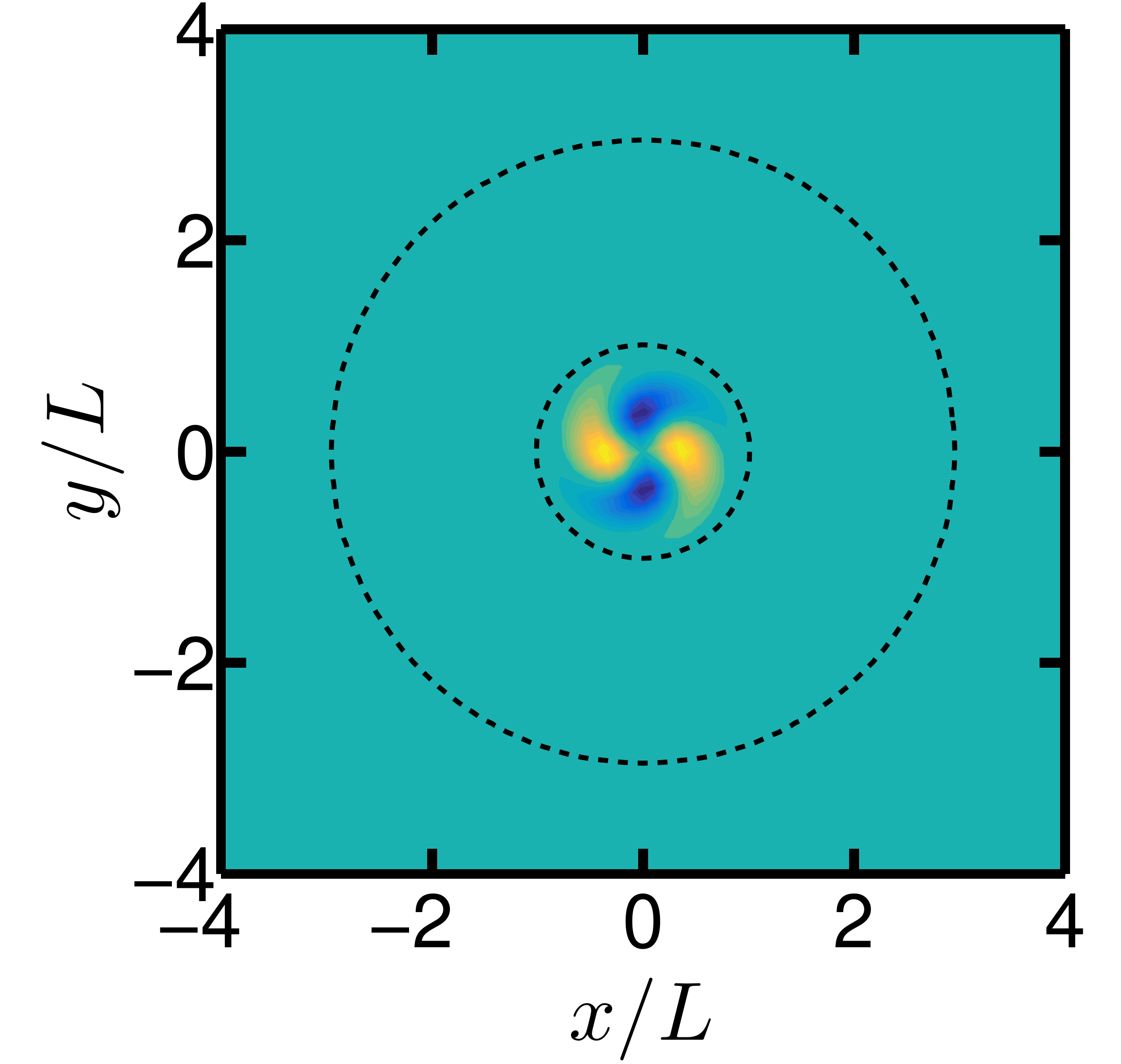}}
\subfloat{\includegraphics[trim={90mm 1mm 115mm 7mm},clip,width=0.217\textwidth]{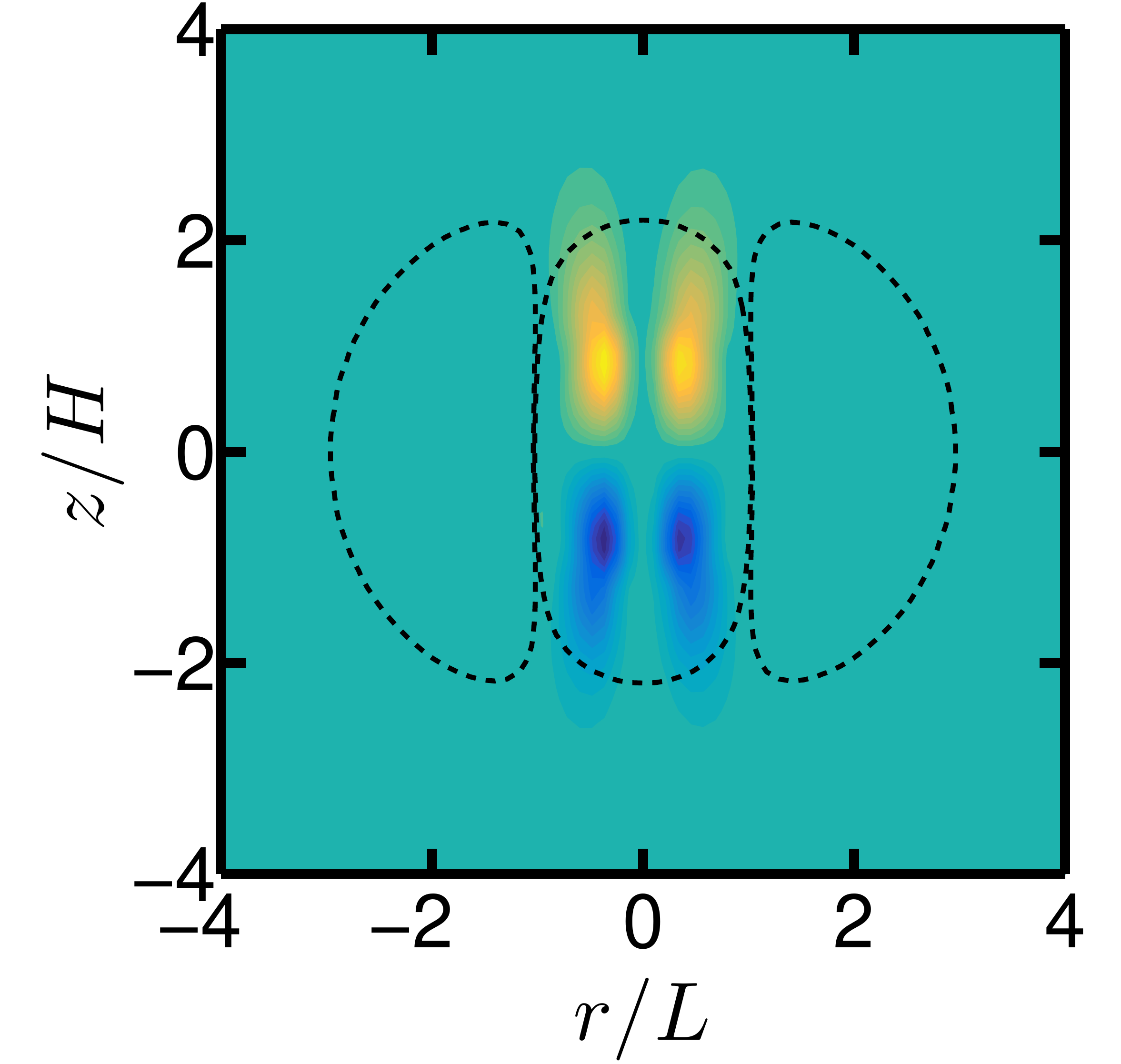}}
$~~~~~$\subfloat{\includegraphics[trim={90mm 1mm 115mm 7mm},clip,width=0.217\textwidth]{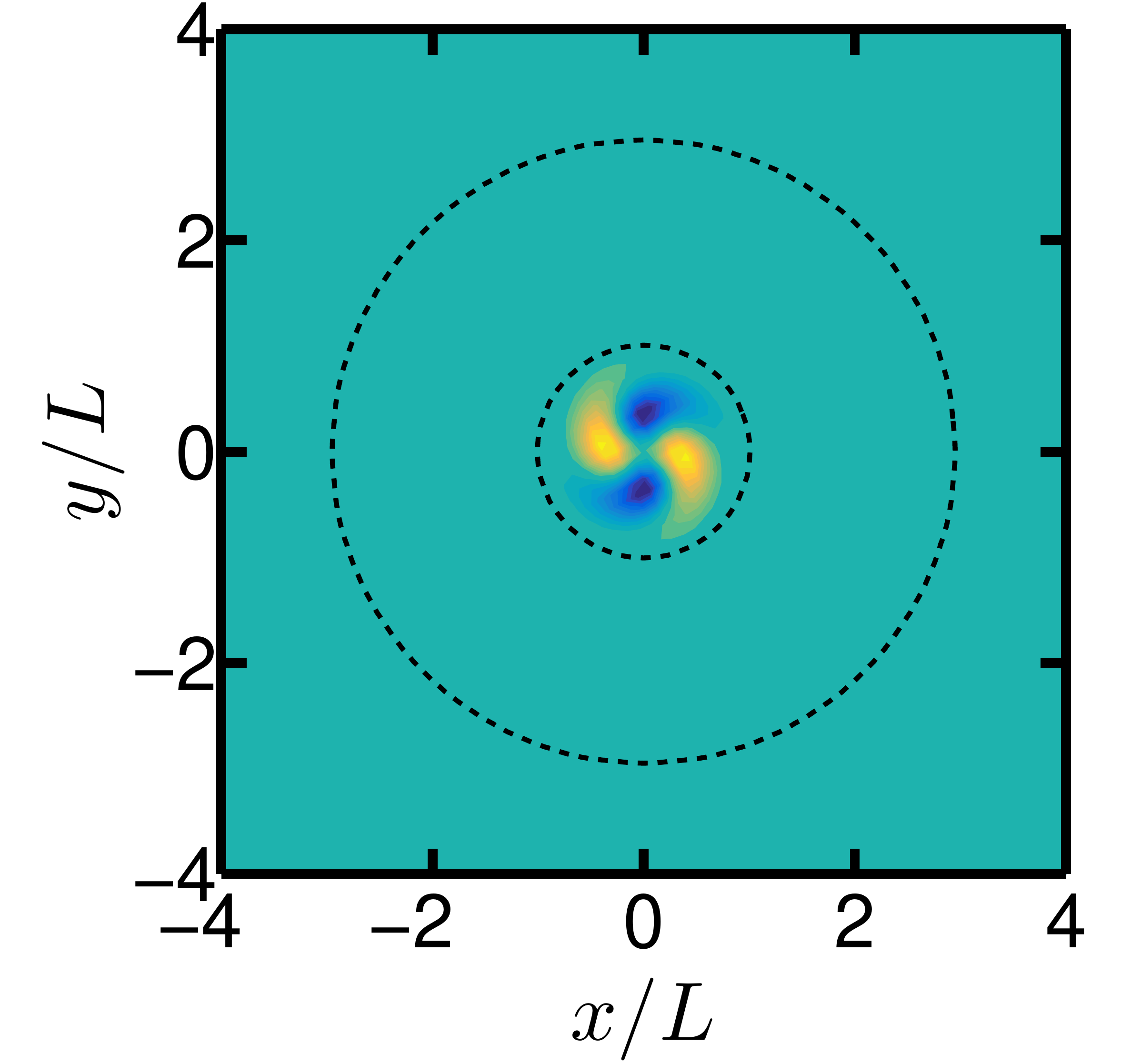}}
\subfloat{\includegraphics[trim={90mm 1mm 115mm 7mm},clip,width=0.217\textwidth]{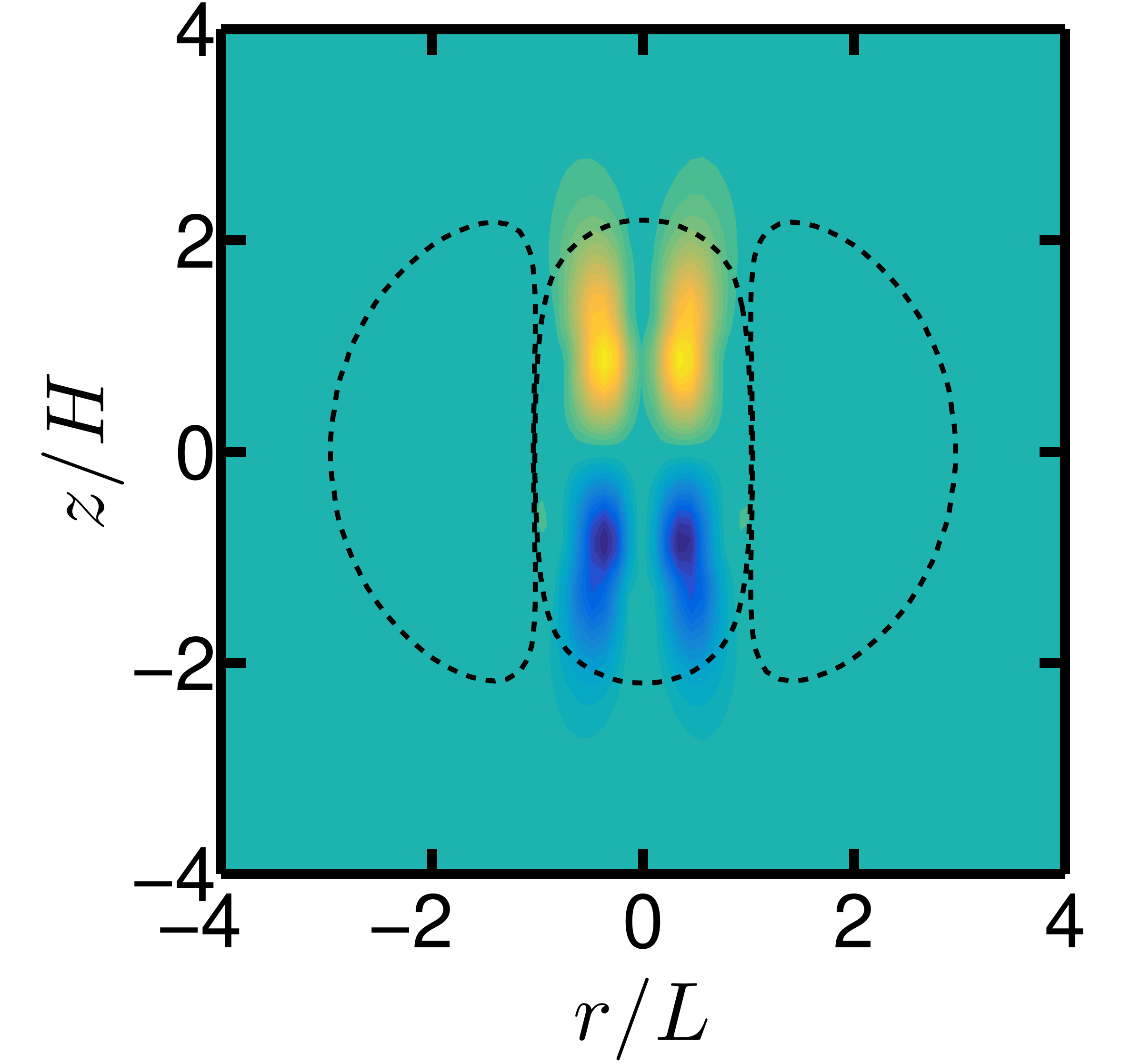}}\\
\subfloat{\includegraphics[trim={55mm 85mm 55mm 95mm},clip,width=0.275\textwidth]{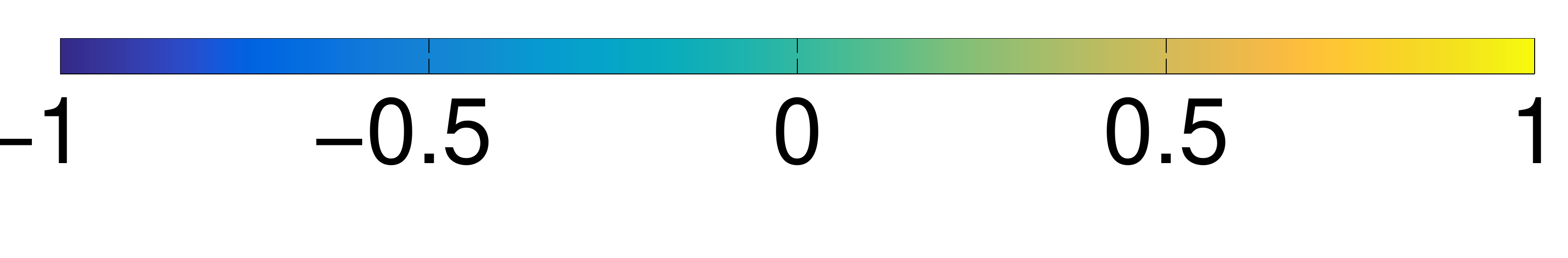}}
\caption{(Colour online) Vertical vorticity of eigenmodes normalized such that the maximum value of $|\omega|$ is 1. The eigenmodes are virtually indistinguishable for $f/\bar{N}=0.1$ and $f/\bar{N}=0.01$. The four rows from top to bottom correspond to the first four rows in table~\ref{tab1}. The broken lines denote the boundaries of the core and shield of the unperturbed Gaussian vortex. In each row, the left two panels are for $f/\bar{N}=0.1$ and the right two are for $f/\bar{N}=0.01$. The first and third panels in each row show the eigenmodes in the $x$-$y$ plane for a fixed $z$. For the $z$-symmetric eigenmode in the second row, this fixed value is $z=0$. For the anti-symmetric eigenmodes in rows 1, 3, and 4, the fixed value of $z$ is the positive value of $z$ at which $|\omega|$ of the eigenmode obtains its maximum value. The second and fourth panels in each row show the eigenmodes in the $r$-$z$ plane for fixed azimuthal angle $\phi$. In all cases, $\phi$ is chosen so that it is the angle at which $\omega$ of the eigenmode obtains its maximum value.}
\label{fcompare}
\end{figure}

Exploiting our semi-analytic method that enables us to accurately and efficiently deal with large $f\Delta t$ and $\bar{N} \Delta t$, we have repeated over $40$ of the simulations with $f/\bar{N}=0.01$. Table~\ref{tab1} shows the linear growth rate and the spatial symmetry of the fastest-growing eigenmode of several Gaussian vortices for $f/\bar{N}=0.1$ and $f/\bar{N}=0.01$. The symmetries are the same in all cases, as are the growth rates (in units of $\tau^{-1}$) within $4\%$. Figure~\ref{fcompare} shows examples of the most unstable eigenvectors (with dimension in $z$ scaled by $H$, and dimensions of $r$, $x$, and $y$ scaled by $L$). The eigenmodes are nearly indistinguishable  for $f/\bar{N}=0.1$ and $f/\bar{N}=0.01$.
   
\setul{4pt}{0.5pt}
\renewcommand{\arraystretch}{1.05}
\begin{table}
 \begin{center}
  \begin{tabular}{llcccc}
             &         &  \multicolumn{2}{c}{\ul{$~~~~~~~f/\bar{N}=0.1~~~~~~~$}} & \multicolumn{2}{c}{\ul{~~~~~~~$f/\bar{N}=0.01$~~~~~~~}} \\[5.5pt]
$~~Ro$ & $Bu$ & Symmetry & $\sigma $ & Symmetry & $\sigma$ \\[4pt]
$+0.45$   & $0.3$  & A3 & 1.5        &  A3 & 1.5 \\
$+0.4$     & $1.2$  & S2 & 0.14      &  S2 & 0.14 \\
$+0.4$     & $1.6$  & A1 & 0.13      &  A1 & 0.13 \\
$+0.2$     & $0.15$ & A2 & 1.1       &  A2 & 1.1 \\
$+0.2$     & $1.0$  & S2 & 0.058     &  S2 & 0.058 \\
$+0.2$     & $2.0$  & A1 & 0.097     &  A1 & 0.098 \\
$+0.05$   & $1.4$  & A1 & 0.040     &  A1 & 0.039 \\
$+0.02$   & $0.5$  & S2 & 0.029     &  S2 & 0.029 \\
$-0.02$    & $1.4$  & A1 & 0.028     &  A1 & 0.028 \\
$-0.18$    & $0.15$ & S2 & 0.024     & S2 & 0.023 \\ 
$-0.2$      & $0.45$ & -   & $<0.02$ &  -   & $<0.02$ \\
$-0.2$      & $2.0$  & A1 & 0.042     &  A1 & 0.043 \\
$-0.3$      & $1.6$  & - & $<0.02$     &  - & $<0.02$ \\
$-0.4$      & $1.72$ & - & $<0.02$    &  - & $<0.02$ \\
$-0.4$      & $1.8$  & A1 & 0.021    &  A1 & 0.021 \\
  \end{tabular}
  \caption{Comparison of the linear growth rates  (in units of $\tau^{-1}$) and symmetries of the most unstable eigenmode of selected Gaussian vortices in the $Ro-Bu$ space for $f/\bar{N}=0.1$ and  $f/\bar{N}=0.01$.}{\label{tab1}}
 \end{center}
\end{table}

The insensitivity to $f/\bar{N}$ is easily explained by non-dimensionalizing the equations of motion (\ref{eq:1}) with $4\pi/\omega_c\equiv\tau$ as the unit of time, $L$ as the unit of horizontal length, $H$ as the unit of vertical length, $L/\tau$ as the unit of horizontal velocity, $H/\tau$ as the unit of vertical velocity, $ \rho_o f L^2/\tau$ as the unit of pressure, $\rho_o$ as the unit of density, and $f L^2/(H\tau)$ as the unit of buoyancy. In the following equations,
asterisk  superscripts  indicate the non-dimensionalized quantity or operator

\begin{eqnarray}
\left(\frac{Ro}{2\pi}\right)\left[\frac{\partial{{v}^*_r}}{\partial t^*}+v^*_r\frac{\partial{v^*_r}}{\partial r^*}+\frac{v^*_{\phi}}{r^*}\frac{\partial{v^*_r}}{\partial \phi}+v^*_z\frac{\partial{v^*_r}}{\partial z^*}-\frac{{v^*}_\phi^2}{r^*}\right] &=& -\frac{\partial p^*}{\partial r^*}+v^*_{\phi},\\
\left(\frac{Ro}{2\pi}\right)\left[\frac{\partial{{v}^*_\phi}}{\partial t^*}+v^*_r\frac{\partial{v^*_\phi}}{\partial r^*}+\frac{v^*_{\phi}}{r^*}\frac{\partial{v^*_\phi}}{\partial \phi}+v^*_z\frac{\partial{v^*_\phi}}{\partial z^*}+\frac{v^*_rv^*_\phi}{r^*}\right] &=& -\frac{1}{r^*}\frac{\partial p^*}{\partial \phi}-v^*_{r}, \\
\left(\frac{Ro}{2\pi}\right)\left(Bu\right)\left(\frac{f}{\bar{N}}\right)^2\left[\frac{\partial{{v}^*_z}}{\partial t^*}+v^*_r\frac{\partial{v^*_z}}{\partial r^*}+\frac{v^*_{\phi}}{r^*}\frac{\partial{v^*_z}}{\partial \phi}+v^*_z\frac{\partial{v^*_z}}{\partial z^*}\right] &=& -\frac{\partial p^*}{\partial z^*}+b^*, \label{only} \\
\left(\frac{Ro}{2\pi Bu}\right)\left[\frac{\partial{b^*}}{\partial t^*}+v^*_r\frac{\partial{b^*}}{\partial r^*}+\frac{v^*_{\phi}}{r^*}\frac{\partial{b^*}}{\partial \phi}+v^*_z\frac{\partial{b^*}}{\partial z^*}\right] &=& -v^*_z, \label{4.4} \\
\frac{v^*_r}{r^*}+\frac{\partial v^*_r}{\partial r^*}+\frac{1}{r^*}\frac{\partial v^*_\phi}{\partial \phi}+\frac{\partial v^*_z}{\partial z^*} &=& 0.
\end{eqnarray}
Only (\ref{only}) depends on $f/\bar{N}$. For $f/\bar{N} \le 0.1$ and for Burger numbers of order unity or less, the left side of (\ref{only}) is of order $10^{-3}$, whereas the two terms on the right side are both of order unity {\it if} we have chosen ``proper'' units of length, time, and mass in our non-dimensionalization such that
the dimensionless quantities denoted with asterisk superscripts and their derivatives with respect to the dimensionless length and time inside the square brackets are of order unity or less. Thus those two terms nearly cancel each other, or
\begin{equation}
\frac{\partial p^*}{\partial z^*} = b^* + {\it O}(10^{-3}). \label{replace}
\end{equation}
So, hydrostatic equilibrium is enforced to one part in a thousand. Thus, replacing the dynamic equation (\ref{only}) with the kinematic equation (\ref{replace}) is a very good approximation, and with the replacement, the equations of motion are formally independent of $f/\bar{N}$. 
However, the argument above is not particularly useful because there is no {\it a priori} way of knowing that we chose ``proper'' units, and, in fact, for many types
of waves, with this choice of units, the dimensionless expressions inside the square brackets are much greater than unity, and the waves are not in hydrostatic balance and the value of $f/\bar{N}$ is important.

However, with the choice of units above, the dimensionless form of our initial Gaussian equilibrium vortices is
\begin{eqnarray}
\hat{p}^* &=& (-\pi)(1+Ro)\chi^*(r^*,z^*), \\
\hat{v}^*_{\phi} &=& \left(\frac{\pi}{Ro}\right)(r^*)\left(-1+\sqrt{1+4Ro(1+Ro)\chi^*(r^*,z^*)}\right),~~~~\hat{v}^*_r = \hat{v}^*_z = 0,  \\
\hat{b}^* &=& (2\pi)(1+Ro)z^*\chi^*(r^*,z^*),
\end{eqnarray}
where $\chi^* \equiv \exp{[-(r^*)^2 -  (z^*)^2]}$. Note that the vortices depend on $Ro$, but not on $f/\bar{N}$ or $Bu$. Also note that as $Ro \rightarrow 0$, the equilibrium velocity $\hat{v}^*_{\phi} \rightarrow 2 \pi r^* \chi^*(r^*,z^*)$ and remains of order unity or less. The equilibrium $p^*$ and $b^*$ are also of order unity or less for $|Ro|$ of order unity or less. 

The non-dimensional equations linearized
around the non-dimensional Gaussian vortex are (after 
dropping the asterisk superscripts  and writing $\boldsymbol{v}=\boldsymbol{\hat{v}}+\boldsymbol{\tilde{v}}$,  $p=\hat{p}+\tilde{p}$, and $b=\hat{b}+\tilde{b}$, where tilde denotes the linear eigenmode)
\begin{eqnarray}
\left(\frac{Ro}{2\pi}\right)\Big{[}\frac{\partial{{\tilde{v}}_r}}{\partial t}+\Big{(}\frac{\hat{v}_{\phi}}{r}\Big{)}\frac{\partial{\tilde{v}_r}}{\partial \phi}-\Big{(}\frac{2\hat{v}_{\phi}{\tilde{v}}_\phi}{r}\Big{)}\Big{]} &=& -\frac{\partial \tilde{p}}{\partial r}+\tilde{v}_{\phi}, \label{linbegin} \\
\left(\frac{Ro}{2\pi}\right)\Big{[}\frac{\partial{\tilde{v}_\phi}}{\partial t}+\Big{(}\frac{\hat{v}_{\phi}}{r}\Big{)}\frac{\partial{\tilde{v}_\phi}}{\partial \phi}+\Big{(}\frac{\partial{\hat{v}_\phi}}{\partial r}\Big{)}\tilde{v}_r+\Big{(}\frac{\partial{\hat{v}_\phi}}{\partial z}\Big{)}\tilde{v}_z+\Big{(}\frac{\hat{v}_\phi}{r}\Big{)}\tilde{v}_r\Big{]} &=& -\frac{1}{r}\frac{\partial \tilde{p}}{\partial \phi}-\tilde{v}_{r}, \\
\left(Ro\right)\left(Bu\right)\left(\frac{f}{\bar{N}}\right)^2\left(\frac{1}{2\pi}\right)\Big{[}\frac{\partial{\tilde{v}_z}}{\partial t}+\Big{(}\frac{\hat{v}_{\phi}}{r}\Big{)}\frac{\partial{\tilde{v}_z}}{\partial \phi}\Big{]} &=& -\frac{\partial \tilde{p}}{\partial z}+\tilde{b}, \\
\left(\frac{Ro}{2\pi Bu}\right)\left[\frac{\partial{\tilde{b}}}{\partial t}+\Big{(}\frac{\hat{v}_{\phi}}{r}\Big{)}\frac{\partial{\tilde{b}}}{\partial \phi}+\Big{(}\frac{\partial{\hat{b}}}{\partial r}\Big{)}\tilde{v}_r+\Big{(}\frac{\partial{\hat{b}}}{\partial z}\Big{)}\tilde{v}_z\right] &=& -\tilde{v}_z, \\
\frac{\tilde{v}_r}{r}+\frac{\partial \tilde{v}_r}{\partial r}+\frac{1}{r}\frac{\partial \tilde{v}_\phi}{\partial \phi}+\frac{\partial \tilde{v}_z}{\partial z} &=& 0. \label{linend}
\end{eqnarray}
For the fastest-growing eigenmodes of vortices with $N_c^2 > 0$, we have numerically computed the dimensionless values of the quantities inside the square brackets and found them to be of order unity or less for all of the eigenmodes represented in figure~\ref{f2}. This calculation shows that for vortices whose interior is statically stable, the fastest-growing eigenmodes are in vertical hydrostatic balance and therefore explains why the non-dimensionalized eigenvalues and eigenmodes are insensitive to the value of $f/\bar{N}$ for $f/\bar{N} \lesssim 0.1$. It should be emphasized that we could not assume {\it a priori} that the fastest-growing eigenmodes of our vortices are in hydrostatic balance. Here we have numerically tested and verified the validity of this assumption. It is worth mentioning that non-hydrostatic effects can be important in the dynamics and evolutions of some geophysical and astrophysical vortices; for example, our previous calculations of vortices \citep{marcus2014surprising}, especially the longevity of the Great Red Spot (GRS) of Jupiter (and we remind the reader that longevity of vortices was the motivation of the study), showed that small departures from vertical hydrostatic equilibrium caused large changes to the lifetime of the GRS (albeit, due to nonlinear effects). 

Finally, it is not surprising that for vortices with statically-unstable interiors ($N_c^2<0$), the terms in the square brackets are large and therefore the most unstable eigenmodes are {\it not} in hydrostatic balance. We have not carried out further eigenmode calculations with $f/\bar{N}=0.01$ in this region because they are computationally very expensive.  

\section{Radial and vertical structure of the unstable eigenmodes}  \label{sec:spatial_distribution}
\begin{figure}
    \centering
    \subfloat[\large{(a)}]{\includegraphics[trim={0 1mm 0 -8mm},clip,width=0.95\textwidth]{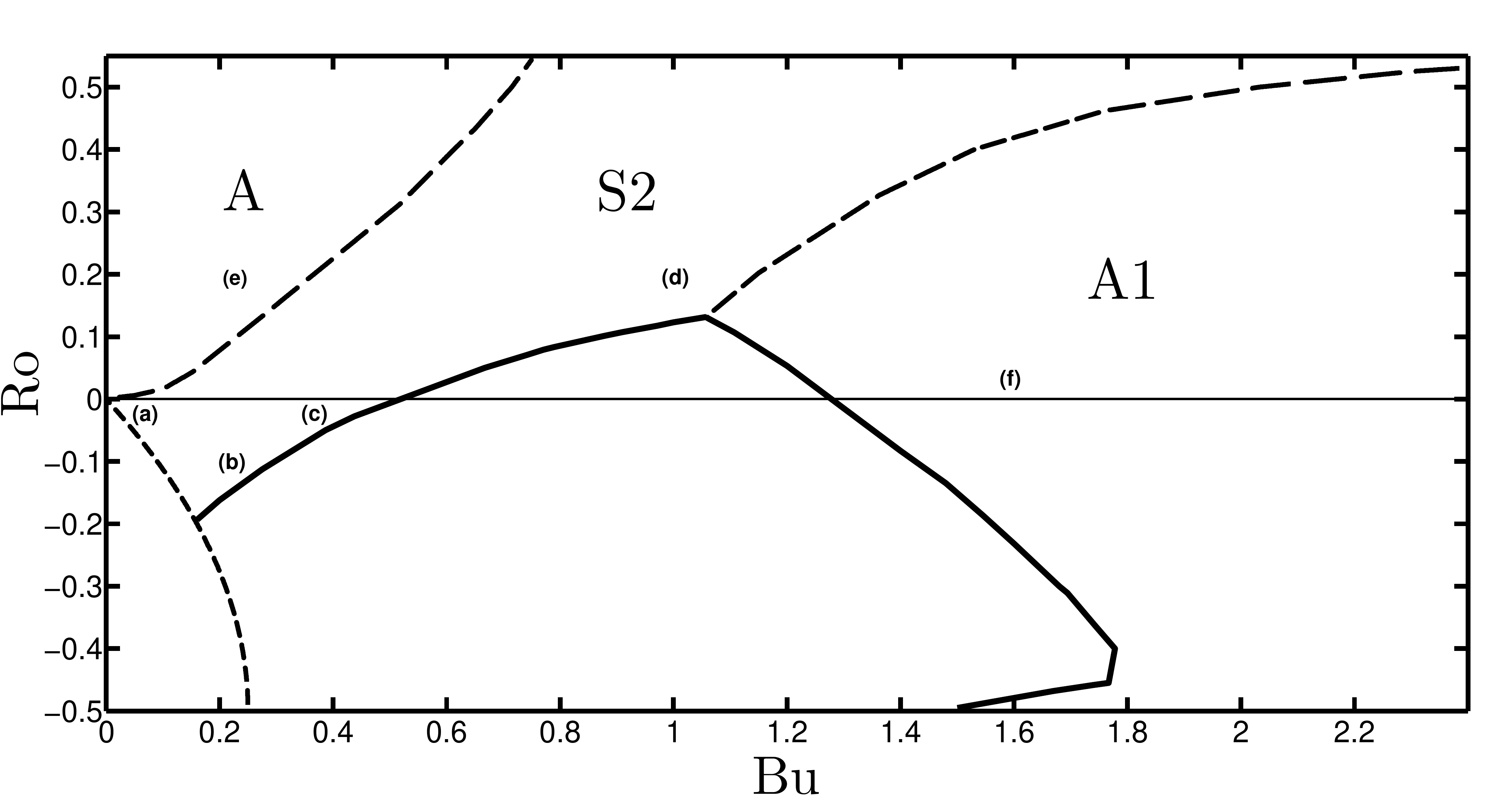}}\\ 
    \subfloat[\large{(b)}]{\includegraphics[trim={26 14mm 10mm 16mm},clip,width=0.94\textwidth]{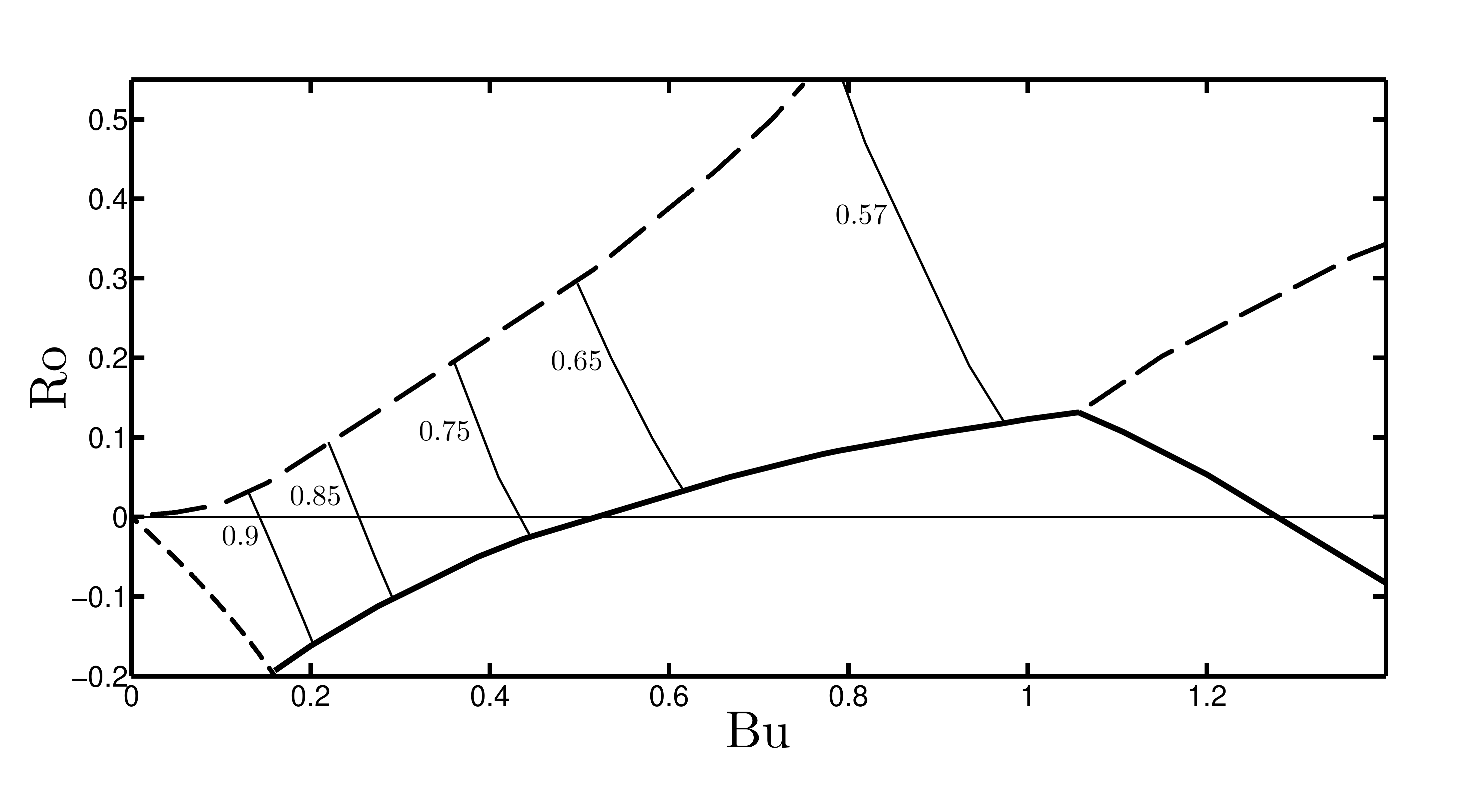}}
    \caption{Panel (a): Simplification of the parameter map in figure~\ref{f2}(a). The fastest-growing eigenmodes in the region labeled ${A}$ are anti-symmetric in $z$ and have azimuthal wave numbers of 1, 2, 3, or 4; otherwise, the fastest-growing eigenmodes have the symmetry of the large labels. The small labels $(a)$-$(f)$ indicate the locations in parameter space of the vortices whose fastest-growing eigenmodes are plotted in figure~\ref{f8}. Panel (b): Blow up of the {S2} region in panel~(a). The thin solid curves are the iso-contours of the enstrophy $S_{shield}$ of the vertical vorticity of the eigenmode in the vortex shield. The value of $S_{shield}$ in the {S2} region decreases from 0.95 to 0.55 with increasing $Bu$. The approximate average value of $S_{core}$ in the {A} and {A1} regions are  0.95 and 0.73, respectively. In the lower left corner of panel~(a) where $N_c^2 < 0$, $S_{core} \simeq 0.99$. } 
    \label{f7}
\end{figure}

\begin{figure}
    \centering
       \subfloat{\includegraphics[trim={90mm 3mm 88mm -8mm},clip,width=0.24\textwidth]{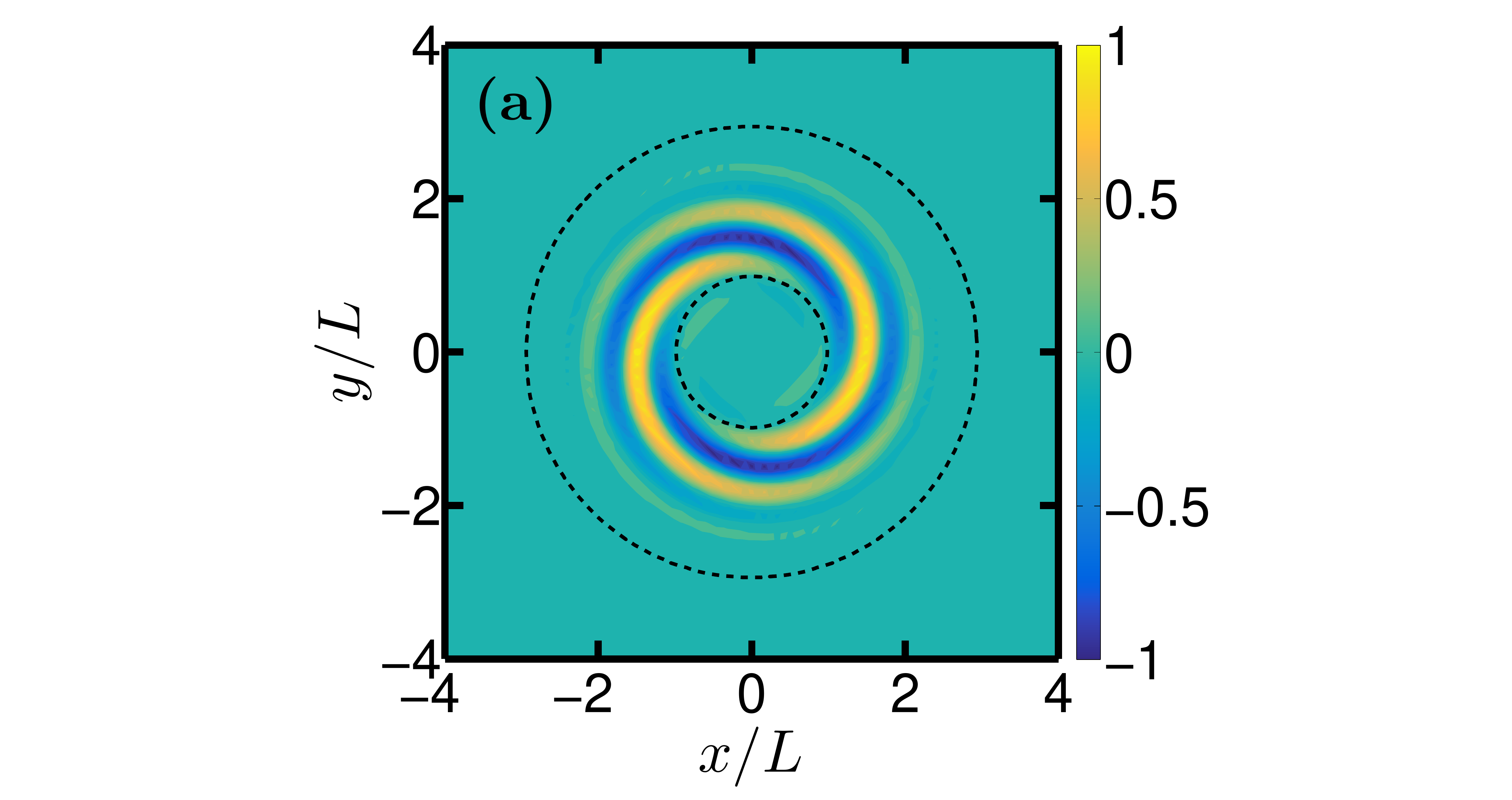}}
       \subfloat{\includegraphics[trim={90mm 3mm 77mm -8mm},clip,width=0.25\textwidth]{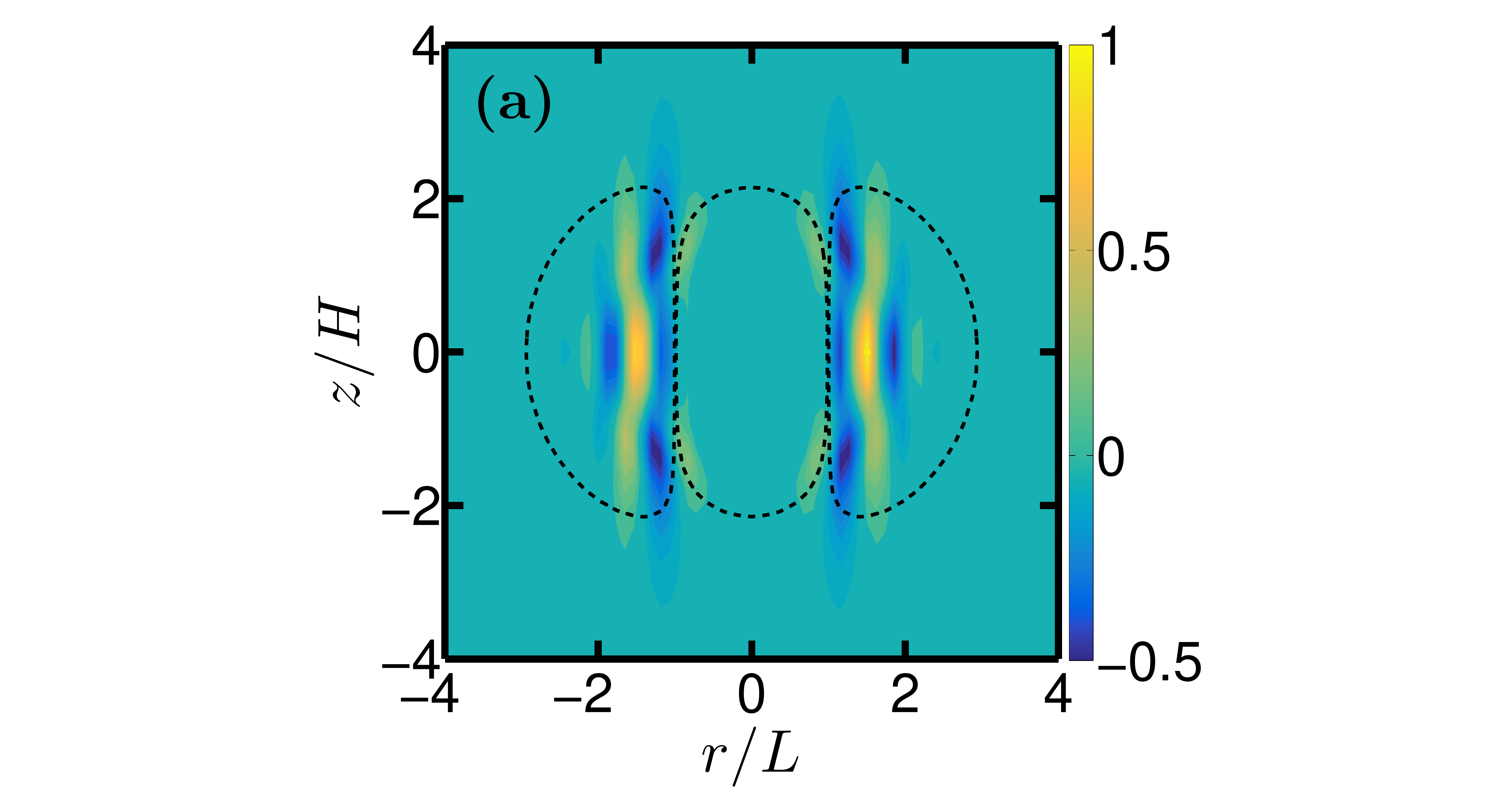}}
       $~~~$\subfloat{\includegraphics[trim={90mm 3mm 88mm -8mm},clip,width=0.24\textwidth]{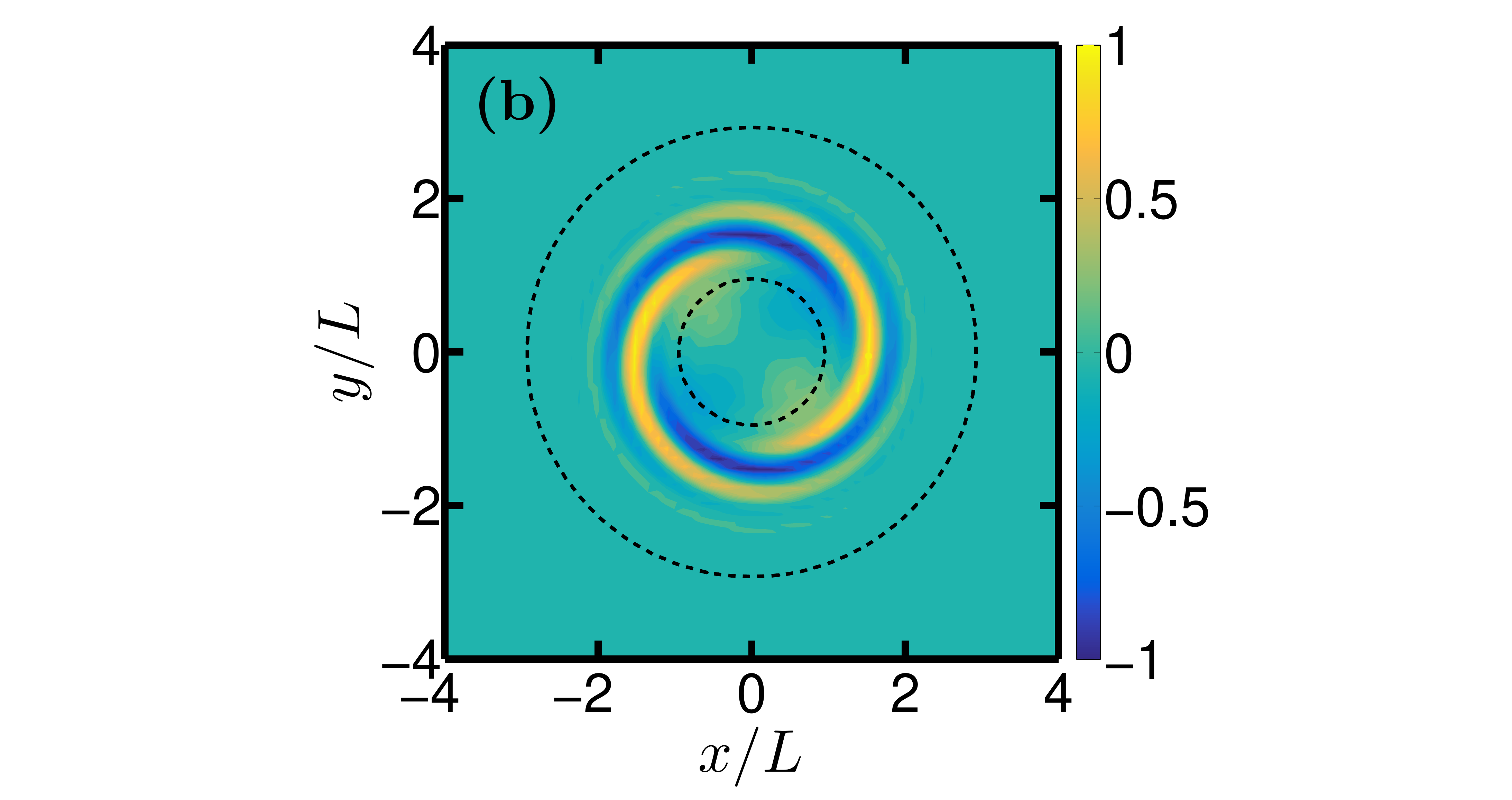}}
       \subfloat{\includegraphics[trim={90mm 3mm 77mm -8mm},clip,width=0.25\textwidth]{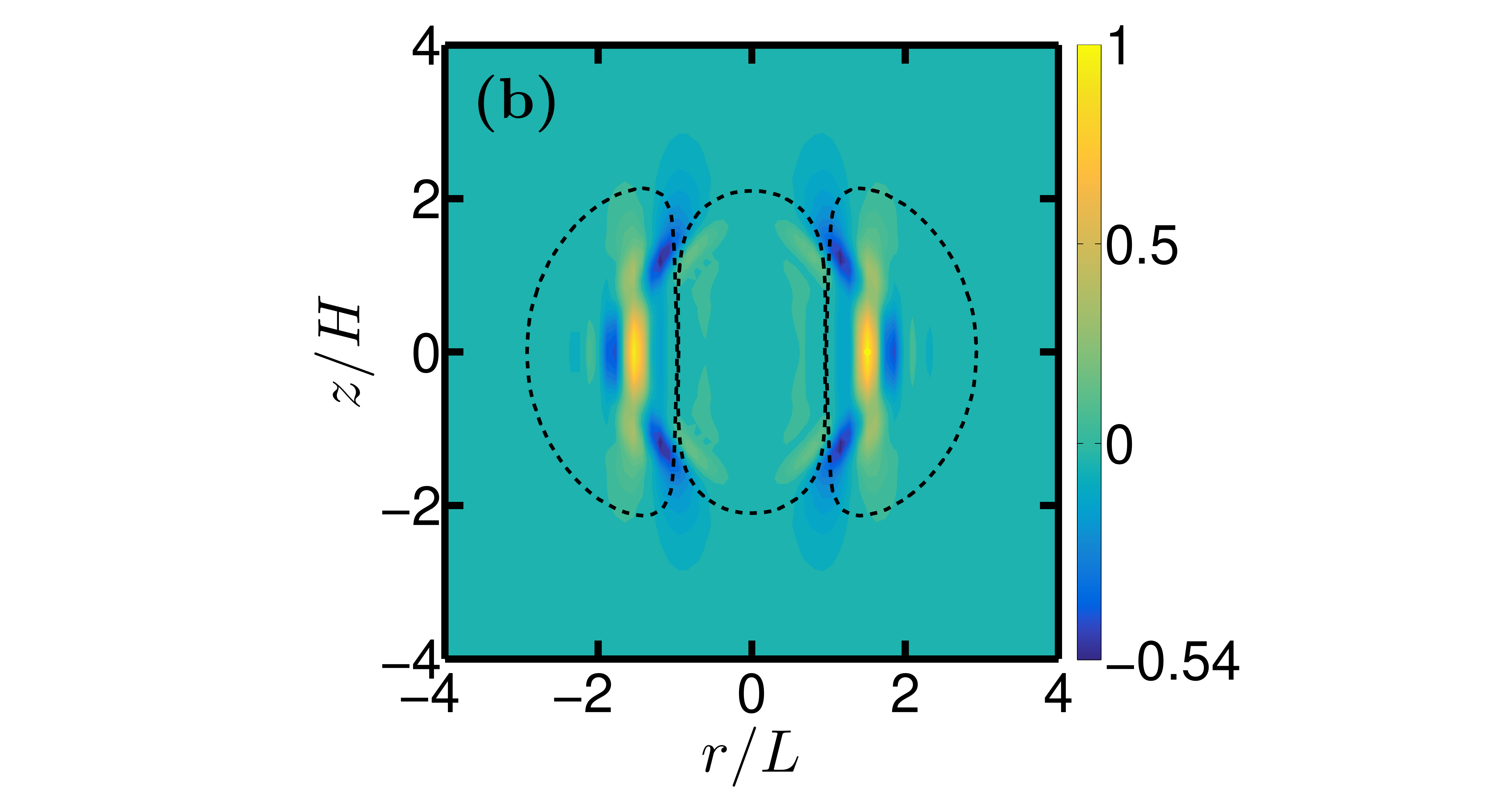}}\\
       \subfloat{\includegraphics[trim={90mm 6mm 88mm 6mm},clip,width=0.24\textwidth]{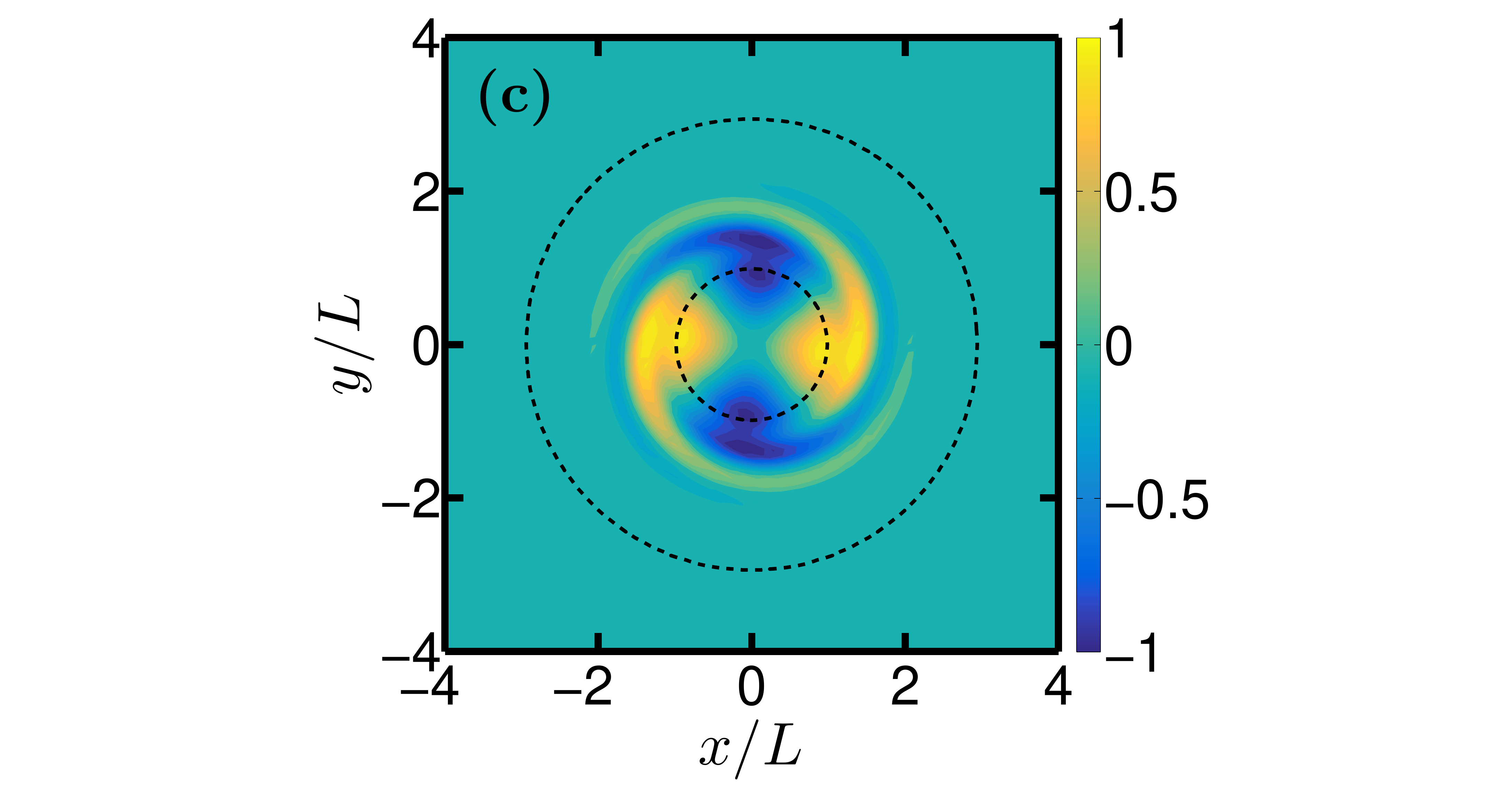}}
       \subfloat{\includegraphics[trim={90mm 6mm 77mm 6mm},clip,width=0.25\textwidth]{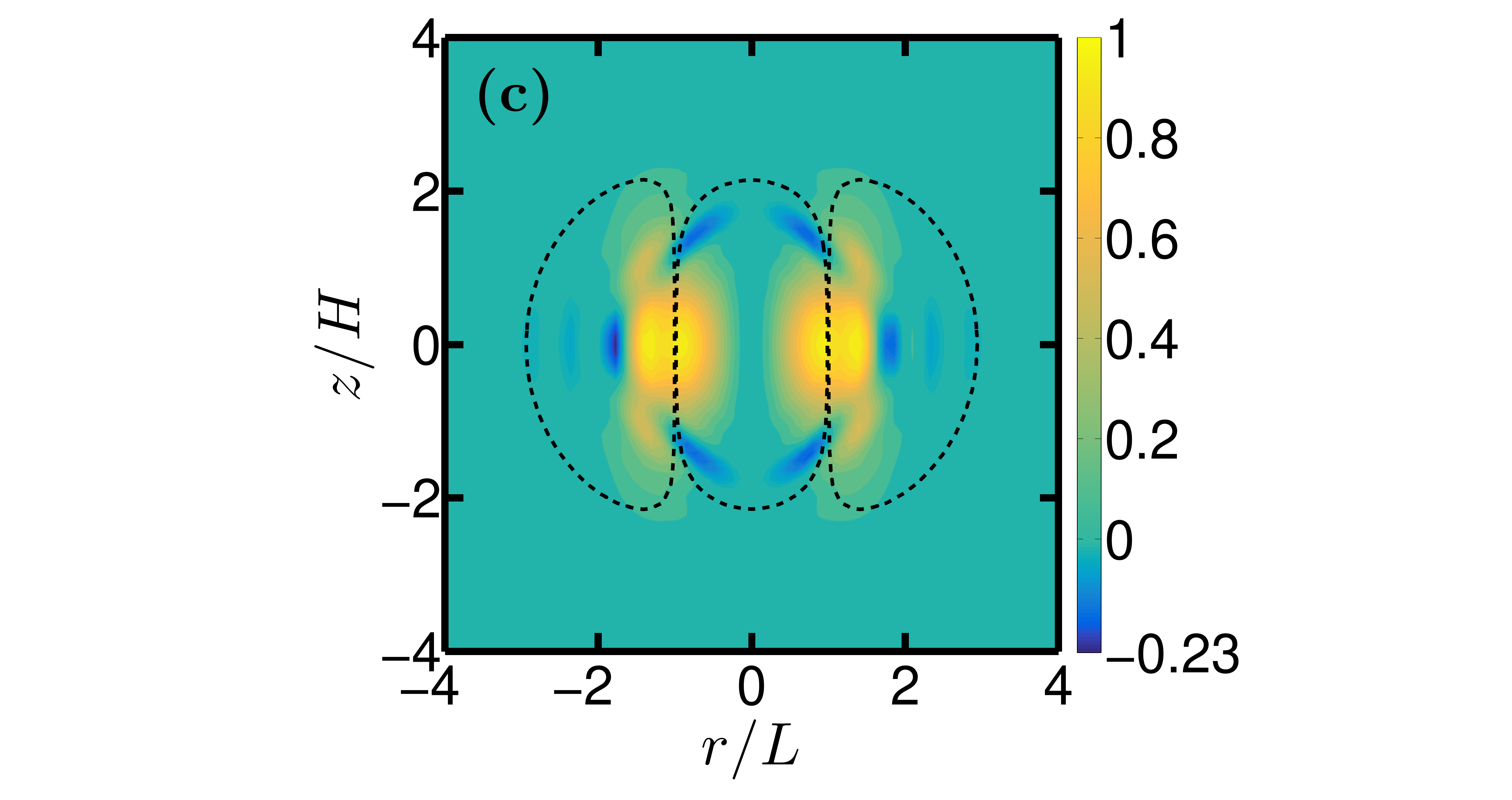}}
       $~~~$\subfloat{\includegraphics[trim={90mm 6mm 88mm 6mm},clip,width=0.24\textwidth]{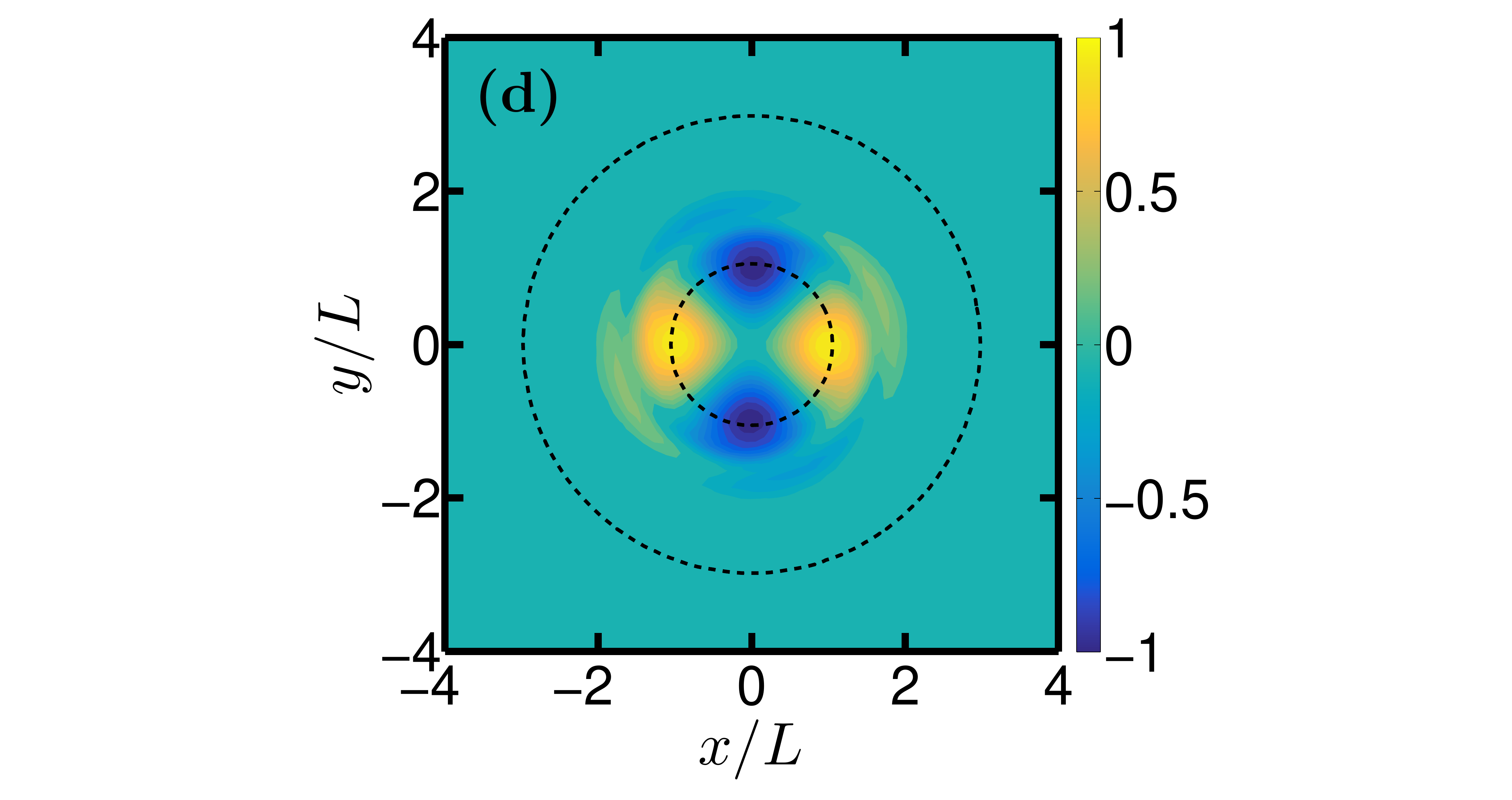}}
       \subfloat{\includegraphics[trim={90mm 5mm 77mm 6.0mm},clip,width=0.25\textwidth]{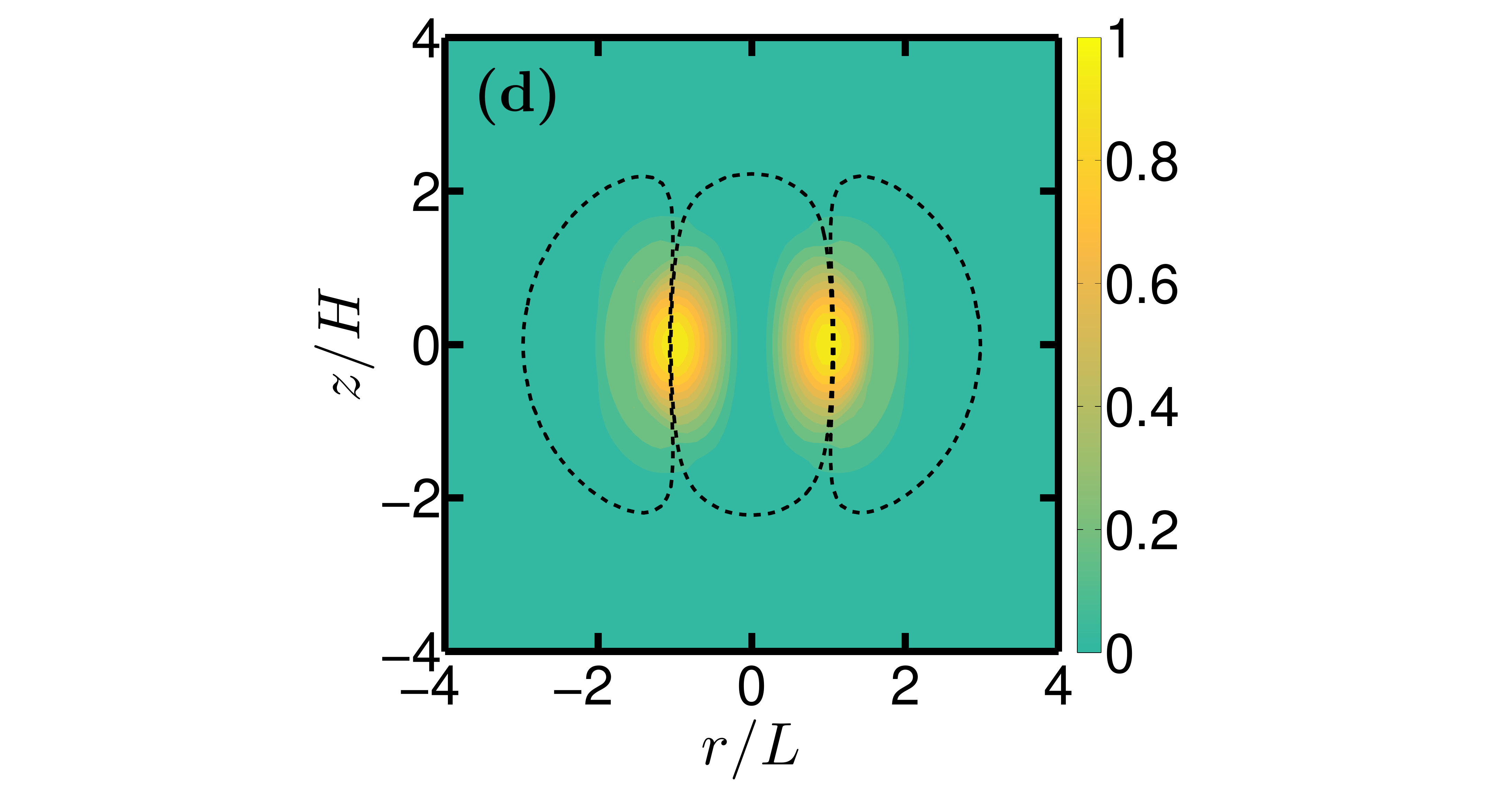}}\\
       \subfloat{\includegraphics[trim={90mm 2.5mm 88mm 8mm},clip,width=0.24\textwidth]{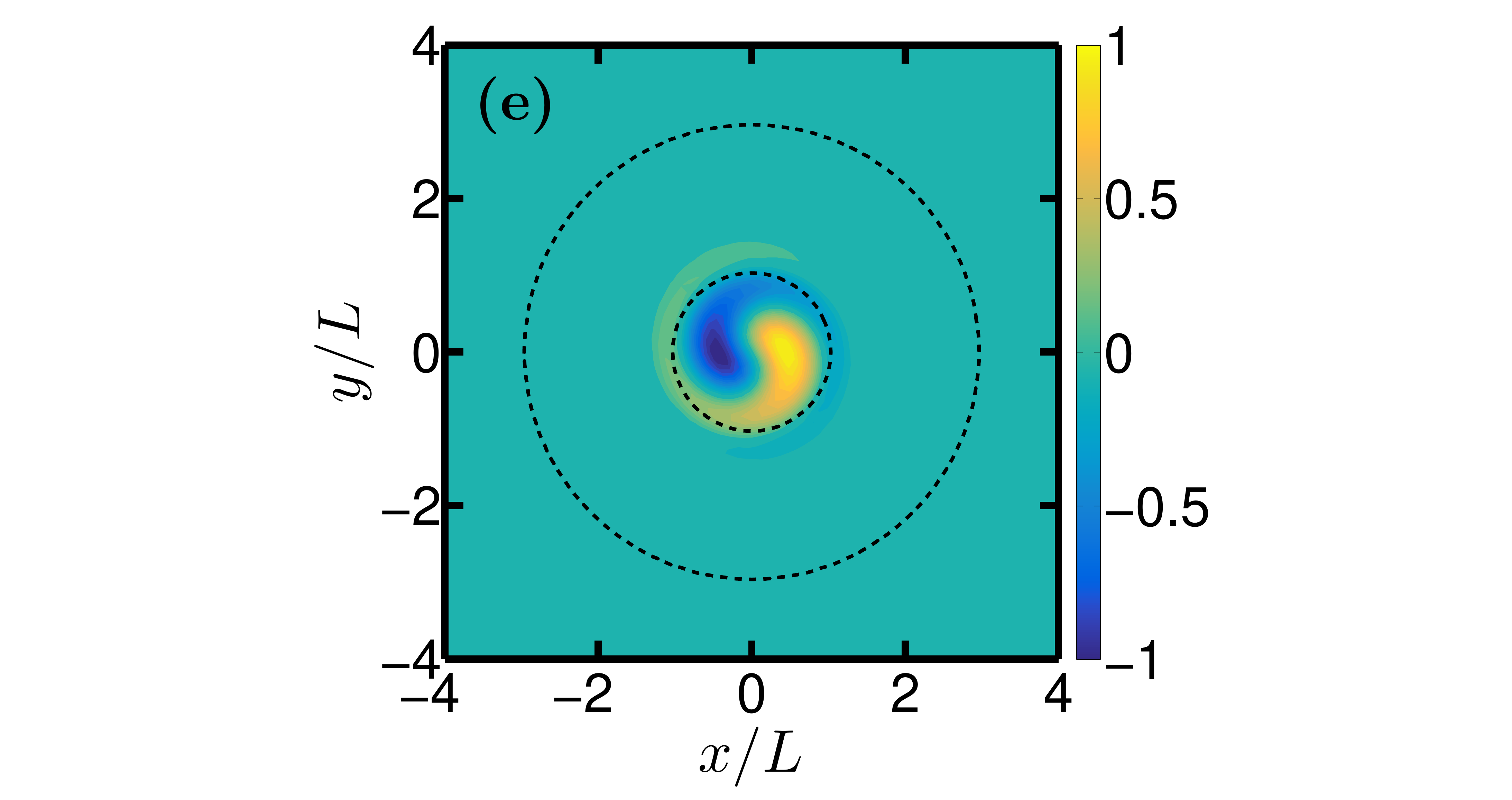}}
       \subfloat{\includegraphics[trim={90mm 2.5mm 77mm 8.5mm},clip,width=0.25\textwidth]{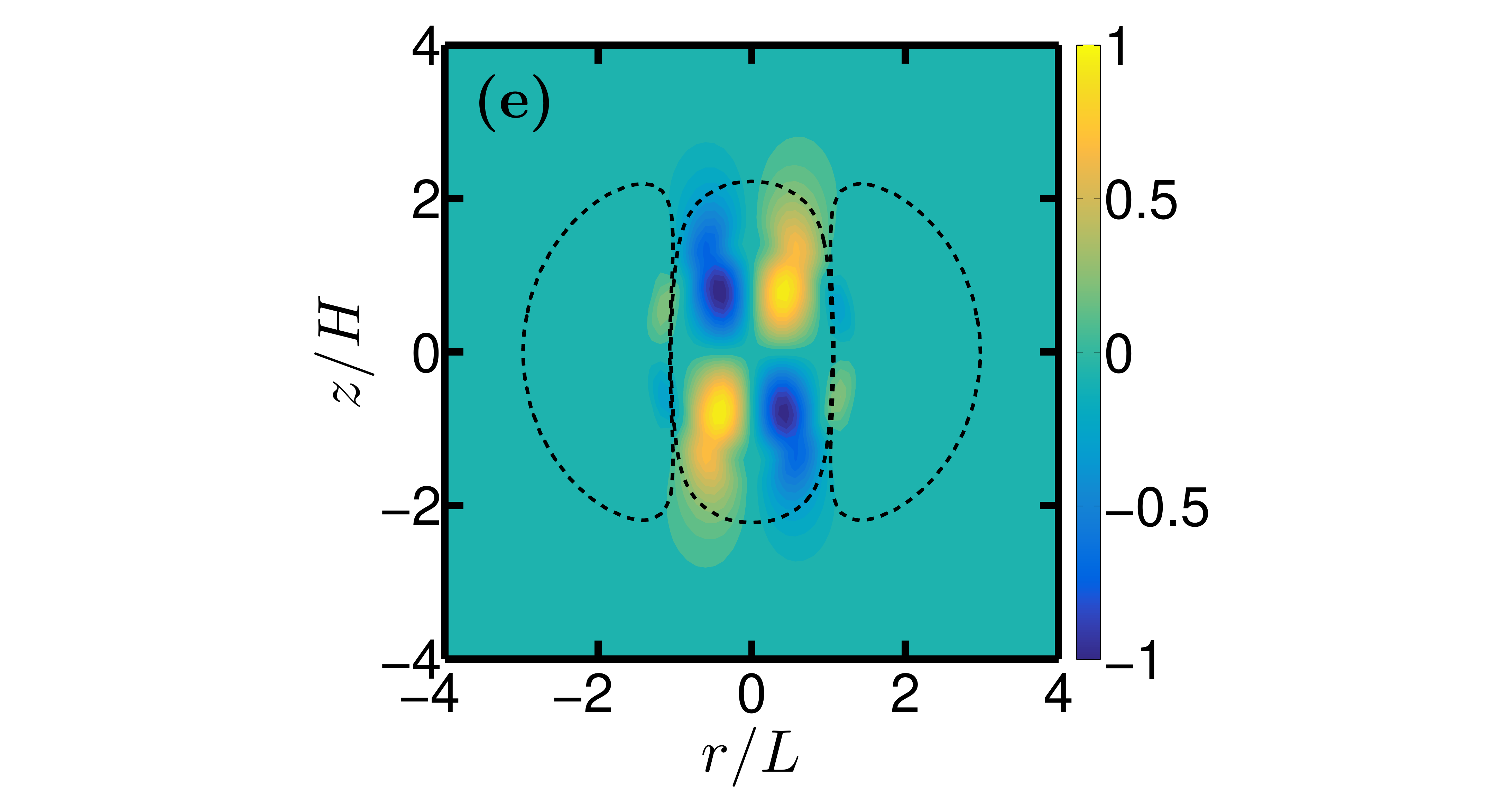}}
       $~~~$\subfloat{\includegraphics[trim={90mm 2.5mm 88mm 8mm},clip,width=0.24\textwidth]{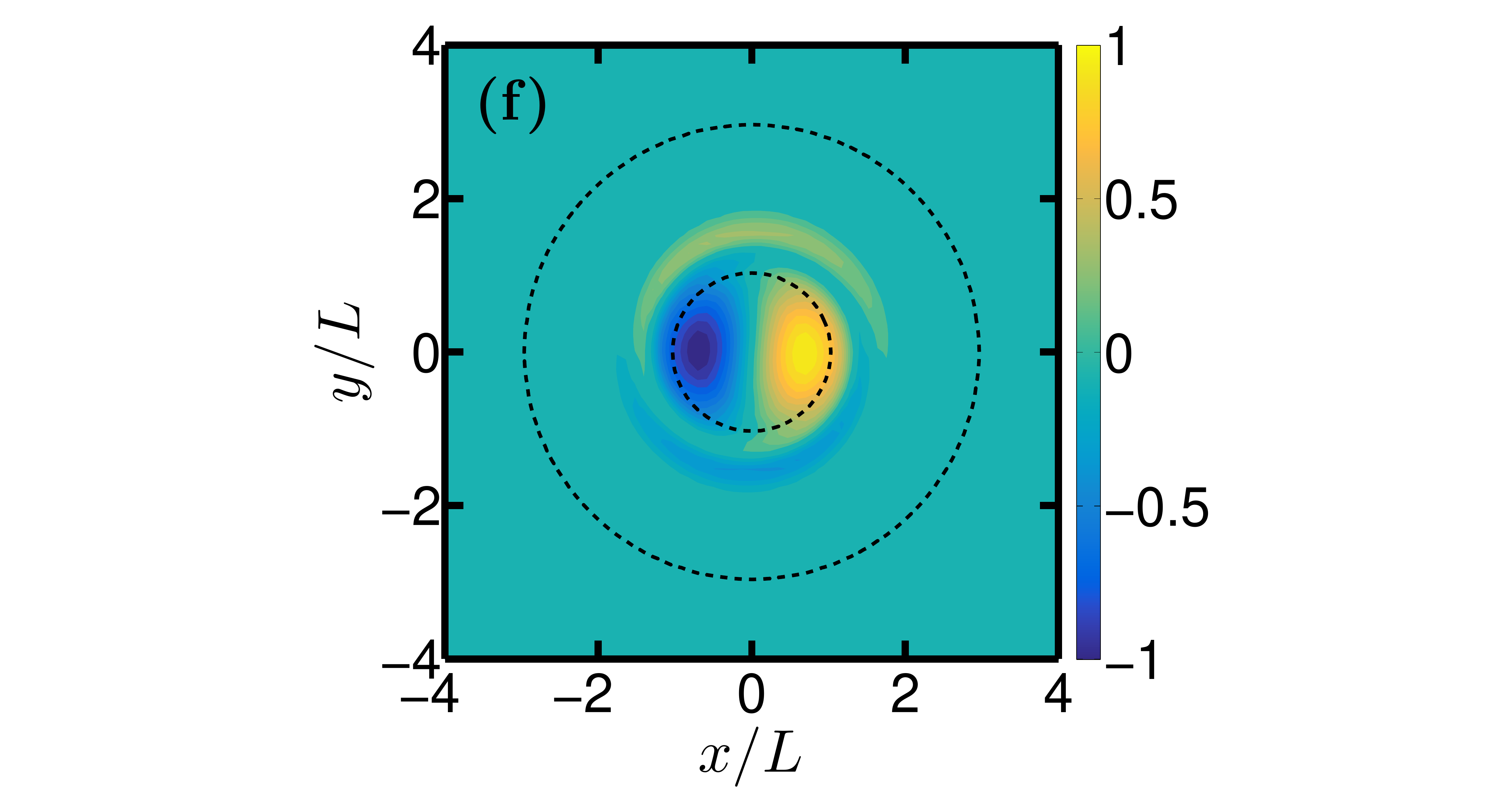}}
       \subfloat{\includegraphics[trim={90mm 2.5mm 77mm 8mm},clip,width=0.25\textwidth]{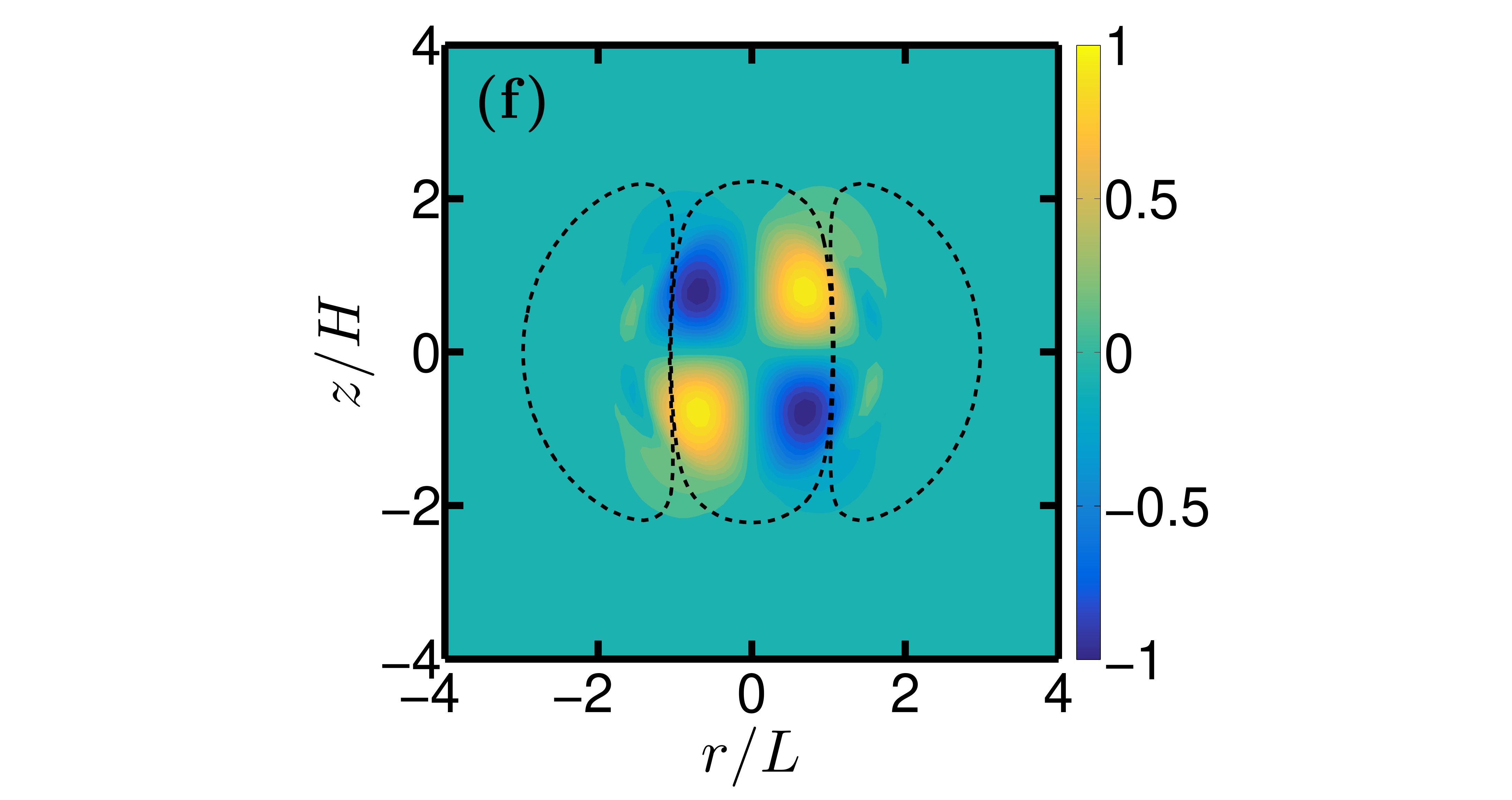}}
    \caption{(Colour online) Vertical vorticity $\omega$ of the eigenmodes of the Gaussian vortices for which the locations in $Ro-Bu$ space are indicated with letters in figure~\ref{f7}(a). The vorticity (normalized as in figure~\ref{fcompare}) and boundaries of the shields and cores are plotted as in figure~\ref{fcompare}. The first and third columns of panels show  eigenmodes in the $x$-$y$ plane, and the second and fourth columns show them  in the $r$-$z$ plane. Consistent with figure~\ref{f7}, the A and A1 eigenmodes [i.e., (e) and (f)] are mainly confined to the cores of the unperturbed vortices. The S2 eigenmodes with low $Bu$ [i.e., (a) and (b)] are mainly confined to the shield. The S2 eigenmodes with higher $Bu$ [i.e., (c) and (d)] are spread over the core and the shield.}
    \label{f8}
\end{figure}

In this section we investigate the radial distribution of vorticity in the fastest-growing eigenmodes. The spatial distribution of these eigenmodes can be characterized quantitatively by determining the fractional amounts of its vertical enstrophy that are within the Gaussian vortex's core $S_{core}$ and within its shield $S_{shield}$, where we use the definitions of {\it core} and {\it shield} given in Appendix~B:
\begin{eqnarray}
S_{core} &\equiv& {\frac{\int_{core} \, |\omega_{eig}|^2 \, d^3 x}{\int \, |\omega_{eig}|^2 \, d^3 x}}, \label{eq:S1} \\ 
S_{shield} &\equiv& {\frac{\int_{shield} \, |\omega_{eig}|^2 \, d^3 x}{\int \, |\omega_{eig}|^2 \, d^3 x}}, \label{eq:S2}
\end{eqnarray}
where $\omega_{eig}$ is the vertical vorticity of the eigenmode, the integrals in the numerators of (\ref{eq:S1}) and (\ref{eq:S2}) are over the core and shield respectively of the unperturbed vortex, and where the integrals in the denominators are taken over the entire computational domain. Not surprisingly, $S_{core} + S_{shield} > 0.95$, meaning that eigenmodes do not effectively extend radially beyond the shield of the unperturbed vortex. Figures~\ref{f7}~and~\ref{f8} show that the radial structure of the fastest-growing mode depends in a simple way on its vertical and azimuthal symmetry. Figure~\ref{f7}(a) is a simplified version of figure~\ref{f2}(a) and divides the $Ro-Bu$ space into 5 regions. The  two unlabeled regions correspond to the region with $N_c^2 < 0$, and to the region of slow growth with $\sigma\le0.02\:\tau^{-1}$. 

The three regions labeled ${A1}$, ${S2}$, and ${A}$, correspond accordingly to the vertical-azimuthal symmetry of the fastest-growing eigenmodes with the region labeled {A} having fastest-growing eigenmodes that are anti-symmetric in $z$ with an azimuthal wave number $m$ of 1, 2, 3 or 4. The fastest-growing eigenmodes in the ${A1}$ region are always [that is, for the vortices illustrated in figure~\ref{f2}(a)] concentrated radially in the core with $ 0.71 \le S_{core} \le 0.75$. The A1 eigenmode indicated by the label (f) in figure~\ref{f7} is shown in the two panels labeled (f) in figure~\ref{f8}, which clearly show the radial concentration of the eigenmode in the core. The fastest-growing eigenmodes in the ${A}$ region of figure~\ref{f7}(a) are even more strongly concentrated in the core and have  $S_{core} > 0.87$. The A1 eigenmode of the cyclone indicated by the label (e) in figure~\ref{f7} is shown in the two panels labeled (e) in figure~\ref{f8}, which show the concentration in the core. In contrast, the fastest-growing eigenmodes in the {S2} region are either radially concentrated in the shield or are spread throughout the core and shield. Figure~\ref{f7}(b) is a blow-up of figure~\ref{f7}(a) and shows iso-contours of $S_{shield}$, which varies in the region of $Ro-Bu$ space that we examined from 0.95 at low $Bu$ to $0.55$ at high $Bu$. Thus, for low values of $Bu$, the fastest-growing S2 eigenmodes are very concentrated in the shields, and as $Bu$ increases, the radial structure spreads into the core such that for the largest values of $Bu$ that we examined, the eigenmode is approximately equally spread between the shield and core. The radial dependence on $Bu$ of the S2 eigenmodes is illustrated in panels (a)-(d) in figure~\ref{f8}. The implications of the spatial structure of the most unstable eigenmodes will be discussed in a subsequent publication that is focused on the nonlinear evolution of these vortices and is outlined in the Discussion.  

\section{Discussion and summary} \label{sec:discussion}
We have studied the linear stability of 3D axisymmetric Gaussian vortices as a function of their Rossby number, $Ro$, and Burger number, $Bu$, over the wide range of values where long-lived geophysical and astrophysical vortices are often observed ($-0.5<Ro<0.5$ and $0.02<Bu<2.3$). For each $(Ro,Bu)$, the growth rate, $\sigma$, and the eigenvector of the most unstable eigenmode have been calculated by numerically solving the 3D non-hydrostatic Boussinesq equations. 

The results of the stability analysis are summarized in the $Ro-Bu$ parameter map (figure~\ref{f2}). These results show that neutrally-stable (i.e., $\sigma=0$) cyclones only exist over a small region of the parameter space where $Ro \sim 0.02-0.05$ and $Bu \sim 0.85-0.95$; we do not find any neutrally-stable anticyclone. On the other hand, the most unstable eigenmodes of anticyclones generally have slower growth rates compared to those of the cyclones. Over a large region of the $Ro-Bu$ parameter space (mainly $Ro < 0$ and $0.5 \lesssim Bu \lesssim 1.3$), the maximum growth rates of the anticyclones are smaller than $50$ turn-around time ($\tau$) of the vortex. For $Bu \gtrsim 1.3$, the maximum growth rate of anticyclones increases (decreases) with increasing $Bu$ ($|Ro|$). In this region, the eigenvector of the most unstable modes is anti-symmetric with respect to the $z=0$ plane and has $m=1$ azimuthal wave number (denoted as A1 mode), and the vertical vorticity ($\omega$) of the most unstable modes is mainly confined to the core of the initial (i.e., unperturbed) anticyclone (similar to figure~\ref{f8}(f), but for an anticyclone). Preliminarily investigation of the nonlinear evolution of these vortices shows that, in addition to the growth rate, the structure of the most unstable mode is also important in determining how the nonlinearly-equilibrated vortex compares with the initial vortex (nonlinear evolution will be addressed in a subsequent publication). For $Bu \lesssim 0.5$, the maximum growth rate of anticyclones increases with decreasing $Bu$ or $|Ro|$. In this region, the eigenvector of the most unstable modes is symmetric with respect to the $z=0$ plane and has $m=2$ azimuthal wave number (S2 mode). The vertical vorticity of these modes is mainly confined to the shield or spread over the core and the shield of the initial anticyclone depending on the Burger number (see figures~\ref{f8}(a)-(c)). For anticyclones if $Bu < -Ro(1+Ro)$, the interior of the vortex is statically unstable. The growth rates of the most unstable mode for these anticyclones are much larger (by factors up to several thousand or more) compared to those of the anticyclones outside this region (see figures~\ref{f4}~and~\ref{f5}).    

For cyclones, the region of small growth rate ($\sigma < 0.02 \, \tau^{-1}$) is much smaller and confined to $Ro < 0.1$ and $0.5 \lesssim Bu \lesssim 1.3$. For $Bu \gtrsim 1$, the maximum growth rate of cyclones increases with increasing $Bu$ or $Ro$. As was the case for anticyclones with large $Bu$, in this region the eigenvector of the most unstable modes is (generally) an A1 mode, and the vertical vorticity of these modes is mainly confined to the core of the initial cyclone (see figure~\ref{f8}(f)). For $Bu \lesssim 1$, the maximum growth rate of cyclones increases with decreasing $Bu$ or increasing $Ro$. In this region, for moderate values of $Bu$, the eigenvector of the most unstable modes is a S2 mode, and its vertical vorticity is spread over the core and the shield of the initial cyclone (see figure~\ref{f8}(d)). For smaller values of $Bu$, the eigenvector is anti-symmetric with respect to the $z=0$ plane and has $m=1,2,3$ or $4$, and its $\omega$ is confined to the core of the initial cyclone. Further analysis shows that although the fastest-growing eigenmodes of cyclones are A1 for both small and large values of $Bu$, the families of these eigenmodes are in fact distinct and have different spatial structures (see figures~\ref{f3_new_after_2}--\ref{criticallayer_3}~and~\ref{f8}).            

The findings described above are compared and contrasted with the relevant published work in \S\ref{sec:comp}. In particular, in the QG limit, \citet{Nguyen2012} found that the fastest-growing mode changes from S2 to A1 around $Bu = 1$, which along with the general dependence of the growth rate of the most unstable mode on $Bu$ agrees with our results for small $Ro$. However, there are differences at the limit of small $Bu$ ($\lesssim 0.05$): the QG analysis showed the dominance of modes with higher $m$, while our analysis using the non-hydrostatic Boussinesq equations shows anticyclones to be statically-unstable with A4 modes dominating, and cyclones to be unstable with A1 modes dominating at low $Ro$ and A2 or A3 modes dominating at moderate $Ro$. We have also investigated critical layers in the eigenmodes of unstable and neutrally-stable vortices (see \S\ref{sec:criticallayers}), and have found them at the periphery of the vortex core for a wide range of $Bu$, in agreement with the QG analysis of \citet{Nguyen2012}.

We have also examined how the vortex profile affects the stability properties by comparing our results for the family of vortices with Gaussian pressure anomaly with those of \citet{Yim2016} who studied the linear stability of a family of vortices with Gaussian angular velocity using non-hydrostatic Boussinesq equations. While for $Bu \gtrsim 1$ both families of vortices have most unstable modes with A1 symmetries, for $Bu \lesssim 1$, there are notable similarities and differences: \citet{Yim2016} found that both cyclones and anticyclones can become statically-unstable at low $Bu$ (while we found that only for anticyclones); they found that anticyclones are neutrally stable for moderate Burger numbers $0.5 \lesssim Bu \lesssim 1$ (while we found them weakly unstable) and are unstable with S2 modes dominating for smaller $Bu$ (which is consistent with our results); \citet{Yim2016} found similar stability properties for cyclones as reported here although they found a much larger neutrally-stable region compared to what we found. 

Most of the calculations reported in this paper have been done for $f/\bar{N}=0.1$. This value, which is approximately 10 times larger than the value in ocean at mid-latitudes, was
commonly used in studies of vortices in rotating stratified flows because at smaller values
the equations of motion are computationally stiff and therefore  computationally expensive to compute because small time steps are necessary.
Focusing on vortices whose interiors are statically stable (i.e., $N^2_c \geq 0$), we have repeated some of the calculations with $f/\bar{N}=0.01$ and found the results to remain quantitatively the same (see Table~\ref{tab1} and figure~\ref{fcompare}). We have further shown that the insensitivity of the growth rate and eigenvector of the most unstable modes to $f/\bar{N}$ can be explained from the non-dimensionalized equations of motion. This is because the most unstable eigenmodes are found to be approximately in the hydrostatic balance, which could not be assumed {\it a priori}. As a result, the dynamics of these modes are nearly independent of $f/\bar{N}$ (as long as this ratio is small, e.g., $\lesssim 0.1$) given that this ratio only appears on the left-hand side of the vertical momentum equation (see \S\ref{sec:effect_of_foN} for details). Note that such insensitivity to $f/\bar{N}$ is not expected in the region where the vortex interior is statically unstable (i.e., $N^2_c < 0$). 

The results of this paper improve the understanding of the generic stability properties
of 3D vortices in rotating stratified flows, and as discussed in \S\ref{sec:introduction}, extend the analyses of the previous studies in several ways, including: using the full 3D non-hydrostatic Boussinesq equations, which extends the stability analysis well beyond the usually-used QG and shallow-water approximations; focusing on a widely-used model of geophysical and astrophysical vortices, i.e., 3D Gaussian vortices with continuous vorticity and density profiles, which, for many applications, is more appropriate than 2D models, Taylor columns, and/or PV patches that are often used to simplify the numerical or analytical stability analysis; and performing the linear stability analysis on vortices that are exact equilibrium solutions of the full 3D non-hydrostatic Boussinesq equations. 

The results also have implications for the two problems that have motivated many studies of vortex stability in the past: the observed stability of long-lived, axisymmetric vortices in the oceans and the observed predominance of anticyclones over cyclones in the oceans (at the mesoscales) and planetary atmospheres (see \S\ref{sec:introduction} for more details). As described above, while neutrally-stable vortices are found only in a very small region of the $Ro-Bu$ parameter space, the maximum (linear) growth rates in a large region of the parameter space, particularly for anticyclones, are small compared to the vortex turn-around time, which means that these vortices can remain nearly axisymmetric for months and even years despite being linearly unstable. This might explain the observations of long-lived axisymmetric vortices in the oceans, given that the slowly-growing non-axisymmetric flow can be difficult to detect in the satellite or ship-based observations and in time-averaged measurements (but also see the next two paragraphs for several caveats). Furthermore, we found the region of slow growth rates for anticyclones to be much larger than that of the cyclones; whether this offers an explanation for the observed cyclone-anticyclone asymmetry in the oceans (at the mesoscales) and atmospheres requires further studies (see below). 

Of course for both problems, the nonlinear stability and nonlinear evolution of these vortices are very important as well, and will be the subject of a subsequent publication. In particular, we will discuss that small linear growth rate  is neither a necessary nor a sufficient condition for a vortex to survive long to be observed. It is not necessary because our nonlinear simulations show that vortices with eigenmodes with very fast growth rates can have very large Landau coefficients \citep{drazin2004hydrodynamic}. Thus, even though the original Gaussian vortex becomes quickly unstable, the instability quickly saturates, and a new equilibrium that looks very similar to the initial unstable Gaussian vortex is established. A slow (linear) growth rate of the fastest-growing eigenmode is not sufficient because the equilibrium vortex may be hard to create from realistic initial conditions, or because nonlinear, finite-amplitude instabilities destroy it.  

The limitations and several important caveats of our analysis, discussed in \S\ref{sec:introduction}, should be again emphasized. The exclusion of background shear, compressible effects, and vertical variation of $\bar{N}$ limit the direct application of the results to vortices in the atmospheres and protoplanetary disk, while using an unbounded domain (hence the absence of free surface, bottom topography, lateral boundaries) and vertical variation of $\bar{N}$ limit the direct applicability of the current analysis to most oceanic eddies. The results are most relevant to the stability of interior oceanic vortices such as Meddies. Still, while our results for stability properties and slow growth rates might explain the observations of long-lived nearly axi-symmetric Meddies, our results for cyclone-anticyclone asymmetry are not relevant to the dominance of anticyclones among Meddies, which has been suggested to be a result of how Meddies form \citep{McWilliams1985}. 

Nonetheless, the results of this paper provide a steppingstone to study the more complicated problems of the stability of geophysical and astrophysical vortices, and the framework developed here can be readily extended to include further complexities such as the meridional dependence of $f$ (i.e., the $\beta$-effect), compressible effects (e.g., by using the anelastic approximation), and the $z$-dependence of $\bar{N}$, for example to account for the thermocline. The framework can be also extended to study the linear and nonlinear stability of vortices in rotating stratified shearing flows such as Jovian vortices, vortices in protoplanetary disks, and oceanic eddies in the Gulf Stream and Antarctic Circumpolar Current. For example, planetary anticyclones on Jupiter appear to have $|Ro| < 0.3$ and $Bu \sim 1$, which gives them a very slow linear growth rate of instability (according to figure~\ref{f2}). Understanding how the Jupiter's strong shear influences the growth rate and the most unstable eigenmode is of great interest and can be studied in the modified framework. 

\section*{Acknowledgments}  \label{sec:acknowledgments}
We thank three anonymous reviewers for insightful comments and suggestions. This work was supported in part by NSF grants AST-0905801 and AST-1009907, and by NASA PATM grants NNX10AB93G and NNX13AG56G. Part of the computational work used an allocation of computer resources from the Extreme Science and Engineering Discovery Environment (XSEDE), which was supported by National Science Foundation Grant No. OCI-1053575, and in part was supported by NASA-HEC. P. H. was supported by a Ziff Environmental Fellowship from Harvard University Center for the Environment.  

\appendix
\section{Numerical sponge layer}\label{appA}
To compute unbounded flows in a triply-periodic computational domain, we 
added an artificial  ``sponge layer" far from the vortices that were initially centered at the origin. This is accomplished by adding  Rayleigh drag and Newtonian cooling terms  in the form of $-f_{bd}\boldsymbol{v}$ and $-f_{bd}b$ to the right sides of the momentum and buoyancy equations in (\ref{eq:1}), respectively, where $f_{bd}$ is a function that smoothly varies from zero inside a cylindrical surface to a value of one outside of the cylinder, i.e.,
\begin{eqnarray}
f_{bd} = \left[1-T(z,L_{z,bd},s_z)T(r,L_{r,bd},s_r)\right]/\tau_{bd},
\end{eqnarray}
where $L_{r,bd}$ is the cylinder diameter, $L_{z,bd}$ is the height, $s_r$ and $s_z$ are the steepness in $r$ and $z$, $\tau_{bd}$ is the damping time scale, $r=(x^2+y^2)^{1/2}$, and
\begin{eqnarray}
T(\gamma,w,s) \equiv 1/2\left(\tanh\left[(\gamma+w)/s\right]-\tanh\left[(\gamma-w)/s\right]\right),
\end{eqnarray}
is top hat function. $T$ smoothly drops from a value of 1 to 0 for $|\gamma|>w/2$ over a distance $s$. We use $\tau_{bd}=20\Delta t$, $s_{r,bd}=0.01(L_x^2+L_y^2)^{1/2}$, $L_{r,bd}=0.85(L_x^2+L_y^2)^{1/2}$, $s_{z,bd}=0.01L_z$, and $L_{z,bd}=0.85L_z$ for the numerical calculations that are carried out here.

\section{Definitions of {\it shield} and {\it core} }\label{appB}

We qualitatively defined {\it core} and {\it shield} in $\S$\ref{subsec:initial_equilibria}.
To prevent our definitions of the {\it core} and {\it shield} from including weak-amplitude vorticity that is far from the vortex itself, we need
to define ``cut-off'' values in order to exclude regions with low-amplitude vorticity. For a cyclone, we define the {\it core} as the contiguous cyclonic region  that includes the vortex center where 
$\omega$ is greater than a cut-off value of $0.01 \Omega_{max}$,  where $\Omega_{max}$ is the maximum vorticity of the vortex. The {\it shield} is defined as the region(s) where $\omega < 0$ and $|\omega| > 0.01 |\Omega_{min}|$, where $\Omega_{min}$, is the minimum value of $\omega$ in the vortex. Our choice of $0.01$ in these two cut-off values  is arbitrary, but the computed values of the enstrophies $S_{core}$ and $S_{shield}$ are insensitive to the exact choice 
of cut-off value because the integrands  in the definitions ((\ref{eq:S1}) and (\ref{eq:S2})) are, by definition, very small in regions where $\omega$ is near the cut-off value. The major influence of the choice of cut-off value is 
qualitative and aesthetic as in figure~\ref{f1}(b). With a bad choice of cut-off value, the  core and/or shield can  extend outward  toward infinity (and therefore do not look like
our intuitive pictures of what  a ``core'' and ``shield'' should look like).

\section{Eigenmode solver and symmetrizer}\label{appC}

We calculate  the fastest-growing eigenmodes of the vortices by modifying our initial-value code into a ``power method'' analogous to the iterative method  used for finding the eigenvector of a matrix whose eigenvalue has the  greatest absolute value \citep{press2007numerical}, but we do not use the pre-conditioners developed by \citet{Tuckerman1988} to speed-up  convergence. Rather, we use a spatial symmetrizer to speed up convergence. The rate of convergence of the power method to the fastest-growing eigenmode depends on the difference between the growth rate of the fastest-growing eigenmode and the growth rate of the second fastest-growing eigenmode. By examining only one spatial symmetry class at a time, we generally increase the difference between the growth rates of the fastest-growing and second fastest-growing eigenmodes, and thereby obtain faster convergence.

The easiest way to limit the solutions of the eigenmode solver to modes that are symmetric or anti-symmetric in $z$ is to limit the initial-value solver used in the power method to those symmetries. Using our spatially triply periodic code, the $z$-dependence of the solutions are represented here with Fourier modes $e^{i 2 \pi k z/L_z}$, where $-L_z/2 \le z < L_z/2$, and where $k$ is an integer. Therefore, it is  easy to compute ``$z$-symmetric'' solutions, where $v_x$, $v_y$, and $p$ are symmetric about $z=0$ and $\rho$, $b$, and $v_z$ are anti-symmetric about $z=0$ by restricting the former three variables  to a cosine series $\cos(2 \pi k z/L_z)$ and the latter three variables to a sine series $\sin(2 \pi k z/L_z)$. For ``$z$-anti-symmetric'' solutions we swap sines with cosines. 

When computing solutions in a cylindrical coordinate system $(r, \phi, z)$ with a spectral code, it is trivial to restrict solutions to have only one value of azimuthal wave number $M$ along with its harmonics. With a spectral method, the velocity, pressure, buoyancy, and density are each represented with a truncated series of  basis functions in which the $\phi$ dependence is expressed in terms of Fourier modes $e^{i m \phi}$, and the $r$ dependence is expressed in terms of the eigenmodes of a Sturm-Liouville equation chosen such that the truncated series converges exponentially and such that all of the basis functions are analytic at the origin (see, for example, the spectral expansions used by \citeauthor{Matsushima1995} \citeyear{Matsushima1995} and by \citeauthor{matsushima1997spectral} \citeyear{matsushima1997spectral}). Solutions can be forced to be $M$-fold symmetric in $\phi$ about the $z$-axis by restricting the basis functions $e^{im\phi}$ in the spectral expansion to
wave numbers $m$ that are divisible by $M$.

However, because we plan to add Cartesian shear to our future calculations, say, for example to represent the Great Red Spot of Jupiter embedded in a shearing zonal flow, we chose here to compute in  Cartesian, rather than cylindrical, coordinates. None the less, it is still possible to force solutions to have only azimuthal wave numbers that are odd, or that are even and divisible by 4, or that are even and not divisible by 4. We can do this efficiently when the grid of collocation points of the Fourier modes in the horizontal direction is made of square cells and the horizontal computational domain is  square. In this case, the grid of collocation points is invariant under rotations of $90^{\circ}$ around the $z$-axis. To restrict the solution to azimuthal wave numbers that are even and divisible by 4 -- without interpolation (which causes errors), and without dividing or multiplying by $r$ (which is problematic near
the origin), we do the following operations after each time step of an initial value code: 
\begin{enumerate}
\item Compute $v_r$ and $v_{\phi}$ at each grid point from the values of $v_x$ and $v_y$ at the grid point. 
\item Compute a new value $v_{\phi}^{NEW}$ at each grid point $(x,y, z)$ by ``averaging'' such that 
\begin{equation}
v_{\phi}^{NEW}(x, y, z) \equiv [v_{\phi}(x, y, z) + v_{\phi}(-y, x, z) + v_{\phi}(-x, -y, z) + v_{\phi}(y, -x, z)]/4. \label{4}
\end{equation}
\item Do the same type of averaging to create new values $v_r^{NEW}$, $v_z^{NEW}$, $\rho^{NEW}$, $b^{NEW}$, and $p^{NEW}$. 
\item Compute $v_{x}^{NEW}$ and $v_{y}^{NEW}$ at each grid point from $v_r^{NEW}$ and $v_{\phi}^{NEW}$ at the grid point. 
\item Compute the flow at the next time using the initial-value solver using the $NEW$ values of all of the variables.
\end{enumerate}

To  restrict the solution to azimuthal wave numbers that are even and {\it not} divisible by 4, we carry out the same procedure as above, but we replace the averaging in (\ref{4}) with
\begin{equation}
v_{\phi}^{NEW}(x, y, z) \equiv [v_{\phi}(x, y, z) - v_{\phi}(-y, x, z) + v_{\phi}(-x, -y, z) - v_{\phi}(y, -x, z)]/4. \label{not4}
\end{equation}

To  restrict the solution to azimuthal wave numbers that are odd, we carry out the same procedure as above, but we replace the averaging in (\ref{4}) with
\begin{equation}
v_{\phi}^{NEW}(x, y, z) \equiv [v_{\phi}(x, y, z) -  v_{\phi}(-x, -y, z)]/2. \label{odd}
\end{equation}

\section{Growth rate and symmetry of selected vortex eigenmodes}\label{appD}
The growth rate $\sigma$ and symmetry of the fastest-growing eigenmode of vortices with $\sigma>0.02$ ($\tau^{-1}$) and $N_c^2>0$ shown by symbols in figure~\ref{f2}(a) are presented in Table~\ref{tabC1}.

\renewcommand{\arraystretch}{1}
\begin{table}
 \begin{center}
  \begin{tabular}{llcccllcc}
$~~Ro$ & $Bu$ & Symmetry & $\sigma$       & $~~~~$  &      $~~Ro$ & $Bu$ & Symmetry & $\sigma$   \\[4pt]
$+0.5$   & $0.65$ & A1 & 0.26      &  &   $+0.1$  & $2.0$  & A1 & 0.081 \\
 $+0.5$   & $0.75$ & S2 & 0.21      &  &   $+0.1$  & $2.3$  & A1 & 0.091 \\
$+0.5$   & $1.0$  & S2 & 0.20       &  &    $+0.05$  & $0.05$  & A1 & 1.0  \\
$+0.5$   & $1.4$  & S2 & 0.18       &  &    $+0.05$  & $0.1$  & A1 & 0.44  \\
$+0.5$   & $1.6$  & S2 & 0.17       &  &    $+0.05$  & $0.125$  & A1 & 0.28   \\
$+0.5$   & $2.0$  & S2 & 0.16       &  &    $+0.05$  & $0.15$  & A1 & 0.16  \\
$+0.5$   & $2.3$  & A1 & 0.16      &  &     $+0.05$  & $0.25$  & S2 & 0.072 \\
$+0.45$ & $0.3$  & A3 & 1.5        &  &     $+0.05$  & $0.3$  & S2 & 0.067  \\
$+0.45$ & $2.3$  & A1 & 0.15      &  &     $+0.05$  & $0.4$  & S2 & 0.054  \\
$+0.4$   & $0.45$  & A1 & 0.4       & &     $+0.05$  & $0.6$  & S2 & 0.028  \\
$+0.4$   & $0.55$  & A1 & 0.24      & &    $+0.05$  & $1.4$  & A1 & 0.040   \\
$+0.4$   & $0.65$  & S2 & 0.18     &  &    $+0.05$  & $1.6$  & A1 & 0.054  \\
$+0.4$   & $0.75$  & S2 & 0.17     &  &    $+0.02$  & $0.02$  & A1 & 1.21  \\
$+0.4$   & $1.2$  & S2 & 0.14       & &     $+0.02$  & $0.05$  & A1 & 0.58  \\
$+0.4$   & $1.4$  & S2 & 0.13      &  &     $+0.02$  & $0.5$  & S2 & 0.029   \\
$+0.4$   & $1.6$  & A1 & 0.13      & &      $+0.02$  & $1.3$  & A1 & 0.025  \\
$+0.35$ & $0.65$  & S2 & 0.16     & &      $+0.02$  & $1.4$  & A1 & 0.034 \\
$+0.3$  & $0.25$  & A1 & 0.77     &  &     $+0.02$  & $1.6$  & A1 & 0.049 \\
$+0.3$  & $0.65$  & S2 & 0.13      & &      $-0.02$  & $0.05$  & S2 & 0.065   \\
$+0.25$ & $0.65$  & S2 & 0.11     & &      $-0.02$  & $0.4$   & S2 & 0.029 \\
$+0.25$ & $1.4$  & A1 & 0.082     & &     $-0.02$  & $1.4$  & A1 & 0.028 \\
$+0.25$ & $1.6$  & A1 & 0.092     & &     $-0.02$  & $1.6$  & A1 & 0.043 \\
$+0.2$  & $0.1$  & A4 & 2.9         & &      $-0.05$  & $0.15$  & S2 & 0.053 \\
$+0.2$  & $0.15$  & A2 & 1.1      & &      $-0.05$  & $0.25$  & S2 & 0.041 \\
 $+0.2$  & $0.18$  & A1 & 0.76   & &       $-0.05$  & $0.3$  & S2 & 0.032 \\
$+0.2$  & $0.225$  & A1 & 0.54   & &      $-0.05$  & $1.4$  & A1 & 0.024 \\
$+0.2$  & $0.26$  & A1 & 0.41    & &       $-0.1$  & $0.1$  & S2 & 0.048 \\
$+0.2$  & $0.3$  & A1 & 0.29      & &       $-0.1$  & $0.15$  & S2 & 0.041 \\
$+0.2$  & $0.45$  & S2 & 0.11      & &      $-0.1$  & $1.6$  & A1 & 0.032 \\
$+0.2$  & $0.55$  & S2 & 0.099     & &     $-0.15$  & $0.2$  & S2 & 0.023 \\
$+0.2$  & $0.65$  & S2 & 0.089     & &    $-0.18$  & $0.15$  & S2 & 0.024  \\
$+0.2$  & $0.75$  & S2 & 0.079     & &    $-0.2$  & $1.6$  & A1 & 0.023 \\
$+0.2$  & $0.85$  & S2 & 0.070     & &    $-0.2$  & $2.0$  & A1 & 0.042 \\
$+0.2$  & $1.0$   & S2 & 0.058     & &    $-0.2$  & $2.3$  & A1 & 0.051 \\
$+0.2$  & $1.2$  & A1 & 0.054     & &     $-0.3$  & $1.7$  & A1 & 0.021 \\  
$+0.2$  & $1.4$  & A1 & 0.070    & &     $-0.35$  & $1.75$  & A1 & 0.021 \\  
$+0.2$  & $1.6$  & A1 & 0.081     & &    $-0.4$  & $1.8$  & A1 & 0.021 \\ 
$+0.2$  & $2.0$  & A1 & 0.097     & &    $-0.4$  & $2.0$  & A1 & 0.028 \\   
$+0.15$  & $1.0$  & S2 & 0.033    & &   $-0.4$  & $2.3$  & A1 & 0.035 \\   
$+0.13$  & $1.06$  & A1 & 0.02  & &    $-0.495$  & $1.5$  & A1 & 0.020 \\ 
$+0.1$  & $0.6$  & S2 & 0.049   & &     $-0.495$  & $1.7$  & A1 & 0.029\\  
$+0.1$  & $0.8$  & S2 & 0.027   & &     $-0.495$  & $1.9$  & A1 & 0.036  \\
$+0.1$  & $1.6$  & A1 & 0.062  & &     \\ 
  \end{tabular}
  \caption{The growth rate $\sigma$  (in units of $\tau^{-1}$) and symmetry of the fastest-growing eigenmode of vortices with $\sigma>0.02\:\tau^{-1}$  and $N_c^2>0$ shown by symbols in figure~\ref{f2}(a).}{\label{tabC1}}
 \end{center}
\end{table}

\newpage

\bibliographystyle{jfm}
\bibliography{Final6}

\end{document}